\documentclass[fleqn,usenatbib]{mnras}

\usepackage{newtxtext,newtxmath}

\usepackage[T1]{fontenc}

\DeclareRobustCommand{\VAN}[3]{#2}
\let\VANthebibliography\thebibliography
\def\thebibliography{\DeclareRobustCommand{\VAN}[3]{##3}\VANthebibliography}

\usepackage{graphicx}	
\usepackage{amsmath}	

\usepackage{booktabs} 
\usepackage{longtable}

\title[Strong Lens Modelling of SPT-CLJ1150$-$2805]{SLICE: Unveiling the multi-component merging core of SPT-CLJ1150$-$2805 through strong-lensing mass modelling}

\author[J. R. Betts et al.]{
James R. Betts,$^{1, 2}$\thanks{E-mail: james.r.betts@durham.ac.uk}
Benjamin Beauchesne,$^{1, 2}$
Mathilde Jauzac,$^{3, 1, 2, 4, 5}$
David Lagattuta,$^{6, 1, 2}$
\newauthor
Guillaume Mahler,$^{7}$
Jean-Paul Kneib,$^{8}$
Gavin Leroy,$^{1, 2}$
Marceau Limousin,$^{9}$
Priyamvada Natarajan,$^{10, 11}$
\newauthor
Sameeksha Saini,$^{12, 1}$
Keren Sharon,$^{13}$
Stephane V. Werner,$^{1, 2}$
Catherine Cerny,$^{13}$
Isaque Dutra,$^{11}$
\newauthor
Dominique Eckert,$^{14}$
Alaina Einsig,$^{13}$
Michael D. Gladders,$^{15, 16}$
Gourav Khullar,$^{17}$ and
Manon Regamey$^{14}$
\\
$^{1}$Centre for Extragalactic Astronomy, Department of Physics, Durham University, South Road, Durham DH1 3LE, UK\\
$^{2}$Institute for Computational Cosmology, Department of Physics, Durham University, South Road, Durham DH1 3LE, UK\\
$^{3}$Univ Toulouse, CNES, CNRS, IRAP, Toulouse, France\\
$^{4}$Astrophysics Research Centre, University of KwaZulu-Natal, Westville Campus, Durban 4041, South Africa\\
$^{5}$School of Mathematics, Statistics and Computer Science, University of KwaZulu-Natal, Westville Campus, Durban 4041, South Africa\\
$^{6}$Centre for Astrophysics Research, Department of Physics, Astronomy and Mathematics, University of Hertfordshire, Hatfield AL10 9AB, UK\\
$^{7}$STAR Institute, Quartier Agora - Allee du six Ao\^ut, 19c B-4000 Liege, Belgium\\
$^{8}$Institute of Physics, Laboratory of Astrophysics, Ecole Polytechnique Federale de Lausanne (EPFL), Observatoire de Sauverny, Versoix, Switzerland\\
$^{9}$Aix Marseille Univ, CNRS, CNES, LAM, Marseille, France\\
$^{10}$Department of Astronomy, Yale University, New Haven, CT 06511, USA\\
$^{11}$Department of Physics, Yale University, New Haven, CT 06511, USA\\
$^{12}$Centre for Advanced Instrumentation, Department of Physics, Durham University, South Road, Durham, DH1 3LE, UK\\
$^{13}$Department of Astronomy, University of Michigan, 1085 South University Avenue, Ann Arbor, MI 48109, USA\\
$^{14}$Department of Astronomy, University of Geneva, Ch. d’Ecogia 16, CH-1290 Versoix, Switzerland\\
$^{15}$Kavli Institute for Cosmological Physics, University of Chicago, 5640 South Ellis Avenue, Chicago, IL 60637, USA\\
$^{16}$Department of Astronomy and Astrophysics, University of Chicago, 5640 South Ellis Avenue, Chicago, IL 60637, USA\\
$^{17}$Department of Astronomy, University of Washington, Physics-Astronomy Building, Box 351580, Seattle, WA 98195-1700, USA
}

\date{Accepted XXX. Received YYY; in original form ZZZ}

\pubyear{\the\year{}}

\begin{document}
\label{firstpage}
\pagerange{\pageref{firstpage}--\pageref{lastpage}}
\maketitle

\begin{abstract}
We present a new strong-lensing and X-ray mass model of the merging galaxy cluster SPT-CLJ1150$-$2805 ($z = 0.383$), combining \textit{JWST} NIRCam, \textit{HST}, \textit{Chandra}, and VLT/MUSE data to robustly decompose its collisionless and collisional components. Exploiting \textit{JWST}, we expand the multiple-image catalogue with 66 new images (30 comprising 6 new systems). Our baseline model uses 118 spectroscopically confirmed images from 20 systems, achieving an image-plane root-mean-square offset of $0.56^{\arcsec}$. An expanded model incorporating 31 additional photometric images from 8 additional systems yields a near-identical mass distribution, confirming excellent model stability. We resolve a complex multi-axis post-merger system requiring five cluster-scale dark matter haloes, including a bimodal core with distinct haloes coincident with each brightest cluster galaxy. The remaining mass components are independently supported by kinematics of 291 cluster members ($\sigma_v = 1960^{+84}_{-92}\,\mathrm{km\,s^{-1}}$), which reveal two new galaxy substructures - the northernmost of which we tentatively identify as the baryonic counterpart to our fifth dark matter halo. The cluster is an extraordinarily efficient lens, possessing an effective Einstein radius $\theta_{\mathrm{E}} = 42.41^{+0.05}_{-0.04}\arcsec$ and an enclosed mass $M(< \theta_{\mathrm{E}}) = 3.35^{+0.04}_{-0.03} \times 10^{14} M_{\odot}$ for a source at $z_s = 2$. For a high-redshift source at $z_s = 9$, it produces a magnification cross-section of $A(>10) \simeq 0.15~\mathrm{arcmin}^{2}$ and $A(>30) \simeq 0.020~\mathrm{arcmin}^{2}$. This global lensing power surpasses all \textit{Hubble} Frontier Fields clusters. In a regime where systematic uncertainties between mass models dominate high-redshift magnification errors, the stability of our reconstruction sets this lens apart. This combination of extreme lensing power and a robust mass model establishes SPT-CLJ1150$-$2805 as a premier cosmic telescope for high-redshift studies.
\end{abstract}




\section{Introduction}

Galaxy clusters constitute the most massive and recently assembled gravitationally bound structures in the observable universe \citep[][]{Kravtsov_2012, Planelles_2015}. Within the Lambda Cold Dark Matter ($\Lambda$CDM) cosmological paradigm, the formation of cosmic structure follows a hierarchical scenario, proceeding `bottom-up' as smaller structures coalesce to form larger ones over time \citep[e.g.][]{Zeldovich_1982, Springel_2005}. On the largest scales, the Universe is structured into a `cosmic web' composed of interconnected dark matter filaments, sheets, and expansive voids. Galaxy clusters, residing at the intersecting nodes of this web, represent the pinnacle of this evolutionary sequence. Their continuous growth via the accretion of surrounding matter and major mergers with smaller galaxy groups and substructures makes them exceptionally sensitive probes of the underlying cosmology \citep[][]{Jullo_2010, Acebron_2017}, the evolution of cosmic structure \citep{Vikhlinin_2009, Allen_2011}, and potential deviations from General Relativity \citep[][]{Clowe_2006, Lam_2012, Cataneo_2018}. Furthermore, in the context of $\Lambda \rm CDM$, they represent some of the best natural laboratories to study the fundamental nature of dark matter \citep[DM;][]{Navarro_1997,Natarajan_2002,Clowe_Douglas_Gonzalez_2004, Markevitch_2004, Natarajan_2017} and test for any divergences from the collisionless model \citep[][]{Harvey_2015, Meneghetti_2020, Limousin_2022}. Realising the full potential of these cosmic laboratories requires precise measurements of their total mass distributions, a requirement for which gravitational lensing stands as a unique tool.

Rooted in General Relativity, gravitational lensing occurs when a massive foreground object deflects the light emitted by background sources. In the context of massive galaxy clusters, this phenomenon typically manifests in two observable regimes. At larger radii, weak lensing induces subtle statistical distortions in the shapes of background galaxies and is mainly used to study galaxy clusters on a large scale \citep[][]{Bartelmann_2001, Hoekstra_2008, Jauzac_2012}. In contrast, within the extremely dense cluster core, the strong lensing regime dominates, producing highly magnified, tangentially sheared, and multiply-imaged background sources. Because the degree of light deflection depends exclusively on the gravitational potential of the lens, strong lensing provides a purely geometric and precise measurement of the total enclosed projected mass density, entirely independent of the cluster's dynamical state \citep[e.g.,][]{Bartelmann_2010, Kneib_and_Natarajan_2011,Natarajan+2024}. Consequently, strong lensing allows one to bypass the limiting assumptions that inherently complicate mass estimates in merging systems - specifically, the assumption of hydrostatic equilibrium required for X-ray analyses \citep[][]{Braspenning_2025} and the assumption of dynamical relaxation required for kinematic techniques \citep[][]{Ettori_2010}.

Despite the success of gravitational lensing in measuring the total mass in the cores of galaxy clusters, the individual contributions from their distinct physical components remain blended within this overall mass measurement \citep[][]{Kneib_and_Natarajan_2011, Treu_2010}. While dark matter dominates the mass content \citep[accounting for $\sim 80 - 85 \%$ of the total mass; e.g.,][]{Kravtsov_2012}, lensing alone cannot isolate its distribution; achieving this requires independent measurements of the baryonic components, specifically the intra-cluster medium (ICM), the intra-cluster light (ICL) and the stellar mass of cluster galaxies. To overcome this limitation, X-ray observations are utilised to map the thermal Bremsstrahlung emission of the hot ICM. Simultaneously, the deep optical photometry and spectroscopy inherently required for strong lensing analyses can be leveraged to catalogue cluster member galaxies and scale their individual masses. By combining these multi-wavelength datasets, one can effectively decompose the total mass budget into its distinct physical components: the large-scale dark matter halo, the hot gas, and the cluster member galaxies. In dynamically complex systems like merging clusters, mapping the spatial displacements between these components provides crucial constraints on the self-interaction cross-section of dark matter particles \citep[e.g.,][]{Markevitch_2004, Harvey_2015}. Conversely, in relaxed systems, comparing these multi-wavelength mass estimates offers a direct test of the hydrostatic equilibrium assumption \citep[e.g.,][]{Cerini_2023, Beauchesne2024}. Furthermore, requiring consistency across multiple independent datasets enables the construction of more robust and physically motivated mass models than any single method could achieve alone \citep{Bonamigo_2018, Granata_2022, Beauchesne2024}. 

Until recently, the benchmark for strong lensing analyses of massive galaxy clusters has been set almost exclusively by \textit{HST} imaging. However, the advent of \textit{JWST} has fundamentally revolutionised strong lensing analyses \citep[][]{Treu_2022, Bergamini_2023}. With its unprecedented spatial resolution and exquisite sensitivity in the near-infrared, \textit{JWST} can uncover fainter, higher-redshift lensed sources and resolve internal structures within giant arcs that were previously invisible or unresolved, allowing for tighter constraints on lens models \citep[e.g.,][]{Furtak_2023, Mahler_2023}. Capitalising on these capabilities is the Strong LensIng and Cluster Evolution (SLICE) programme \citep{Cerny_2026}, a large \textit{JWST} Cycle 3 Treasury survey (Programme ID: GO-5594; PI: G. Mahler) designed to map the dark and luminous components of massive galaxy clusters over $\sim8$ billion years of cosmic history. SLICE has obtained NIRCam imaging for a total of 124 galaxy clusters. Leveraging these new data, this work focuses on one of the standout targets of the SLICE programme: the massive, merging galaxy cluster PLCK G287.0+32.9 (also known as SPT-CLJ1150$-$2805, hereafter J1150$-$2805) at $z=0.383$. 

J1150$-$2805 was initially discovered in the Planck Early Release Catalogue, where it stood out as the second most significant newly detected cluster candidate via the Sunyaev-Zeldovich (SZ) effect \citep{Planck_Collaboration_2011}. Subsequent X-ray, optical and radio follow-up by \citet[][]{Bagchi_2011} confirmed it as an exceptionally hot ($T_{X}\sim13~\rm keV$), massive ($\rm M_{500}=1.5~\times~10^{15}~\rm M_{\odot}$) post-merger system. Radio observations have revealed a massive central radio halo bounded by giant asymmetric double relics \citep{Bagchi_2011, Bonafede_2014}, embedded within a record-breaking 20 million light year envelope of energized plasma \citep{Rajpurohit_2025}. X-ray analyses have identified internal shock structures and cold fronts, indicating that the cluster core has been violently reheated by recent merger activity \citep{Cerini_2023, Rajpurohit_2025}. In the optical regime, weak lensing analyses have corroborated evidence that the cluster hosts a very complex structure \citep[][]{Gruen_2014, Finner_2017}. The first strong lensing mass model of the cluster was developed by \citet[][]{Zitrin_2017} relying on \textit{HST} imaging. A more recent strong lensing analysis by \citet{DAddona_2024} updated the mass model through the addition of spectroscopic data from the Very Large Telescope's (VLT) Multi Unit Spectroscopic Explorer (MUSE). 

Building upon these foundational works, we present a new, state-of-the-art strong-lensing mass model of J1150$-$2805, leveraging newly acquired \textit{JWST} NIRCam imaging from the SLICE programme. The depth and resolution of these data allow us to substantially enlarge the strong-lensing constraint set, adding 66 newly identified multiple images including 6 previously unknown systems to the existing catalogue. To more accurately disentangle the cluster's collisionless and collisional mass components, we explicitly incorporate the ICM into our lensing reconstruction using archival \textit{Chandra X-ray Observatory} data, utilising the \textsc{LensTool} X-ray extension introduced by \citet{Beauchesne2024}. To rigorously assess the stability of our reconstruction, we build two parallel models: a baseline constrained purely by spectroscopically confirmed systems and an expanded model augmented with photometric candidates. With this framework, we resolve a complex post-merger potential with a bimodal core, and we exploit our multi-component modelling to place stricter constraints on the cluster's ongoing merger dynamics, which we independently cross-validate against the kinematics of its member galaxies. Finally, we characterise J1150$-$2805 as a gravitational telescope, quantifying its Einstein radius, magnification power, and galaxy-galaxy strong-lensing cross-section, and benchmarking these against the Hubble Frontier Fields \citep[HFF][]{Lotz_2017} to situate the cluster among some of the most powerful known cluster lenses.

The structure of this paper is as follows: in Section \ref{sec: Observations and Data}, we detail the multi-wavelength observations and our data reduction procedures. In Section \ref{sec: Mass Modelling}, we describe the methodology used to select cluster members, identify new strongly lensed multiple images, and construct our joint lensing and X-ray mass models. Finally, we present our results and discuss their physical implications in Section \ref{sec: Results and Discussion}. Throughout this paper, we assume a standard flat $\Lambda$CDM cosmology with $\Omega_m = 0.3$, $\Omega_\Lambda = 0.7$, and $H_0 = 70 \text{ km s}^{-1} \text{ Mpc}^{-1}$. At the redshift of J1150$-$2805 ($z = 0.383$), $1\arcsec$ corresponds to a physical scale of $5.23 \text{ kpc}$.

\begin{table*}
  \centering
  \caption{List of observations of SPT-CLJ1150$-$2805 available. In this work, we
  use the \textit{JWST}/NIRCam \textit{F}150\textit{W}2 and
  \textit{F}322\textit{W}2 imaging, the archival \textit{HST}
  \textit{F}475\textit{W}, \textit{F}606\textit{W} and \textit{F}814\textit{W}
  imaging, the VLT/MUSE integral field spectroscopy and the \textit{Chandra}
  X-ray imaging. Columns 1 and 2 give the right ascension and declination of the observed field in degrees. Column 3 lists the instruments and filters (or instrument configurations). Columns 4 and 5 give the total exposure time in seconds and the observation date, respectively; where a band was observed at more than one epoch, column 5 gives the first and last dates. For the VLT/MUSE pointings column 5 gives the date of the first observing block; each pointing was built from four observing blocks of 2760 s (five for the westernmost pointing) obtained between March 8th and May 3rd 2019, and column 4 gives the total integration time on target. Finally, columns 6 and 7 provide the programme ID and the PI respectively. For the \textit{Chandra} observations, column 6 lists the ObsID.}
  \label{tab:observations}
  \begin{tabular*}{\textwidth}{@{\extracolsep{\fill}}lllrlll}
    \hline
    RA (J2000) & Dec. (J2000) & Instrument/Filter & Exp. Time (s) & Obs. Date & Prog. ID & PI \\
    \hline
    \multicolumn{7}{l}{\textit{JWST}} \\
    177.7090458 & $-$28.0822361 & NIRCam/\textit{F}150\textit{W}2 & 1835.988 & 2025-07-01 & 5594 & Mahler \\
    177.7090458 & $-$28.0822361 & NIRCam/\textit{F}322\textit{W}2 & 1835.988 & 2025-07-01 & 5594 & Mahler \\
    \\
    \multicolumn{7}{l}{\textit{HST}} \\
    177.6950000 & $-$28.0773611 & WFC3/\textit{F}110\textit{W} & 5223.502 & 2016-05-18 & 14\,165 & Seitz \\
    177.7220833 & $-$28.0860000 & WFC3/\textit{F}110\textit{W} & 5223.502 & 2016-05-18 & 14\,165 & Seitz \\
    177.7133333 & $-$28.0835556 & ACS/\textit{F}475\textit{W}  & 2160.000 & 2016-08-03 & 14\,165 & Seitz \\
    177.7133333 & $-$28.0835556 & ACS/\textit{F}606\textit{W}  & 2320.000 & 2016-08-03 & 14\,165 & Seitz \\
    177.7133333 & $-$28.0835556 & ACS/\textit{F}814\textit{W}  & 4680.000 & 2016-08-03 & 14\,165 & Seitz \\
    \\
    177.7116754 & $-$28.0811611 & ACS/\textit{F}435\textit{W}  & 2125.000 & 2017-02-21 & 14\,096 & Coe \\
    177.7107576 & $-$28.0805343 & WFC3/\textit{F}105\textit{W} & 1361.740 & 2017-02-21, 2017-03-18 & 14\,096 & Coe \\
    177.7107576 & $-$28.0805343 & WFC3/\textit{F}125\textit{W} & 711.744  & 2017-02-21, 2017-03-18 & 14\,096 & Coe \\
    177.7107576 & $-$28.0805343 & WFC3/\textit{F}140\textit{W} & 711.744  & 2017-02-21, 2017-03-18 & 14\,096 & Coe \\
    177.7107576 & $-$28.0805343 & WFC3/\textit{F}160\textit{W} & 1961.745 & 2017-02-21, 2017-03-18 & 14\,096 & Coe \\
    \\
    177.7117261 & $-$28.0811471 & WFC3-IR/\textit{F}105\textit{W}   & 2504.039 & 2022-04-25, 2022-08-12 & 16\,729 & Kelly \\
    177.7116667 & $-$28.0811667 & WFC3-UVIS/\textit{F}606\textit{W} & 5320.000 & 2022-04-25, 2022-08-12 & 16\,729 & Kelly \\
    \\
    \multicolumn{7}{l}{VLT} \\
    177.6912800 & $-$28.0876310 & MUSE/WFM-AO-N & 13\,800.000 & 2019-04-01 & 0102.A-0640(A) & Mercurio \\
    177.7077220 & $-$28.0791500 & MUSE/WFM-AO-N & 11\,040.000 & 2019-03-08 & 0102.A-0640(A) & Mercurio \\
    177.7204330 & $-$28.0914320 & MUSE/WFM-AO-N & 11\,040.000 & 2019-05-02 & 0102.A-0640(A) & Mercurio \\
    \\
    \multicolumn{7}{l}{\textit{Chandra}} \\
    177.7090458 & $-$28.0822361 & ACIS-I & 59\,300.000 & 2015-08-17 & 17\,494 & Bonafede \\
    177.7090458 & $-$28.0822361 & ACIS-I & 54\,400.000 & 2015-11-17 & 17\,165 & Bonafede \\
    177.7090458 & $-$28.0822361 & ACIS-I & 20\,800.000 & 2015-11-24 & 17\,166 & Bonafede \\
    177.7090458 & $-$28.0822361 & ACIS-I & 32\,200.000 & 2016-03-22 & 17\,495 & Bonafede \\
    177.7090458 & $-$28.0822361 & ACIS-I & 28\,700.000 & 2016-03-23 & 18\,807 & Bonafede \\
    \hline
  \end{tabular*}
\end{table*}

\section{SPT-CLJ1150\texorpdfstring{$-$}{-}2805: Observations and Data} \label{sec: Observations and Data}

\begin{figure*}
	\includegraphics[width=\textwidth]{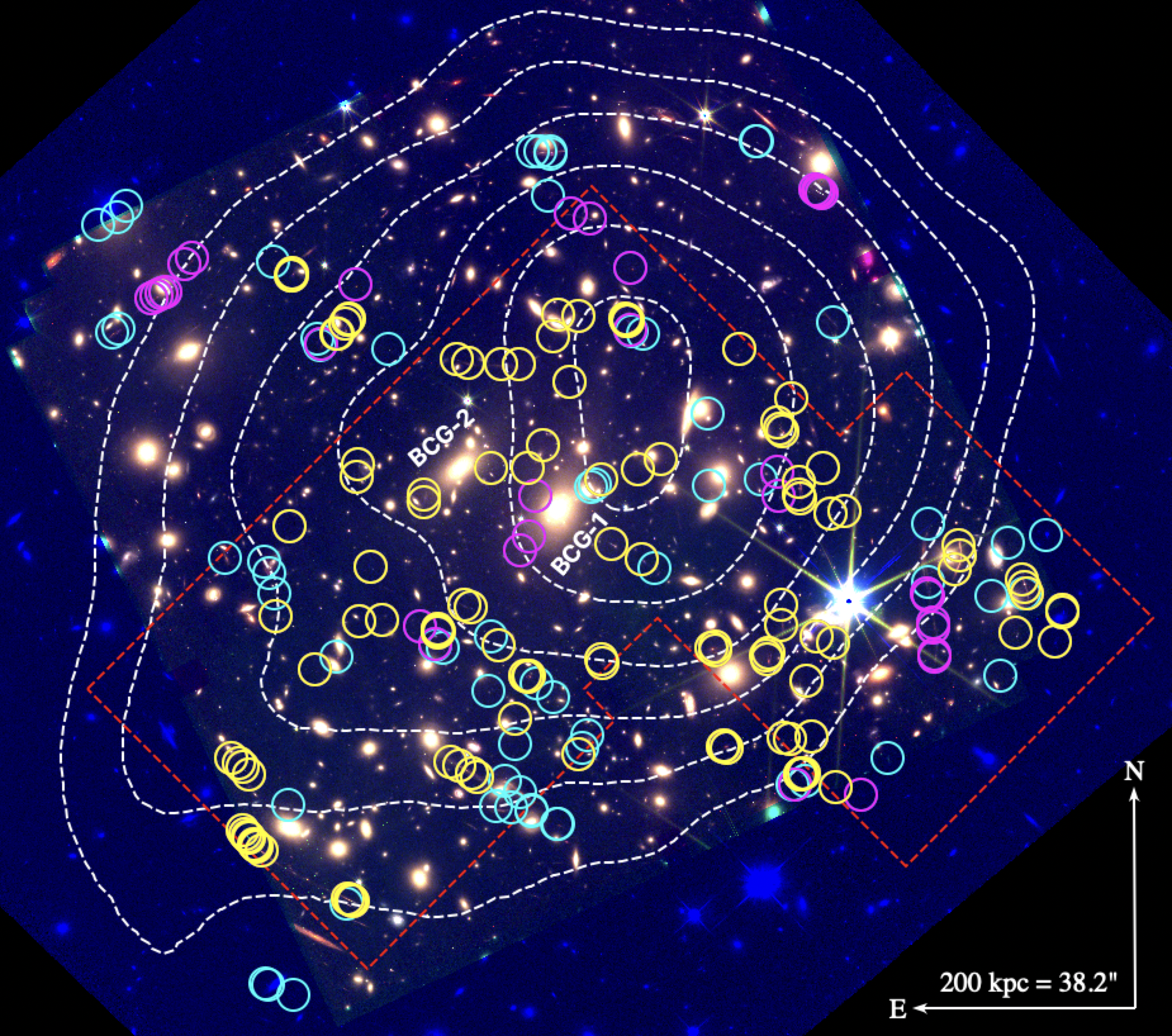}
    \caption{Colour-composite image of the J1150$-$2805 galaxy cluster core, created using \textit{JWST} F322W2 (red), F150W2 (green), and \textit{HST} F814W (blue) filters. The coloured circles mark the observed positions of multiple images. Yellow points denote the constraints used in our Spec model (spectroscopically confirmed), magenta points represent the supplementary images incorporated into our Phot model (not spectroscopically confirmed but confirmed through other means - see Section \ref{sec:multiple_images}), and cyan points indicate additional multiple image candidates that are excluded from the current modelling. White dashed contours display the X-ray surface brightness (smoothed over for visualisation using \textsc{SAOimage DS9} level 13 smoothing factor) as observed by the \textit{Chandra X-ray Observatory}. The red dashed polygon highlights the footprint of the MUSE spectral data cubes. The two BCGs are labelled with BCG-1 being the brightest in F322W2.}
    \label{fig:RGB_and_data}
\end{figure*}

Our strong lensing analysis of the J1150$-$2805 galaxy cluster is primarily driven by newly acquired high-resolution imaging from \textit{JWST}. To construct a robust and comprehensive mass model, we supplement these new observations with an extensive suite of archival data, including optical and near-infrared imaging from the \textit{Hubble Space Telescope} (\textit{HST}), integral field spectroscopy from VLT/MUSE, and X-ray imaging from the \textit{Chandra} X-ray Observatory. This section summarises these data and our reduction methodologies. We also provide a full log of observations of the cluster in Table \ref{tab:observations}.

\subsection{\textit{JWST} Observations}

J1150$-$2805 was observed with \textit{JWST} in Cycle 3 (Programme ID: GO-5594; PI: G. Mahler) using the Near Infrared Camera \citep[NIRCam;][]{Rieke_2023}. These observations were taken in both F150W2 and F322W2 simultaneously, covering the wavelength range between $1$ and $4~\mu\mathrm{m}$. Exposure times and observation dates are listed in Table~\ref{tab:observations}. 

The \textit{JWST} NIRCam data reduction largely follows the procedures outlined in \citet[][]{Cerny_2026}, which builds upon the methodology developed for the TEMPLATES Early Release Science programme \citep[][]{Rigby_2024}. Starting with uncalibrated raw data (Level 1b products) downloaded from MAST, the data are processed using a custom script adapted from the official \textit{JWST} Science Calibration Pipeline. In contrast to the reduction described in \citet[][]{Cerny_2026}, the processing relies instead on the updated \textit{JWST} pipeline build 12.0 and software version 1.19.1 \citep[][]{Bushouse_2025}. During the reduction process, a custom destriping algorithm is applied to the count-rate images (Level 2A data products) to accurately correct for correlated 1/f read noise and jumps between amplifiers. These corrected frames are then processed through the remainder of the pipeline to produce the final science-ready dithered mosaics. The resulting images are oriented North-up and East-left, with their World Coordinate System (WCS) astrometrically matched to the \textit{Gaia} reference frame \citep[][]{Gaia_Mission_2016, Gaia_Collaboration_2023}. To ensure a consistent spatial baseline across our entire analysis, all supplementary archival observations are astrometrically anchored to this \textit{JWST} reference frame. A false-colour composite image of the cluster, constructed from the final \textit{JWST} mosaics and archival \textit{HST} F814W data, is presented in Fig.~\ref{fig:RGB_and_data}.

\subsection{\textit{HST} Archival Observations}

J1150$-$2805 was previously observed with \textit{HST} as part of the Reionization Lensing Cluster Survey \citep[RELICS; P.I.: D. Coe,][Programme ID 14096]{Coe_2019}. RELICS built upon earlier Cycle 23 imaging of the cluster \citep[ P.I.: S. Seitz,][Programme ID 14165]{Seitz_2016}. The cluster was subsequently revisited by the Cycle 29 snapshot programme GO-16729 (P.I.: P. L. Kelly); we list these data in Table~\ref{tab:observations} for completeness but do not use them here, as the WFC3/UVIS \textit{F}606\textit{W} imaging is considerably shallower than the RELICS data. Filters, exposure times, observation dates, programme IDs and pointing centres for all archival \textit{HST} observations are given in Table~\ref{tab:observations}. The reduced RELICS data are provided on the Mikulski Archive for Space Telescopes (MAST)\footnote{\url{https://archive.stsci.edu/prepds/relics/}} at two distinct pixel scales, 30 mas and 60 mas. To complement our \textit{JWST} observations, we utilised the reduced 30 mas \textit{HST} optical imaging in the F475W, F606W, and F814W filters, which we astrometrically aligned to the NIRCam reference frame. These specific bands were chosen for their larger field of view compared to the NIRCam footprint, allowing for a wider spatial mapping of the cluster, as shown in Fig. \ref{fig:RGB_and_data}.

\subsection{Spectroscopic Data}

J1150$-$2805 was observed with VLT/MUSE and the \textit{Adaptive Optics Facility} (\textit{AOF}) working in Wide Field Mode \citep[][]{Arsenault_2008, Strobele_2012} over nine nights between 2019 March 8 and May 3 (P.I.: A. Mercurio, ESO programme 0102.A-0640(A)).

These observations cover the centre of the galaxy cluster, as shown in Fig. \ref{fig:RGB_and_data}. The pointings cover a total area of $\sim 3~\mathrm{arcmin}^2$. We reduced the data frames using the methodology outlined in Section 2.2 of \citet[][]{Lagattuta_2022}. This procedure largely follows the standard MUSE data reduction pipeline \citep[][]{Weilbacher_2020} but incorporates an additional auto-calibration step to mitigate flux variations between integral field unit (IFU) image slices and also applies the Zurich Astrophysical Purge (\textsc{ZAP}) software \citep{Soto_2016} to more effectively suppress residual sky emission lines. The final reduced cubes each have a mean spectral resolution of $R = 3000$, with total integration times of $11040~\mathrm{s}$ for the central and eastern pointings and $13800~\mathrm{s}$ for the westernmost pointing (Table~\ref{tab:observations}).

Although we independently reduced the archival MUSE observations, we do not present a new spectroscopic catalogue in this work. Instead, we utilise the catalogue produced recently by \citet[][]{DAddona_2024} from these data. We used our independently reduced data cubes to verify the spectroscopic redshifts of known multiple-image systems, to reassess existing configurations in light of the new \textit{JWST} observations, and to search for previously undetected counter-images.

\subsection{X-ray Data}

J1150$-$2805 was observed by the \textit{Chandra} X-ray Observatory between August 2015 and March 2016 (P.I.: A. Bonafede) in five ACIS-I pointings totalling $195.4$~$\rm ks$ (Table~\ref{tab:observations}). The observations are available in the \textit{Chandra} Data Collection (CDC) $584$~\href{https://doi.org/10.25574/cdc.584}{doi:10.25574/cdc.584}. The data have been reduced and analysed following the procedure detailed in \citet{Beauchesne2024} with \textsc{CIAO}\footnote{\url{https://cxc.cfa.harvard.edu/ciao/}} $4.17$ \citep{ciao} and \textsc{CALDB} $4.12.0$. To measure the cluster temperature, we limit our analysis to the $[0.5,12.5]$~${\rm keV}$ energy range, and we use a hydrogen column density, $n_{\rm H}$, of $8.993\times10^{20}$~${\rm atoms/cm^2}$. We calculate $n_{\rm H}$ based on the neutral hydrogen ($n_{\rm H,I}$) measurement from \citet{HI4PI} where we have accounted for molecular hydrogen following \citet{Willingale2013}. To visualise the macroscopic gas distribution in Fig. \ref{fig:RGB_and_data}, we overlay X-ray surface brightness contours onto the optical data. Note that these contours have been smoothed for display purposes.

\section{Mass Modelling} \label{sec: Mass Modelling}

In this section, we describe the selection of cluster members and multiple images, as well as the strong lensing and X-ray mass modelling for the cluster. We used the publicly available parametric lens mass modelling software \textsc{LensTool}\footnote{\url{https://projets.lam.fr/projects/lenstool}} \citep[][]{Kneib_1996,Jullo_2007} which adopts a Bayesian approach to mass model optimisation. 

\subsection{Cluster Members} \label{sec:cluster_members}

\begin{figure*}
	\includegraphics[width=\textwidth]{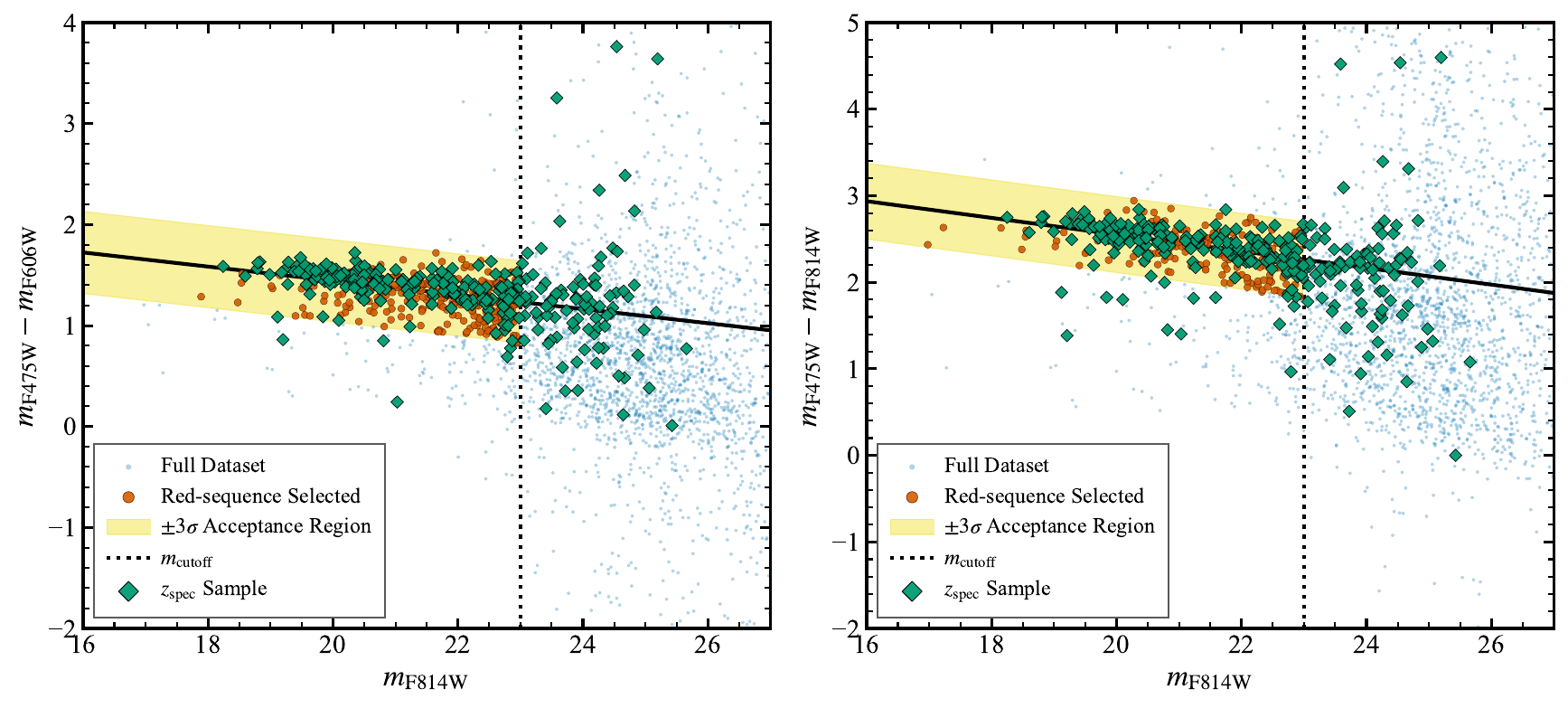} 
    \caption{Left panel: colour-magnitude diagram of ($m_{\mathrm{F475W}} - m_{\mathrm{F606W}}$) versus $m_{\mathrm{F814W}}$. Green squares represent the spectroscopically confirmed sample of cluster members while the orange large circles show the red-sequence selected galaxies (excluding the spectroscopic sample). Small blue points represent the full photometric dataset generated from \textsc{SExtractor}. The yellow shaded region marks the $\pm3\sigma$ selection boundaries around the fitted red-sequence slope (solid black line) from the final iteration of the sigma-clipping algorithm, and the vertical dashed black line indicates the apparent magnitude limit of $m_{\mathrm{F814W}}=23$. Right panel: the same as the left panel but for the colour-magnitude diagram of ($m_{\mathrm{F475W}} - m_{\mathrm{F814W}}$) versus $m_{\mathrm{F814W}}$.}
    \label{fig:red_sequence}
\end{figure*}

We construct the cluster member catalogue by combining spectroscopic confirmation with a photometric selection based on the red-sequence technique of \citet[][]{Gladders_2000}. For our red-sequence fitting, we utilise the \textit{HST} filters F475W, F606W, and F814W. At the cluster redshift (${z_{c} = 0.383}$), these filters optimally bracket the 4000 \AA\ break, with the F475W band sampling the spectral energy distribution blueward of the break, and the F606W and F814W bands sampling the continuum redward. The photometric catalogues for each of these filters were generated using \textsc{SExtractor} \citep[][]{Bertin_1996} in dual-image mode, with the F814W band serving as the detection image.

Using the spectroscopic catalogue provided by \citet[][]{DAddona_2024}, we identify $N_{\rm CM} = 291$ cluster members within their specified redshift range of ${0.360 \leq z_{c} \leq 0.405}$. For further validation, we ran our own iterative bi-weight sigma clipping algorithm \citep[][]{Beers_1990} and found the same result. This, therefore, corresponds to a peculiar velocity range of $-5000 \le v_{{pec}} \le +4700$ $\mathrm{km~s^{-1}}$ relative to the cluster rest-frame centered at $z_{c} = 0.383$. To initialise our red-sequence fit, we first determine the best-fit linear relation for the spectroscopically confirmed members in the F475W-F606W and F475W-F814W colour spaces. This provides a robust fit for the red-sequence slope and intercept, which we then use to identify cluster members within the broader photometric catalogue. To ensure a robust census of the cluster population, we apply two cleaning steps to the photometric catalogues. First, we utilise the \textsc{SExtractor} \textsc{CLASS\_STAR} stellarity index to remove likely foreground stars. Second, we reject any candidates with existing spectroscopic redshifts  \citep[determined using the spectroscopic catalogue from][]{DAddona_2024} falling outside the cluster redshift range defined at the beginning of this section. To construct our final cluster member sample for the lens model, we impose an apparent magnitude cut such that $m_{{F814W}} \leq 23$. At the cluster redshift, the characteristic magnitude is estimated to be ${m^{*}_{\rm F814W} \approx 19.8}$ \citep[based on passive galaxy evolution models; e.g.,][]{Mancone_2010, Hennig_2017}. Our selection, therefore, reaches a depth of ${m^{*}_{\rm F814W} +3.2}$, ensuring a high level of completeness for the member population while excluding background noise. We find final red-sequence slopes of $-0.070 \pm 0.006$ and $-0.096 \pm 0.007$ for the F475W-F606W and F475W-F814W colour spaces respectively. The corresponding intercepts are $2.85 \pm 0.13$ and $4.48 \pm 0.15$.

Fig. \ref{fig:red_sequence} presents the colour-magnitude diagrams for the ($m_{\mathrm{F475W}} - m_{\mathrm{F606W}}$) and ($m_{\mathrm{F475W}} - m_{\mathrm{F814W}}$) colour spaces. In each panel, the yellow shaded region indicates the $3\sigma$ red-sequence boundaries, with the photometrically selected members shown as orange large circles and the spectroscopic member sample overlaid as green squares. To maximize sample purity, we perform an independent selection in both colour spaces and retain only those sources satisfying the red-sequence criteria in both diagrams simultaneously. The ($m_{{F475W}} - m_{{F606W}}$) selection yields $441$ candidates, while the ($m_{{F475W}} - m_{{F814W}}$) cut identifies $379$. This dual-colour selection strategy minimises contamination from interlopers that serendipitously align with the red-sequence but are not true cluster members. Our final cluster member catalogue thus comprises $N_{\rm CM} = 317$ galaxies after cross-matching the candidates. 

To assign luminosities and morphological parameters to these galaxies for the lens model, we cross-match our selection with the \textit{JWST}/NIRCam F322W2 \textsc{SExtractor} catalogue. Using this infrared imaging is advantageous because it traces the rest-frame near-infrared, providing a more reliable proxy for total stellar mass than optical imaging. As the NIRCam field of view is smaller than our \textit{HST} footprint, we construct two separate catalogues: one utilising F322W2 parameters for members within the NIRCam footprint, and a second utilising \textit{HST} F814W parameters for the remaining peripheral members. In our mass modelling, these are implemented as two independent populations, each with its own scaling relation. This approach ensures that the inner cluster potential is constrained by the most direct mass tracers available, whilst incorporating the large-scale cluster environment without introducing biases from mixing two wavebands.

We note that our use of the red-sequence introduces a colour-based selection bias, excluding blue cluster members from our final catalogue. However, massive red ellipticals overwhelmingly dominate the local strong lensing deflection field. The omission of these typically lower-mass blue galaxies is, therefore, expected to have a negligible impact on the final mass reconstruction \citep[][]{Richard_2014}.

\subsection{Multiple Images} \label{sec:multiple_images}

Gravitationally lensed multiple image positions serve as the primary constraints for our modelling of the total cluster mass. We build upon multiple image catalogues reported in previous strong lensing analyses of this cluster \citep[][]{Zitrin_2017,DAddona_2024} by leveraging the unprecedented sensitivity of \textit{JWST} NIRCam imaging which allows us to probe the lens plane to greater depths and at higher resolution than previously possible \citep[][]{Cerny_2026}. New multiple images have been identified through an iterative process combining visual inspection with model predictions. Candidate systems exhibiting consistent morphologies, matching colours, and identical MUSE redshifts within measurement uncertainties were iteratively tested within lens models to ensure their geometry and parity inversions agreed with the predicted lensing configuration. All of the spectroscopically confirmed systems were previously presented in \citet[][]{DAddona_2024}, however, in some cases, images of lensed galaxies are so highly magnified that we can resolve several compact, knot-like, star-forming regions within them. Such regions are, themselves, strongly lensed and are seen in all images of their host galaxy. With our \textit{JWST} data, these features are much more prevalent among a number of already identified systems, and, in our analysis, we treat each knot as an independent source belonging to the same system, as was also done by \citet{Cerny_2026}. Multiple images of the same knot are henceforth termed a ‘clump’, such that a single system may contain multiple families, each providing a constraint on the lens deflection field. Indeed, for a given source at an intrinsic angular position, $\boldsymbol{\beta}$, the positions of its multiple images, $\boldsymbol{\theta_{\mathrm{i}}}$, must satisfy the lens equation,

\begin{equation}
    \boldsymbol{\beta} = \boldsymbol{\theta}_i - \frac{D_{\mathrm{LS}}}{D_{\mathrm{S}}} \boldsymbol{\alpha}(\boldsymbol{\theta}_i),
    \label{eq:lens_equation}
\end{equation}

where $\textit{}{D_{\mathrm{S}}}$ and $\textit{}{D_{\mathrm{LS}}}$ are the angular diameter distances between the observer and the source, and the lens and the source respectively, and $\boldsymbol{\alpha}(\boldsymbol{\theta}_{\mathrm{i}})$ is the lens deflection angle at position $\boldsymbol{\theta_{\mathrm{i}}}$. 

To remain consistent with previous catalogues while simultaneously abiding by the naming convention defined in the SLICE collaboration, we label multiple images using the format \textit{X.Y.Z}. Here, \textit{X} is an integer that identifies the background source system, \textit{Y} is an integer that indicates a specific clump, and \textit{Z} is an integer that distinguishes between the different multiple images of the system. We use the same \textit{X} integer as \citet[][]{DAddona_2024} to describe each system, such that multiple images named 5.1.1 and 5.1.2 in this work correspond directly to the images 5.1a and 5.1b in their work. 

Our catalogue contains 196 multiple images from 41 background sources corresponding to 52 separate clumps. This includes candidate systems which have no spectroscopic confirmation but have been identified based on other selection criteria as described earlier in this section. To systematically quantify the reliability of our strong lensing constraints, we classify all multiple image candidates using a three-tier ranking system, akin to that used by the Hubble Frontier Fields \citep[][]{Jauzac_2014, Jauzac_2016}. "Gold" images are defined as our most secure constraints, possessing robust spectroscopic redshifts and exhibiting unambiguous signs of a strongly lensed source. "Silver" images lack spectroscopic confirmation but demonstrate compelling photometric and geometric evidence, making them highly probable strong lensing features. Finally, "Bronze" systems represent lower-confidence multiple image candidates; they rely strictly on photometric data and often exhibit significant blending or less distinct parity, making their definitive classification less secure. The full catalogue of multiple images is provided in Appendix \ref{sec:multiple_images_appendix}.

We construct two distinct multiple image catalogues to act as strong lensing constraints for our mass modelling. The first consists strictly of our highest-confidence Gold images, utilising 118 of the 135 available candidates with secure spectroscopic confirmation. We hereafter refer to the model based on this catalogue as the "Spec model". The second configuration incorporates the same 118 Gold constraints but introduces an additional 31 Silver images which lack spectroscopic redshifts but are clearly lensed features, bringing the total to 149 multiple images. We refer to the model based on this catalogue as the "Phot model". These two models are not designed to be in competition, but rather to complement one another: the Spec model acts as our most robust baseline, while the Phot model allows us to explore the mass distribution with a higher density of spatial constraints.

\subsection{Lens Model} \label{sec:lens_model}

\textsc{LensTool} is a parametric lens modelling algorithm, such that it assumes the lens plane can be constructed from a linear combination of individual parametric mass haloes. Therefore, we assume that the cluster's total gravitational potential can be decomposed into a sum of large-scale and small-scale components \citep[][]{Natarajan_1997}:

\begin{equation} 
\label{eq:total_gravitational_potnetial}
    \phi_{\mathrm{tot}} = \sum_{i=1}^{N_{DM}} \phi_i^{\mathrm{DM}} + \sum_{i=1}^{N_{ICM}} \phi_i^{\mathrm{ICM}}  + \sum_{i=1}^{N_{CM}} \phi_i^{\mathrm{CM}},
\end{equation}

where, from left to right, the terms represent the contributions from: the cluster-scale DM distribution; the intra-cluster medium (ICM) - the hot, tenuous, and metal-enriched plasma that fills the cluster’s potential well \citep[][]{Kravtsov_2012}; the cluster member galaxies.

To determine the best-fitting configuration for these potentials, \textsc{LensTool} employs a Bayesian Markov Chain Monte Carlo (MCMC) sampler \citep[][]{Jullo_2007} to explore the parameter space. It does this by sampling the posterior distribution, where the likelihood is defined by the positional $\chi^2_{\rm tot}$ between the observed and predicted multiple images in the image plane:

\begin{equation}
    \chi^2_{\rm tot} = \sum_{j=1}^{N_{\mathrm{sys}}} \sum_{i=1}^{n_{j}} \frac{\left| \boldsymbol{\theta}_{i,j}^{\mathrm{obs}} - \boldsymbol{\theta}_{i,j}^{\mathrm{pred}} \right|^2}{\sigma_{i,j}^2},
\end{equation}

where $N_{\mathrm{sys}}$ is the total number of multiple-image systems, $n_{j}$ is the number of images in the $j$-th system, $\boldsymbol{\theta}_{i,j}^{\mathrm{obs}}$ and $\boldsymbol{\theta}_{i,j}^{\mathrm{pred}}$ are the observed and model-predicted image positions, respectively, and $\sigma_{i,j}$ represents the positional uncertainty, in arcseconds, assigned to each image.

The number of degrees of freedom ($\rm D.o.F.$) available to the model is defined as the difference between the number of observational constraints and the number of free parameters ($N_{\mathrm{free}}$). Because the intrinsic source position for each system is unknown and implicitly solved for during the optimisation, each system provides $2(n_{j} - 1)$ independent constraints. The degrees of freedom are therefore calculated as:

\begin{equation}
\label{eq:degrees_of_freedom_eqn}
    D.o.F. = \sum_{j=1}^{N_{\mathrm{sys}}} 2(n_{j} - 1) - N_{\mathrm{free}}.
\end{equation}

Finally, to quantify the overall macroscopic precision of the reconstructed lens model, we compute the image-plane RMS positional offset:
\begin{equation}
    \Delta_{\mathrm{RMS}} = \sqrt{ \frac{1}{N_{\mathrm{tot}}} \sum_{j=1}^{N_{\mathrm{sys}}} \sum_{i=1}^{n_{j}} \left| \boldsymbol{\theta}_{i,j}^{\mathrm{obs}} - \boldsymbol{\theta}_{i,j}^{\mathrm{pred}} \right|^2 },
\end{equation}
where $N_{\mathrm{tot}}$ is the total number of multiple images utilised as constraints in the fit.

In the following sections, we detail the approaches taken to model each of the individual elements from Equation \ref{eq:total_gravitational_potnetial}.

\subsubsection{Cluster-scale Modelling} \label{sec:dark_matter_modelling}

We model the cluster-scale DM distribution using dual Pseudo-Isothermal Ellipsoidal mass density profiles \citep[dPIE;][]{Limousin_2005}. The dPIE is characterized by seven parameters including the sky position (RA and Dec), the position angle (\textit{$\theta$}), as measured counter-clockwise from the west direction, the ellipticity $e = (a^2 - b^2)/(a^2 + b^2)$, where \textit{a} and \textit{b} are the major and minor semi-axes, the velocity dispersion (\textit{$\sigma_{0}$}), and two scale radii, the core and the truncation radii (\textit{$r_{\rm{core}}$} and \textit{$r_{cut}$}). The mass density of the dPIE is:

\begin{equation}
        \rho(r) = \frac{\rho_0}{(1 + r^2/r_{\mathrm{core}}^2)(1 + r^2/r_{\mathrm{cut}}^2)},
\end{equation}

where \textit{${\rho_0}$} is the central density. As strong-lensing constraints are confined to the central region of the cluster, the truncation radius of the cluster-scale halo is observationally ill-constrained. To avoid parameter degeneracies, we fix \textit{$r_{cut}$} = 2200 kpc. This value represents a typical virial radius for a massive cluster \citep{Richard_2014} and places the truncation scale beyond the strong-lensing regime, ensuring this assumption does not bias our measurement of the core mass distribution.

\subsubsection{Intracluster Medium Modelling} \label{sec:intracluster_medium_modelling}

To incorporate the ICM into our mass model, we utilise the modelling techniques introduced to \textsc{LensTool} by \citet{Beauchesne2024}. While this approach assumes that the gas mass can be represented by a free-form sum of dPIE profiles - sharing the same parametrisation as earlier works by \citet{Bonamigo_2017, Bonamigo_2018} - it differs by performing a joint parameter inference rather than a two-step optimisation. All parameters of the dPIE potentials are free to vary except for the truncation radius, \textit{$r_\mathrm{cut}$}, which, similarly to the cluster-scale DM haloes, is set to 2200 kpc as the parameter is ill-constrained at large radii.

The first step in our approach is to construct a 3D gas density profile for the ICM based on both X-ray surface brightness and spectroscopic temperature and metallicity measurements from the \textit{Chandra} data. Consequently, this gas mass component can be derived via direct volume integration, completely independently of the strong lensing constraints. As discussed in \citet[][]{Beauchesne2024}, we consider a model of the X-ray gas to be reliable if the likelihood of the observed X-ray data lies within a $5\sigma$ credible interval (CI) of the expected likelihood for that model. The X-ray likelihood for each model is calculated using:

\begin{align}
    \mathcal{L}_{X-ray}=\prod_i \frac{\Gamma\left(k_i+\frac{\mu_i^2}{\sigma_{\rm X}^2}\right)}{k_i!\Gamma\left(\frac{\mu_i^2}{\sigma_{\rm X}^2}\right)}\left(\frac{\mu_i}{\mu_i+\sigma_{\rm X}^2}\right)^\frac{\mu_i^2}{\sigma_{\rm X}^2}\left(\frac{\sigma_{\rm X}^2}{\mu_i+\sigma_{\rm X}^2}\right)^{k_i}
\end{align}

where $k_i$ is the observed number of photon counts in the $i$-th bin, and $\mu_i$ is the predicted count from the model for that same bin. The parameter $\sigma_{\mathrm{X}}^2$ represents the systematic uncertainty associated with surface brightness fluctuations in the plasma \citep{Eckert_2017}, which is assumed to be constant across all bins. $\Gamma$ denotes the standard Gamma function, and the product is taken over all bins, $i$, which are assumed to be independent. To achieve a reliable mass model, we systematically explore multiple configurations, altering the number of dPIE potentials and adjusting the prior distributions of their free parameters. We perform this optimisation iteratively until the resulting likelihood falls within the required $5\sigma$ CI. 

Table \ref{tab:dpie_ci_summary} outlines the statistical viability of our mass models as a function of the number of included dPIE profiles. Irrespective of the total number of haloes, all successful models require mass components in two distinct regions to accurately reproduce the macroscopic gas distribution: a primary structure situated $\sim100$~kpc North-West of BCG-1, and a secondary component roughly $340$~kpc to its South-East. Both the 3-halo and 4-halo models satisfy our statistical criteria. Therefore, in the interest of parsimony, we adopt the simpler 3-halo configuration, as the marginal statistical improvement of a fourth halo does not justify the increased complexity.

\begin{figure*}
	\includegraphics[width=\textwidth]{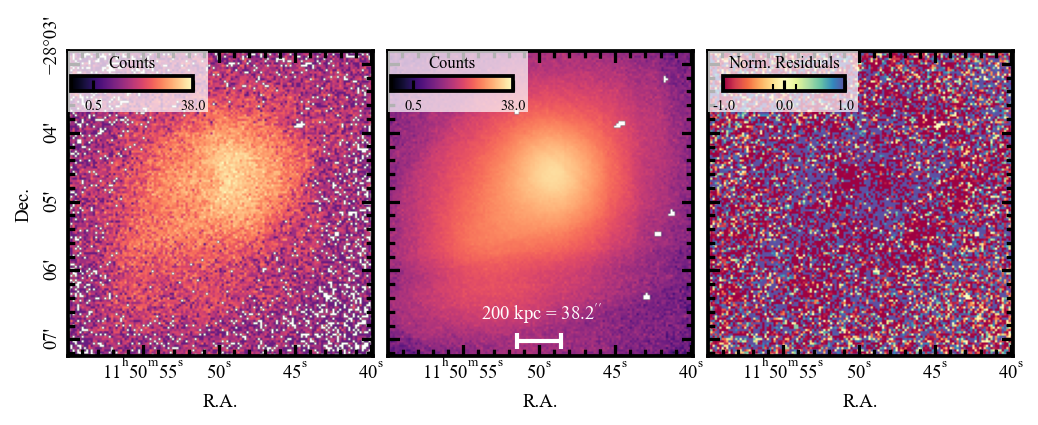}
    \caption{From left to right: 2D maps of the observed X-ray counts, the counts model from our best-fitting model, and the normalised residuals between the two previous maps (i.e. observation minus model). In both the observation and the model, the X-ray point sources have been masked out and excluded from the fit.}
    \label{fig:ICM_mass_model_plot}
\end{figure*}

Fig. \ref{fig:ICM_mass_model_plot} illustrates the resulting ICM mass model, comparing the observed X-ray data (left panel) with the best-fitting model (middle panel) and their normalised residuals (right panel). Furthermore, we opt to fix the ICM mass model during the subsequent strong lensing analysis rather than performing a simultaneous joint optimisation. This sequential approach is justified by \citet{Limousin_2026}, who recently demonstrated that both methods yield comparable results. This is further discussed in Section \ref{sec:impacts_of_icm_model}.

\begin{table}
    \centering
    \caption{Summary of the dPIE mass model configurations and their performance against the X-ray likelihood credible interval (CI) criteria. A checkmark (\checkmark) indicates the optimised model likelihood falls within the specified CI bound, while a cross ($\times$) indicates it does not.}
    \label{tab:dpie_ci_summary}
    \begin{tabular}{lccc}
        \toprule
        Mass Model & \multicolumn{3}{c}{CI Bounds} \\
        \cmidrule(lr){2-4}
        & \textbf{$5\sigma$} & \textbf{$3\sigma$} & \textbf{$1\sigma$} \\
        \midrule
        2 Haloes & $\times$ & $\times$   & $\times$ \\
        3 Haloes & \checkmark & $\times$ & $\times$ \\
        4 Haloes & \checkmark & \checkmark & $\times$ \\
        \bottomrule
    \end{tabular}
\end{table}

\subsubsection{Cluster Member Galaxy Modelling} \label{sec:cluster_member_modelling}

In order to construct an accurate strong lensing model of J1150-2805, the perturbing effects of the galaxy-scale potentials within the cluster must also be considered. As with the cluster-scale haloes described in Sections \ref{sec:dark_matter_modelling} and \ref{sec:intracluster_medium_modelling}, we again use dPIE profiles to describe these potentials. In this case, the dPIE acts as an approximation for the total mass of each cluster member, encapsulating both the stellar and dark matter components within a single effective mass profile \citep[][]{Limousin_2007, Natarajan_2009, Richard_2014}. 

The primary difficulty in modelling cluster substructure is not only the lack of sufficient lensing constraints to independently fit the large number of cluster member galaxies, but more importantly, that these macroscopic constraints lack the local spatial leverage to probe the mass profiles of individual subhaloes. As such, it is necessary to strictly limit the degrees of freedom assigned to these small-scale deflectors.

The inclusion of cluster members adds significant structure to the lensing potential of the cluster. If located close enough, some may also perturb the locations of multiple images in their vicinity and can alter the multiplicity and shapes of the lensed images \citep[][]{Meneghetti_2020, Meneghetti_2022}. We, therefore, adopt two separate modelling techniques dependent on whether cluster members are close to or far from multiple image constraints. 

For cluster members located far from multiple images, we fix the central position, ellipticity and position angle of each dPIE to the corresponding values determined by \textsc{SExtractor} during the construction of our cluster member catalogue (See section \ref{sec:cluster_members}). We also make the assumption that their core radii are small ($0.001 \mathrm{kpc}$), effectively reducing the dPIE mass models to truncated isothermal profiles. Indeed, the inner density profiles of massive early-type galaxies (which dominate cluster cores) have consistently been shown to be consistent with being isothermal \citep[see e.g.][]{Koopmans_2009, Auger_2010, Newman_2013, Lyskova_2018, Wang_2019}. This leaves two parameters of the dPIE profiles remaining; the velocity dispersion, \textit{$\sigma_\mathrm{0}^\mathrm{CM}$}, and the truncation radius, \textit{$r_\mathrm{cut}^\mathrm{CM}$}. As is common practice for \textsc{LensTool} modelling \citep[][]{Kneib_and_Natarajan_2011}, we assume these parameters scale as a function of the Kron luminosity, \textit{L}, in a designated reference band, selected for being the most representative of the stellar content. \textsc{LensTool} implements such scaling relations as follows:

\begin{align} \label{eq: sigma_scaling_relation_equation}
    \sigma_{\mathrm{LT}} &= \sigma^\star_{\mathrm{LT}} \left( \frac{L}{L^\star} \right)^{\alpha}, \\ \label{eq: cut_radius_scaling_relation_equation}
    r_{\mathrm{cut}} &= r_{\mathrm{cut}}^\star \left( \frac{L}{L^\star} \right)^{\beta_{\mathrm{cut}}}, \\
    r_{\mathrm{core}} &= r_{\mathrm{core}}^\star \left( \frac{L}{L^\star} \right)^{\beta_{\mathrm{core}}}.
\end{align}

The parameters denoted by a star superscript serve as the baseline normalisation for the scaling relations, representing the velocity dispersion and radii that a cluster member would possess if its measured Kron luminosity matched the chosen reference luminosity (\textit{$L^{\star}$}). In practice, the luminosity ratio is calculated directly from the observed apparent magnitudes as $L/L^\star = 10^{-0.4(m - m^\star)}$, where $m^\star$ is the designated reference magnitude. $\sigma_{LT}$ is the velocity dispersion measured by \textsc{LensTool}, which relates to the central velocity dispersion of the dPIE model via:

\begin{equation}
    \sigma_0 = \sqrt{\frac{3}{2}}\sigma_{\mathrm{LT}}.
\end{equation}

We fix the velocity dispersion scaling exponent to \textit{$\alpha$} = $1/4$ \citep[consistent with the theoretical Faber-Jackson relation for early-type galaxies][]{Faber_and_Jackson_1976} and the radius scaling exponents to \textit{$\beta_{\mathrm{core}}$} = \textit{$\beta_{\mathrm{cut}}$} = $1/2$. By doing this, the model enforces a constant total mass-to-light ratio ($M/L$) for all cluster members, as the total mass of a dPIE profile scales as $M \propto \sigma^2 r_{\mathrm{cut}}$. Consequently, the vast majority of cluster members are described by just two parameters in total - \textit{$\sigma_\mathrm{0}^\mathrm{CM}$} and \textit{$r_\mathrm{cut}^\mathrm{CM}$}. 

As described in Section \ref{sec:cluster_members}, to account for the limited field of view of our \textit{JWST} F322W2/ F150W2 observations, we use \textit{HST} F814W photometry to trace cluster members in the outer core region and complete our catalogue. To prevent biases in the mass-to-light scaling caused by combining different photometric bands, we implement two independent cluster member catalogues in our \textsc{LensTool} models. Each catalogue uses the brightest member galaxy as the reference for the scaling relations (e.g. equations \ref{eq: sigma_scaling_relation_equation} and \ref{eq: cut_radius_scaling_relation_equation}), such that \textit{$m_{\mathrm{F322W2}}^{\mathrm{\star}}$} = 16.83 and \textit{$m_{\mathrm{F814W}}^{\mathrm{\star}}$} = 18.83. The optimised parameter values are presented in Table \ref{tab:optimised_model_parameters_table}, while the corresponding prior distributions are detailed in Appendix \ref{sec:input_parameters_and_priors} (Table \ref{tab:lens_params_priors}).

Cluster member galaxies located in close proximity to multiple image constraints exert localised deflections. If not explicitly accounted for, these perturbing potentials can introduce geometric tensions that degrade the overall mass reconstruction \citep[][]{Pignataro_2021}. To mitigate this, we decouple these proximal members from the global scaling relations, allowing their mass parameters ($\sigma_{LT}$ and $r_{cut}$) to optimise independently. We individually optimise 7 such members in the Spec model, and 8 in the Phot model. The selection of these independently optimised members is achieved through iteratively identifying regions where multiple images are poorly reproduced in the image-plane of preliminary models. In such regions, we decouple the nearest perturbing cluster members from the scaling relations. A galaxy remains decoupled in subsequent models only if freeing its mass parameters quantitatively improves both the local RMS and the global Bayesian evidence.

\subsubsection{Foreground Galaxy Considerations} \label{sec:foreground_galaxies}

We identify four spectroscopically confirmed foreground galaxies (three at $\mathrm{z} \sim 0.33$ and one at $\mathrm{z} \sim 0.35$) that locally perturb the lensing deflection field. As demonstrated by \citet{Chiriv_2018}, neglecting such line-of-sight (LOS) structures introduces systematic errors in the reconstruction of multiple images. To account for their gravitational influence, we project these galaxies onto the primary cluster plane. We adopt this single-plane approximation because full multi-plane ray-tracing is not natively supported in the current official release of \textsc{LensTool}.

As these LOS galaxies are not gravitationally bound to the cluster, they are modelled as an entirely separate component. We assign an independent dPIE halo to each of the four foreground interlopers, optimising their mass parameters (\textit{$\sigma_\mathrm{0}$} and \textit{$r_\mathrm{cut}$}) individually to capture their localised deflections \citep{Jauzac_2016, Acebron_2025}.

\section{Results and Discussion} \label{sec: Results and Discussion}

In this section, we show the results of the mass modelling and discuss their implications. We present two models here: a Spec model that uses only Gold multiple image constraints and a Phot model that uses both Gold and Silver constraints. Both models have grown through several intermediate steps, with an increasing number of constraints being used, and the model complexity has been gradually increased to account for this. In total, the Spec model is constrained by 118 multiple images from 20 systems, which expands to 149 images in the Phot model from 28 systems. Following Equation \ref{eq:degrees_of_freedom_eqn}, these configurations yield 136 and 172 degrees of freedom, respectively. Both models are characterised by 5 cluster-scale DM haloes and 3 ICM mass haloes, as well as 317 cluster member galaxies.

\subsection{Comparison of Mass Models}
\label{sec:comparison_of_mass_models}

\begin{figure*}
	\includegraphics[width=\textwidth]{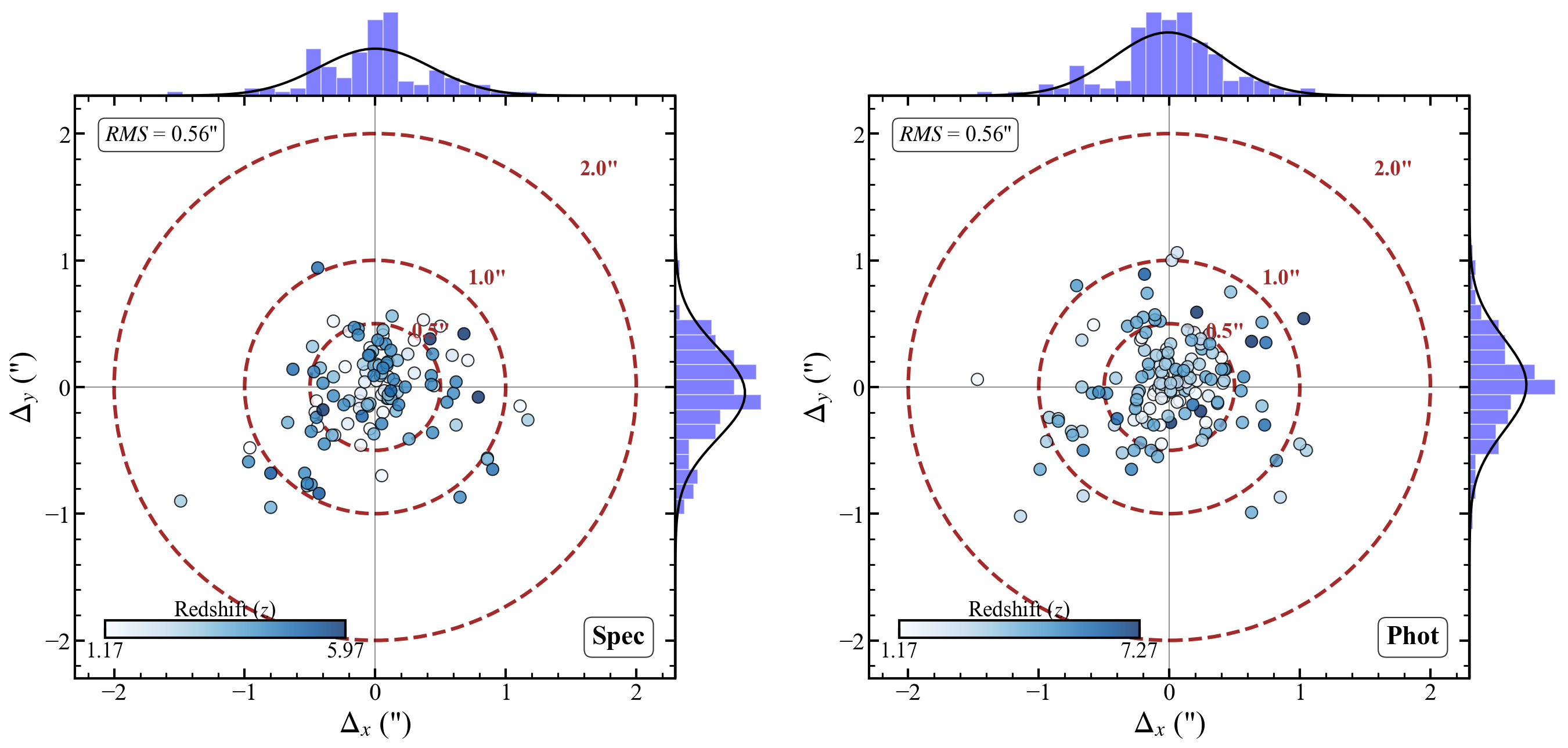}
    \caption{Distribution of image-plane displacements ($\Delta_x, \Delta_y$) between the model-predicted multiple image positions and their respective observed locations for the Spec model (\textit{left}) and the Phot model (\textit{right}). Each data point represents an individual multiple image, colour-coded by its respective redshift ($z_{s}$). The red dashed circles denote constant total radial offsets of $0.5\arcsec$, $1.0\arcsec$, and $2.0\arcsec$. The top and right panels display 1D histograms of the displacements along the $x$ and $y$ axes, respectively, overlaid with best-fit Gaussian probability density functions (solid black lines). The overall root-mean-square error of each best-fit model is given in the top left.}
    \label{fig:image_scatter_spec_model_plot}
\end{figure*}

Fig. \ref{fig:image_scatter_spec_model_plot} compares the image-plane reproduction accuracy of the two mass models by displaying the positional offsets ($\Delta_{\mathrm{x}}$, $\Delta_{\mathrm{y}}$) between the observed and model-predicted multiple images. The left panel shows the Spec model ($\Delta_{\mathrm{RMS}}=0.56\arcsec$), while the right panel illustrates the Phot model ($\Delta_{\mathrm{RMS}}=0.56\arcsec$). In both cases, the marginal distributions demonstrate that the positional errors are Gaussian and tightly centred around the origin (e.g., for the Spec model, $\langle \Delta_x \rangle=0.002\pm 0.428\arcsec$ and $\langle \Delta_x \rangle=-0.048\pm 0.360\arcsec$), indicating no systematic directional bias in the mass reconstructions. However, evaluating the magnitude of the predicted displacement against source redshift (coloured in shades of blue) reveals a positive correlation. For the Spec model, this is weak and only marginally significant (Spearman $\rho=0.183$, $\rm p=0.047$), likely reflecting the increased difficulty in precisely centroiding fainter, higher redshift sources. By contrast, the Phot model exhibits a more significant correlation ($\rho =0.390$, $\rm p < 0.001$). This is largely expected since the inclusion of Silver systems introduced photometric redshift uncertainties, which the model tries to accommodate by absorbing the redshift error into larger image-plane positional offsets. Notably, despite this effect, the inclusion of the Silver systems does not change image-plane RMS. In parametric mass modelling, introducing additional constraints often inflates the RMS, as the finite flexibility of analytical mass profiles struggles to accommodate complex local substructures across wider regions of the cluster field \citep[e.g.,][]{Johnson_Sharon_2016, Lagattuta_2019}. The fact that the RMS instead remains unchanged demonstrates that the spectroscopic constraints alone are sufficient to accurately capture the global cluster potential. Consequently, the addition of the photometric systems primarily refines locally ill-constrained regions without perturbing the global fit.

Introducing photometric multiple images risks a degeneracy between source redshifts and the mass profile slope \citep{Johnson_2014}. However, our underlying spectroscopic sample rigidly constrains the core deflection field, acting to suppress this bias. To verify whether this is the case, we evaluate the posterior covariance of the optimised source redshifts against the core radius ($ r_{ core}$) across all cluster-scale DM haloes. We find a lack of diagonal covariance (as visualised for Halo 1 - the primary DM halo - in Appendix \ref{sec:model_predicted_redshifts}, Fig. \ref{fig:redshift_corner_plot}) and no significant linear dependence. We quantify the extent of any covariance using the Pearson correlation coefficient ($r$). Across all 5 DM haloes we find a median value of $r=0.156$ indicating that the Silver constraints do not systematically bias the global mass fit.

\begin{figure*}
	\includegraphics[width=\textwidth]{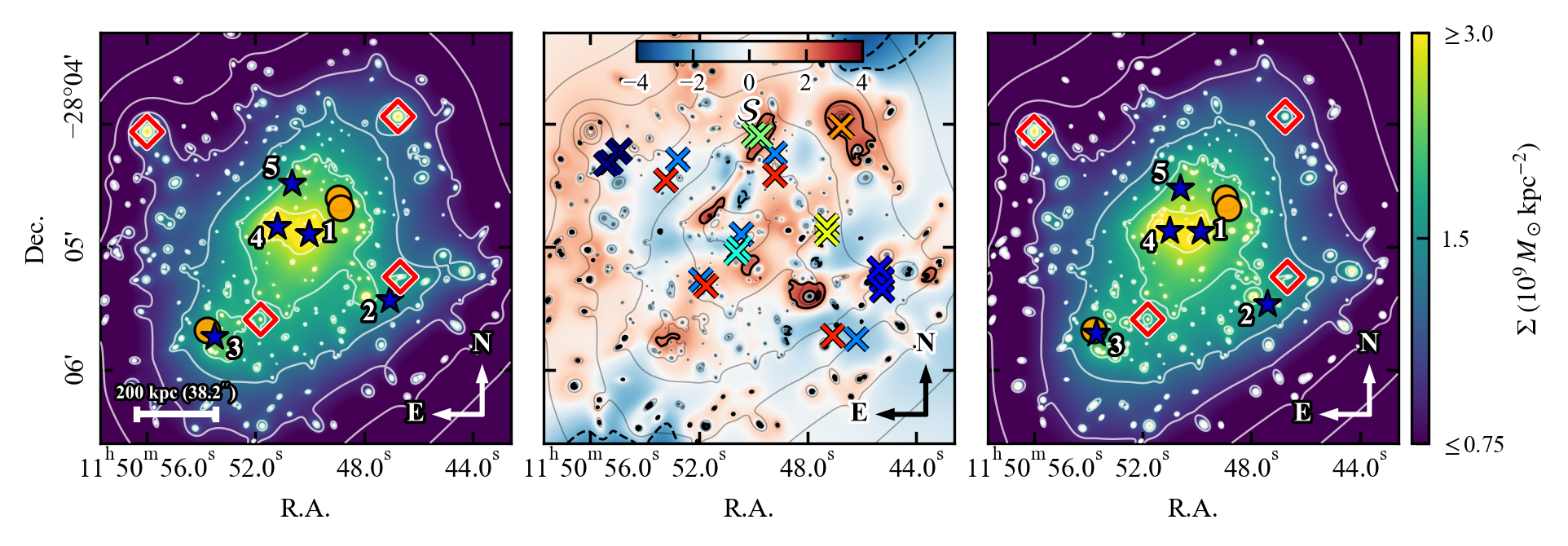}
    \caption{2D surface mass density reconstructions for J1150$-$2805.
Left and right panels: The mass distributions derived from the Spec and Phot models, respectively. White contours indicate logarithmic mass density levels. Overlaid symbols denote the optimised positions of the dark matter haloes (blue numbered stars), ICM gas potentials (orange circles) and foreground galaxies (red open diamonds) for each model. Central panel: The significance ($\mathcal{S}$) of the difference between the two maps, highlighting localised tensions. Grey contours provide the mean mass density between the models, while bold black contours (solid and dashed) delineate regions of significant tension ($\mathcal{S} = \pm 2,3$). The $\times$ symbols indicate the locations of the Silver multiple image constraints added to the Phot model - each colour representing a separate system.}
    \label{fig:mass_map_spec}
\end{figure*}

To further assess how sensitive our mass reconstruction is to the choice of multiple image constraints, we compare the two mass models in Fig. \ref{fig:mass_map_spec}. We show the 2D surface mass density maps derived from the best-fit models in the left (Spec) and right (Phot) panels. Rather than differencing the two maps directly—which would conflate genuine differences with the varying statistical precision of the two models—we construct a significance map of the difference (middle panel of Fig. \ref{fig:mass_map_spec}). This scales the absolute difference by the combined local error budget:

\begin{equation}
\mathcal{S}(x,y) = \frac{\Sigma_{\rm spec}(x,y)-\Sigma_{\rm phot}(x,y)}
{\sqrt{\sigma_{\rm spec}^{2}(x,y)+\sigma_{\rm phot}^{2}(x,y)}},
\label{eq:mass_difference}
\end{equation}

where $\Sigma$ is the projected surface mass density, and $\sigma_{\rm spec}$ and $\sigma_{\rm phot}$ are the per-pixel standard deviations of the projected mass, estimated from the scatter across 500 mass maps generated from random samples of each model's MCMC posterior.

The central panel of Fig. \ref{fig:mass_map_spec} demonstrates remarkable consistency between the two reconstructions, with the significance map ($\mathcal{S}$) remaining below $2\sigma$ across $\simeq95\%$ of the cluster. Where statistically significant differences do emerge ($|\mathcal{S}| \gtrsim 3$, peaking at 3.9), they are highly localised. Overlaying the Silver multiple images shows that these compact deviations - marked by solid contours for Spec model excesses (red) and dashed contours for Phot model excesses (blue) - correlate with the locations of the Silver image constraints. This behaviour is expected and demonstrates that the inclusion of the Silver images in the model refines the mass distribution in their immediate vicinity without destabilising the global cluster potential, which remains highly consistent elsewhere (grey contours).

The blue stars and orange circles denote the optimised spatial coordinates of the five cluster-scale DM haloes and the three cluster-scale ICM haloes respectively. Consistent with the configuration reported by \citet[][]{DAddona_2024}, both optimisations strongly favour the placement of DM mass components in the three distinct regions labelled by Haloes 1, 2, and 3 in Fig. \ref{fig:mass_map_spec}. Notably, these positions also correspond to the NWc, SEc, and Wc haloes that were identified by \citet{Finner_2017} in their weak lensing analysis. In contrast to previous strong lensing analyses, we find that the cluster is best fit as a bimodal system where each of the two BCGs hosts its own distinct DM halo (Haloes 1 and 4). This configuration is not only strongly preferred by our modelling (e.g. see Table \ref{tab:rms_haloes}), but is also physically justified by the small magnitude difference between the two central galaxies ($\Delta m_{\mathrm{F322W2}} = 0.647 \pm 0.001$). This indicates that they possess comparable stellar masses and therefore influence the core as separate macroscopic haloes. Furthermore, our modelling strongly prefers the inclusion of a fifth cluster-scale DM halo situated slightly North (Halo 5) of the bimodal core. As detailed in Table \ref{tab:rms_haloes}, incorporating these additional DM haloes yields a significant improvement in the overall goodness of fit as quantified by the $\Delta_{RMS}$ values for each model variant. 

The close physical proximity of Haloes 1, 4, and 5 within the cluster core explains why the previous weak lensing analysis by \citet{Finner_2017} lacked the spatial resolution to distinguish these individual mass peaks. By leveraging strong lensing constraints, our model successfully maps the complex substructure of this central region. As detailed in Section \ref{sec: cluster_dynamical_state}, this enhanced spatial resolution is critical for accurately reconstructing and interpreting the cluster's merger scenario.

\begin{table}
    \centering
    \caption{Summary of the strong lensing and X-ray mass model configurations evaluated using the Bayesian Information Criterion (BIC), the Akaike Information Criterion (AICc), the minimised $\chi^2_{\rm min}$ statistic, the Bayesian evidence ($\ln(E)$), and the image-plane root-mean-square (RMS) offset. $\rm{H}_{\rm{t}}$ denotes the total number of cluster-scale DM haloes, where the baseline model consists of three haloes. A checkmark (\checkmark) indicates the inclusion of an additional DM halo (Halo 4 or Halo 5 as labeled in Fig. \ref{fig:mass_map_spec}) in the optimisation, while a cross ($\times$) denotes its exclusion.}
    \label{tab:rms_haloes}
    \begin{tabular}{cccccccc}
        \toprule
        $\mathrm{H_{t}}$ & $\mathrm{H_{4}}$ & $\mathrm{H_{5}}$ & BIC & AICc & $\chi^2_{\rm min}$ & $\ln(E)$ & $\rm RMS$ ($\arcsec$) \\
        \midrule
        \multicolumn{8}{c}{Spec Model} \\
        \midrule
        3 & $\times$   & $\times$   & 536.18 & 471.36 & 307.19 & -304.04 & 1.05 \\
        4 & $\times$   & \checkmark & 487.97 & 432.64 & 230.35 & -303.03 & 0.91 \\
        4 & \checkmark & $\times$   & 421.49 & 366.16 & 163.87 & -241.74 & 0.77 \\
        5 & \checkmark & \checkmark & 374.30 & 336.48 & 88.06 & -214.48 & 0.56 \\
        \midrule
        \multicolumn{8}{c}{Phot Model} \\
        \midrule
        3 & $\times$   & $\times$   & 662.78 & 564.58 & 372.54 & -555.38 & 1.03 \\
        4 & $\times$   & \checkmark & 694.79 & 601.59 & 374.54 & -591.81 & 1.03 \\
        4 & \checkmark & $\times$   & 537.75 & 444.55 & 217.50 & -491.46 & 0.79 \\
        5 & \checkmark & \checkmark & 459.37 & 376.53 & 109.09 & -413.50 & 0.56 \\
        \bottomrule
    \end{tabular}
\end{table}

We evaluate the relative goodness-of-fit of our mass models using formal information criteria, and the Bayesian evidence ($\ln (E)$). Generally, an improved model configuration is indicated by a minimisation of the information criteria — which penalise unnecessary parameter complexity \citep{Liddle_2007} — coupled with a maximisation of the Bayesian evidence. To account for the varying degrees of freedom introduced by adding primary dark matter haloes, we compute both the Bayesian Information Criterion (BIC) and the corrected Akaike Information Criterion (AICc), defined as:

\begin{equation}
    \mathrm{BIC} = k\ln(n) - 2\ln(\hat{L})
\end{equation}
and
\begin{equation}
    \mathrm{AICc} = 2k - 2\ln(\hat{L}) + \frac{2k^2 + 2k}{n - k - 1},
\end{equation}

where $k$ is the number of free parameters, $n$ is the number of observational constraints, and $\hat{L}$ is the maximum likelihood of the model. In general, $-2\ln(\hat{L}) = \chi_{min}^{2}$ for our \textsc{LensTool} models. While both criteria effectively penalise unnecessary complexity, the most robust metric for parametric lens modelling is the Bayesian evidence \citep[$\ln (E)$; e.g.,][]{Jullo_2007, Jauzac_2018b}, which integrates the likelihood over the entire parameter space. This effectively scales the maximum likelihood by the ratio of the posterior to prior volumes. Consequently, as useful complexity is added to the models, the evidence increases. However, if unconstrained parameters are introduced - where the posterior simply recovers the prior - the evidence plateaus rather than decreases. The optimal model is identified as the simplest configuration at the onset of the evidence plateau.

To quantify the statistical significance of adding Halo 4 and Halo 5, we evaluate the difference in the Bayesian evidence ($\Delta \ln E$). Following \citet{Trotta_2008}, we map the evidence difference to an equivalent Gaussian significance and require a minimum increase of $\Delta \ln E = 12.5$ (equivalent to a $5 \sigma$ significance) to justify additional model complexity. As demonstrated in Table~\ref{tab:rms_haloes}, advancing from the three-halo baseline to the complete five-halo architecture not only systematically minimises the information criteria, but yields an overwhelming increase in the Bayesian evidence. We find $\Delta \ln E \approx 110$ for the Spec model and $\Delta \ln E \approx 142$ for the Phot model, corresponding to $\sim 14.8 \sigma$ and $\sim 16.8 \sigma$ significance, respectively. These decisive margins rigorously justify our approach, confirming the physical necessity of the multi-component core.

Having established the statistical necessity of the five-halo architecture, we present the complete set of optimised parameters in Table \ref{tab:optimised_model_parameters_table}. Values are reported as posterior medians with corresponding $1\sigma$ confidence intervals, derived from all 60,000 MCMC samples. Crucially, a direct comparison between the Spec and Phot models reveals excellent statistical agreement; every free parameter remains consistent within $2\sigma$. This robust alignment acts as further confirmation that the addition of the photometric systems does not unphysically bias the mass reconstruction. To ensure the full reproducibility of our mass models, the complete set of input parameters and their corresponding prior distributions are detailed in Appendix \ref{sec:input_parameters_and_priors} (Table \ref{tab:lens_params_priors}).

\begin{table*}
\centering
\caption{Optimised parameters for the strong-lensing and X-ray models of J1150$-$2805. The upper panel lists the Spec model, the lower panel the Phot model. Values are posterior medians from the MCMC chain with $+1\sigma/-1\sigma$ (16th/ 84th percentile) uncertainties; quantities in square brackets were held fixed. Coordinates $\Delta x, \Delta y$ are angular offsets from the reference position (RA, Dec) = (177.7089950, $-28.0821673$); $\Delta x$ increases West of the reference. All radii are reported in proper kpc at the lens redshift $z_{\mathrm{lens}} = 0.383$ (1\,arcsec $\approx$ 5.23\,kpc). The cluster member galaxy rows provide the baseline normalisation parameters, $\sigma_{\mathrm{LT}}^{\star}$ and $r_{\mathrm{cut}}^{\star}$, evaluated at the designated reference magnitude, $m^{\star}$.}
\label{tab:optimised_model_parameters_table}
\footnotesize
\setlength{\tabcolsep}{3pt}
\renewcommand{\arraystretch}{1.0}
\begin{tabular*}{\textwidth}{@{\extracolsep{\fill}}llccccccc}
\toprule
Category & Component & $\Delta x$ & $\Delta y$ & $e$ & $\theta$ & $\sigma_{\rm LT}$ & $r_{\rm core}$ & $r_{\rm cut}$ \\
 & & [arcsec] & [arcsec] & & [$^{\circ}$] & [km\,s$^{-1}$] & [kpc] & [kpc] \\
\midrule
\multicolumn{9}{c}{Spec model} \\
\midrule
Cluster-scale DM & Halo 1 & $2.18^{+1.59}_{-1.24}$ & $2.04^{+1.03}_{-1.02}$ & $0.59^{+0.05}_{-0.05}$ & $63.97^{+1.22}_{-1.21}$ & $1139.86^{+41.58}_{-36.22}$ & $148.07^{+6.98}_{-8.69}$ & $[2200.00]$ \\
 & Halo 2 & $42.73^{+3.11}_{-2.99}$ & $-29.95^{+1.37}_{-1.47}$ & $0.68^{+0.07}_{-0.07}$ & $25.50^{+2.80}_{-2.73}$ & $597.93^{+46.06}_{-39.54}$ & $111.61^{+15.52}_{-14.65}$ & $[2200.00]$ \\
 & Halo 3 & $-45.72^{+1.22}_{-1.03}$ & $-49.17^{+1.91}_{-1.34}$ & $0.22^{+0.20}_{-0.08}$ & $62.82^{+18.98}_{-141.28}$ & $437.80^{+47.77}_{-34.02}$ & $46.90^{+10.25}_{-11.82}$ & $[2200.00]$ \\
 & Halo 4 & $-12.18^{+0.82}_{-1.12}$ & $5.02^{+1.06}_{-0.93}$ & $0.51^{+0.08}_{-0.08}$ & $-5.18^{+6.66}_{-4.75}$ & $522.87^{+39.33}_{-40.24}$ & $37.36^{+11.34}_{-8.33}$ & $[2200.00]$ \\
 & Halo 5 & $-5.66^{+2.83}_{-1.07}$ & $25.69^{+7.32}_{-1.35}$ & $0.74^{+0.13}_{-0.21}$ & $65.61^{+8.01}_{-10.73}$ & $303.15^{+46.63}_{-35.76}$ & $29.45^{+18.43}_{-12.73}$ & $[2200.00]$ \\
\addlinespace
Foreground galaxies & NE Foregr. & $[-77.20]$ & $[51.76]$ & $[0.10]$ & $[-76.32]$ & $391.32^{+6.15}_{-17.06}$ & $[0.001]$ & $131.60^{+13.37}_{-22.04}$ \\
 & W Foregr. & $[46.25]$ & $[-18.95]$ & $[0.80]$ & $[-171.46]$ & $154.09^{+30.77}_{-28.37}$ & $[0.001]$ & $98.27^{+32.64}_{-58.00}$ \\
 & NW Foregr. & $[45.14]$ & $[59.21]$ & $[0.07]$ & $[-2.87]$ & $323.72^{+48.52}_{-71.96}$ & $[0.001]$ & $88.50^{+33.70}_{-26.79}$ \\
 & SE Foregr. & $[-21.80]$ & $[-39.67]$ & $[0.23]$ & $[-64.69]$ & $117.44^{+73.79}_{-13.00}$ & $[0.001]$ & $32.26^{+76.51}_{-31.10}$ \\
\addlinespace
BCGs & BCG 1 (ID 328) & $[0.00]$ & $[0.00]$ & $[0.42]$ & $[-121.31]$ & $309.41^{+32.59}_{-23.61}$ & $[0.001]$ & $58.73^{+19.15}_{-19.51}$ \\
 & BCG 2 (ID 559) & $[-16.75]$ & $[7.02]$ & $[0.47]$ & $[-136.84]$ & $262.81^{+36.46}_{-35.59}$ & $[0.001]$ & $132.18^{+12.39}_{-32.25}$ \\
\addlinespace
Cluster members & Gal 1745 & $[29.55]$ & $[-28.45]$ & $[0.16]$ & $[-157.32]$ & $266.58^{+10.03}_{-9.71}$ & $[0.001]$ & $109.62^{+11.40}_{-21.06}$ \\
 & Gal 637 & $[16.02]$ & $[9.01]$ & $[0.01]$ & $[-157.89]$ & $70.60^{+30.23}_{-20.45}$ & $[0.001]$ & $11.06^{+20.71}_{-7.13}$ \\
 & Gal 1684 & $[-38.03]$ & $[32.07]$ & $[0.35]$ & $[-71.99]$ & $254.79^{+41.18}_{-26.14}$ & $[0.001]$ & $12.40^{+5.98}_{-5.36}$ \\
 & Gal 631 & $[-0.23]$ & $[31.96]$ & $[0.44]$ & $[-47.50]$ & $211.24^{+68.04}_{-54.33}$ & $[0.001]$ & $5.03^{+11.75}_{-3.08}$ \\
 & Gal 1322 & $[67.17]$ & $[-8.26]$ & $[0.15]$ & $[-157.50]$ & $183.23^{+9.64}_{-8.36}$ & $[0.001]$ & $83.61^{+30.17}_{-42.22}$ \\
 & Gal 7 & $[77.46]$ & $[-16.16]$ & $[0.22]$ & $[-11.43]$ & $261.66^{+11.95}_{-9.93}$ & $[0.001]$ & $94.09^{+4.15}_{-12.98}$ \\
 & Gal 1272 & $[86.14]$ & $[-18.86]$ & $[0.37]$ & $[-144.76]$ & $122.23^{+12.51}_{-8.47}$ & $[0.001]$ & $55.29^{+34.62}_{-34.79}$ \\
\addlinespace
X-ray potentials & X-ray Halo 1 & $[15.90]$ & $[19.38]$ & $[0.33]$ & $[126.10]$ & $[321.72]$ & $[207.42]$ & $[2200.00]$ \\
 & X-ray Halo 2 & $[17.30]$ & $[14.74]$ & $[0.85]$ & $[39.25]$ & $[382.00]$ & $[534.29]$ & $[2200.00]$ \\
 & X-ray Halo 3 & $[-48.03]$ & $[-44.77]$ & $[0.58]$ & $[115.33]$ & $[338.28]$ & $[494.57]$ & $[2200.00]$ \\
\addlinespace
Scaling relations & \textit{JWST} & ($m^{\star}=16.83$) & & & & $223.50^{+19.53}_{-17.23}$ & $[0.001]$ & $99.42^{+26.59}_{-22.99}$ \\
 & \textit{HST} & ($m^{\star}=18.83$) & & & & $220.15^{+56.19}_{-35.23}$ & $[0.001]$ & $48.89^{+29.80}_{-16.94}$ \\
\midrule
\multicolumn{9}{c}{Phot model} \\
\midrule
Cluster-scale DM & Halo 1 & $3.79^{+0.82}_{-0.91}$ & $2.24^{+1.10}_{-1.07}$ & $0.72^{+0.04}_{-0.04}$ & $63.43^{+1.76}_{-1.46}$ & $1107.98^{+49.89}_{-53.55}$ & $168.30^{+6.73}_{-8.12}$ & $[2200.00]$ \\
 & Halo 2 & $39.77^{+4.17}_{-2.32}$ & $-29.59^{+1.08}_{-0.96}$ & $0.57^{+0.06}_{-0.05}$ & $22.86^{+1.93}_{-2.22}$ & $667.82^{+43.58}_{-48.76}$ & $115.53^{+14.11}_{-14.60}$ & $[2200.00]$ \\
 & Halo 3 & $-47.67^{+1.30}_{-0.89}$ & $-50.10^{+1.97}_{-1.62}$ & $0.42^{+0.17}_{-0.15}$ & $76.35^{+6.21}_{-8.50}$ & $448.96^{+26.97}_{-30.15}$ & $54.18^{+9.50}_{-8.28}$ & $[2200.00]$ \\
 & Halo 4 & $-12.25^{+0.82}_{-1.01}$ & $4.49^{+0.41}_{-0.62}$ & $0.48^{+0.08}_{-0.11}$ & $-4.39^{+6.57}_{-3.68}$ & $599.49^{+59.96}_{-40.88}$ & $52.11^{+5.35}_{-6.50}$ & $[2200.00]$ \\
 & Halo 5 & $-5.20^{+0.95}_{-1.13}$ & $26.08^{+1.97}_{-1.58}$ & $0.54^{+0.16}_{-0.16}$ & $74.07^{+8.94}_{-19.36}$ & $335.95^{+58.83}_{-44.16}$ & $28.67^{+13.48}_{-11.38}$ & $[2200.00]$ \\
\addlinespace
Foreground galaxies & NE Foregr. & $[-77.20]$ & $[51.76]$ & $[0.10]$ & $[-76.32]$ & $385.40^{+10.39}_{-19.43}$ & $[0.001]$ & $133.88^{+11.77}_{-12.94}$ \\
 & W Foregr. & $[46.25]$ & $[-18.95]$ & $[0.80]$ & $[-171.46]$ & $157.86^{+36.56}_{-21.73}$ & $[0.001]$ & $39.98^{+37.53}_{-22.66}$ \\
 & NW Foregr. (ID 13) & $[45.14]$ & $[59.21]$ & $[0.07]$ & $[-2.87]$ & $237.23^{+53.87}_{-51.03}$ & $[0.001]$ & $44.11^{+18.55}_{-18.07}$ \\
 & SE Foregr. (ID 1397) & $[-21.80]$ & $[-39.67]$ & $[0.23]$ & $[-64.69]$ & $130.68^{+32.25}_{-20.59}$ & $[0.001]$ & $59.90^{+33.19}_{-39.96}$ \\
\addlinespace
BCGs & BCG 1 (ID 328) & $[0.00]$ & $[0.00]$ & $[0.42]$ & $[-121.31]$ & $338.10^{+14.83}_{-12.33}$ & $[0.001]$ & $48.15^{+15.48}_{-13.09}$ \\
 & BCG 2 (ID 559) & $[-16.75]$ & $[7.02]$ & $[0.47]$ & $[-136.84]$ & $292.24^{+27.28}_{-23.23}$ & $[0.001]$ & $94.30^{+31.00}_{-25.10}$ \\
\addlinespace
Cluster members & Gal 1745 & $[29.55]$ & $[-28.45]$ & $[0.16]$ & $[-157.32]$ & $283.92^{+32.25}_{-14.32}$ & $[0.001]$ & $48.24^{+44.95}_{-23.15}$ \\
 & Gal 637 & $[16.02]$ & $[9.01]$ & $[0.01]$ & $[-157.89]$ & $52.61^{+16.12}_{-11.87}$ & $[0.001]$ & $29.62^{+21.63}_{-15.89}$ \\
 & Gal 1684 & $[-38.03]$ & $[32.07]$ & $[0.35]$ & $[-71.99]$ & $274.99^{+53.07}_{-68.97}$ & $[0.001]$ & $2.51^{+2.31}_{-0.91}$ \\
 & Gal 631 & $[-0.23]$ & $[31.96]$ & $[0.44]$ & $[-47.50]$ & $145.12^{+26.29}_{-27.47}$ & $[0.001]$ & $15.65^{+14.69}_{-8.04}$ \\
 & Gal 1322 & $[67.17]$ & $[-8.26]$ & $[0.15]$ & $[-157.50]$ & $230.73^{+45.31}_{-23.08}$ & $[0.001]$ & $13.45^{+7.50}_{-5.79}$ \\
 & Gal 7 & $[77.46]$ & $[-16.16]$ & $[0.22]$ & $[-11.43]$ & $276.35^{+12.63}_{-12.67}$ & $[0.001]$ & $78.68^{+14.61}_{-13.77}$ \\
 & Gal 1272 & $[86.14]$ & $[-18.86]$ & $[0.37]$ & $[-144.76]$ & $168.46^{+48.59}_{-30.58}$ & $[0.001]$ & $7.63^{+7.60}_{-3.62}$ \\
 & Gal 1337 & $[-67.86]$ & $[38.27]$ & $[0.12]$ & $[-74.13]$ & $185.13^{+14.35}_{-10.89}$ & $[0.001]$ & $90.18^{+25.11}_{-25.52}$ \\
\addlinespace
X-ray potentials & X-ray Halo 1 & $[15.90]$ & $[19.38]$ & $[0.33]$ & $[126.10]$ & $[321.72]$ & $[207.42]$ & $[2200.00]$ \\
 & X-ray Halo 2 & $[17.30]$ & $[14.74]$ & $[0.85]$ & $[39.25]$ & $[382.00]$ & $[534.29]$ & $[2200.00]$ \\
 & X-ray Halo 3 & $[-48.03]$ & $[-44.77]$ & $[0.58]$ & $[115.33]$ & $[338.28]$ & $[494.57]$ & $[2200.00]$ \\
\addlinespace
Scaling relations & \textit{JWST} & ($m^{\star}=16.83$) & & & & $251.32^{+18.07}_{-18.69}$ & $[0.001]$ & $60.84^{+17.01}_{-12.41}$ \\
 & \textit{HST} & ($m^{\star}=18.83$) & & & & $308.76^{+46.38}_{-45.03}$ & $[0.001]$ & $28.81^{+12.77}_{-10.57}$ \\
\bottomrule
\end{tabular*}
\end{table*}

\begin{figure}
	\includegraphics[width=\columnwidth]{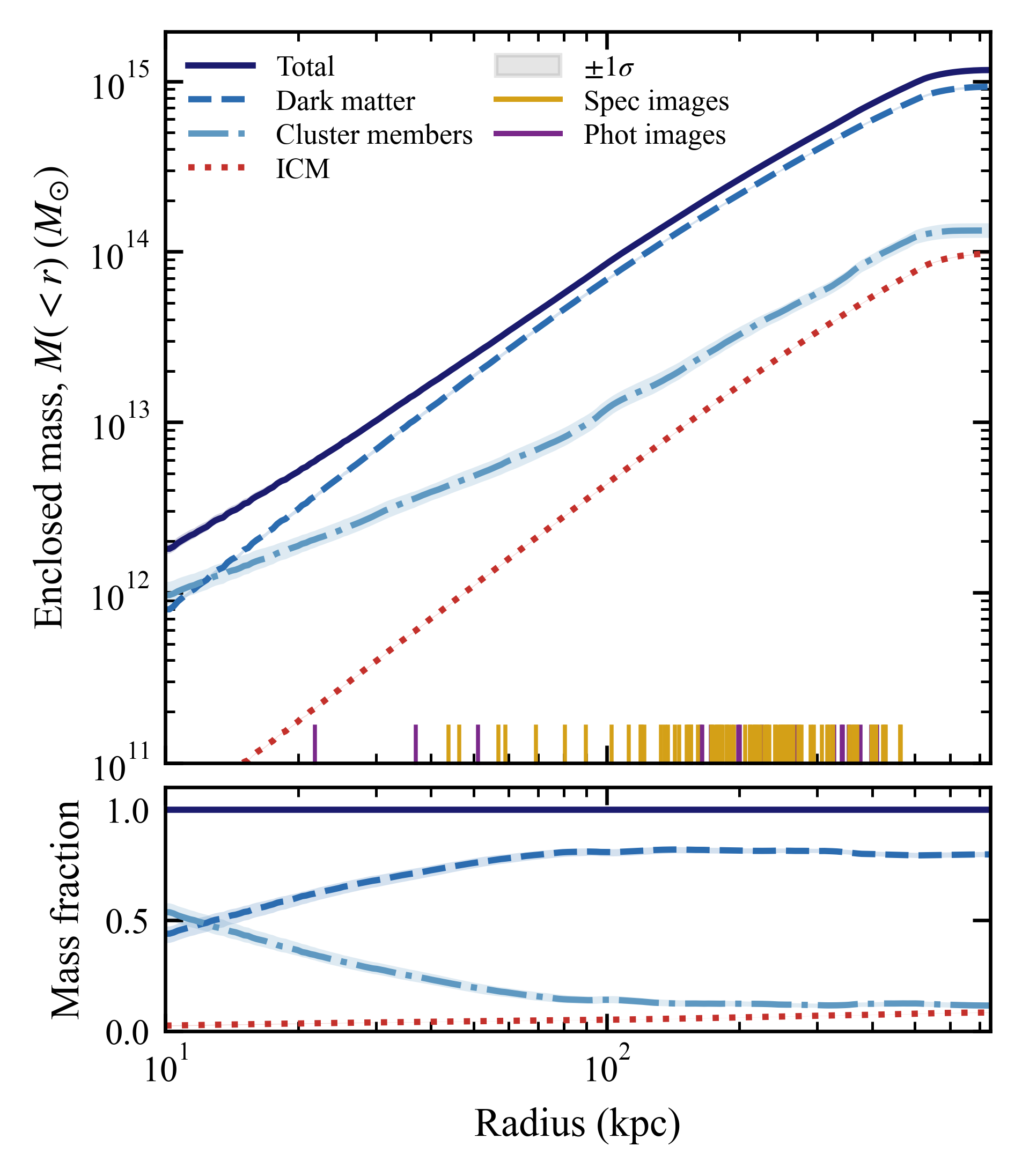}
    \caption{Top panel: Cumulative surface mass density profile ($M_{\odot}$) as a function of distance from BCG-1 (kpc). The total mass is represented by the solid dark blue line, while the constituent mass components are broken down into DM (dashed blue line), cluster member galaxies (dash-dotted light blue line), and the ICM (dotted red line). Shaded bands surrounding the profiles indicate the $1\sigma$ confidence intervals. The vertical tick marks along the lower axis denote the cluster-centric radii of the multiple images, distinguishing between the gold images (yellow), and the additional silver images included in the Phot model (purple). Bottom panel: Cumulative mass fraction of each individual component relative to the total mass as a radial function. The solid dark blue line at unity represents the total mass baseline.}
    \label{fig:enclosed_mass_plot}
\end{figure}

Fig. \ref{fig:enclosed_mass_plot} illustrates the 2D projected enclosed mass profile, $M(<r)$, of the cluster out to a radius of 740~kpc from BCG-1, which corresponds to the maximum spatial extent of our \textsc{LensTool} mass map. The plot details both the total enclosed mass and the relative contributions of the individual structural components. For visual clarity, the enclosed mass profiles presented here are derived exclusively from the Spec model. We note, however, that both configurations return nearly identical radial profiles and recover total enclosed mass estimates that agree well within $1\sigma$. To quantify the statistical uncertainties, we compute the median cumulative mass and corresponding $1\sigma$ credible intervals (shaded bands; 16th and 84th percentiles) by evaluating 500 random realisations drawn from the MCMC chain. While the ICM potentials were held fixed during the \textsc{LensTool} optimisation, we manually propagate the statistical scatter from the independent X-ray modelling into our plotted error budget. As shown in Fig. \ref{fig:enclosed_mass_plot}, the uncertainty is extremely small and, therefore, validates our choice to fix the ICM mass during the lensing fit. The projected mass enclosed within a $1$~Mpc radius around BCG-$1$ is $M(<1~\mathrm{Mpc}) = 1.86 \pm {0.02} \times 10^{15} M_{\odot}$. This exceptional mass firmly places J1150$-$2805 at the high-mass end of the cluster mass function, ranking it among the most massive known structures in the Universe \citep[][]{Tinker_2008, Bocquet_2016}. Within 200~kpc, the mass budget is dominated by the DM ($\sim81 \pm 1\%$), with cluster member galaxies and the BCGs, contributing $\sim12 \pm 1\%$ and the ICM the remaining $\sim6\%$. The ICM fraction grows monotonically to $\sim 9.7\%$ at $1$~Mpc. We also overlay the radial positions of the multiple images in Fig. \ref{fig:enclosed_mass_plot} to highlight the comprehensive spatial coverage of our strong lensing constraints. The images are colour coded such that the yellow ones appear in both models and the purple ones are unique to the Phot model. Ultimately, both models yield highly consistent total enclosed mass profiles, demonstrating that the inclusion of the 31 supplementary photometric multiple images does not artificially bias the global cluster potential as expected (e.g. see \citet{Meneghetti_2017}).

Both of our best-fitting mass models represent significant advancements over previous strong lensing analyses of J1150$-$2805. \citet{DAddona_2024} provided the most recent reconstruction of the cluster, achieving an image-plane RMS of $0.75^{\prime\prime}$ - a substantial improvement over the $1.9^{\prime\prime}$ reported by \citet{Zitrin_2017}. Building upon this progress, our analysis achieves an image-plane RMS of $0.56\arcsec$ even whilst drastically increasing the number of multiple-image constraints in each model. Such an expansion of the multiple image catalogue has been made possible by the exquisite data quality from the \textit{JWST} SLICE observations of the cluster. Whereas \citet{DAddona_2024} utilised a "golden sample" of $47$ multiple images, our Spec model incorporates $118$ multiple image constraints, expanding to $149$ in our Phot model. When evaluated using this "golden sample" of multiple images, our more complex model achieves an image-plane RMS of $0.50\arcsec$, confirming excellent agreement with previously established constraints. Importantly, despite the vast increase in model complexity, our global mass estimates remain highly consistent with those of \citet{DAddona_2024}, and we robustly confirm the existence of the secondary DM structures in the West (Halo 2) and South-East (Halo 3).

\subsection{Impacts of ICM Model}
\label{sec:impacts_of_icm_model}

In this section, we evaluate how incorporating an X-ray-derived ICM mass component impacts the overall strong lensing mass reconstruction. Traditionally, strong lensing models parametrise the large-scale cluster potential using a primary DM halo, implicitly forcing it to absorb the mass contribution of the ICM. However, the hot X-ray emitting gas typically accounts for $\sim10 - 15 \%$ of the total cluster mass. Since the collisional gas may exhibit a distinct spatial distribution from the collisionless DM, neglecting it introduces systematic biases into the inferred parameters of the DM profile, such as its core size and density slope \citep[e.g.,][]{Bonamigo_2018}. This effect is even more prevalent in dynamic cluster systems such as J1150$-$2805. By incorporating an independent, X-ray-constrained ICM component, we can more reliably isolate the true underlying DM distribution.

As outlined in Section \ref{sec:intracluster_medium_modelling}, the ICM mass distribution was kept fixed during the optimisation of our fiducial lens models to restrict the dimensionality of an already extensive parameter space. However, to validate this approach, we additionally perform test runs where the ICM and strong lensing parameters are optimised simultaneously and where the ICM is not included in the model. This allows us to directly compare the performance and stability of the three methods, analogous to the investigation conducted by \citet{Limousin_2026}. 

In Fig. \ref{fig:ICM_model_variations_posteriors}, we compare the effects of the various ICM configurations on both the macro-scale (left panel) and galaxy-scale (right panel) mass distributions. We derive these posteriors exclusively using the Spec model to strictly isolate the effects of the ICM. The Phot model is excluded from this comparison, as the inclusion of photometric systems requires modelling their unknown redshifts as additional free parameters, thereby increasing the dimensionality of the parameter space. Using the Spec model, therefore, provides a cleaner, simplified baseline to strictly isolate the impact of including or not including the ICM mass. Blue contours denote the posteriors for the strong lensing-only model; red contours represent our fiducial model, which pairs strong lensing with a fixed ICM mass profile; and green contours illustrate the results of simultaneously optimising the strong lensing and ICM mass. 

For the macro-scale mass distribution, we focus our comparison on the largest DM halo (Halo 1 in Fig. \ref{fig:mass_map_spec}), which serves as the primary tracer of the overall cluster potential. We find that the primary halo is sensitive to the chosen ICM configuration, although this sensitivity is not uniform across its structural parameters. Most notably, we observe a marked reduction in the $\sigma_{LT,1}$ parameter, indicating a less massive primary DM halo when the X-ray gas is accounted for alongside the strong lensing data. This behaviour is expected since the dPIE profile no longer has to attempt to account for both the DM and the gas. By contrast, the core radius ($r_{core,1}$), ellipticity ($e_{1}$), and position angle ($\theta_{1}$) are considerably more stable, with $r_{core,1}$ shifting by only $-0.38\arcsec$ between the lensing-only and fixed-ICM models. In this respect, our results are broadly consistent with \citet{Limousin_2026}, who reported general agreement between the three model variations for AC 114; the pronounced shift we recover in $\sigma_{LT,1}$ (and hence in the total mass of Halo 1) nevertheless indicates that the primary halo's mass normalisation remains more sensitive to the ICM treatment than its geometry (i.e., the DM slope is not affected). We note that the large number of free parameters in our model provides significantly more flexibility than simpler configurations, which makes isolating the exact impact of the ICM modelling challenging. Despite the mathematical degeneracy between $r_{\rm{core}}$ and $\sigma_{LT}$ for dPIE profiles, we observe no change in $r_{\rm{core}}$ between the model variations, indicating that the DM slope is conserved. Notably, $r_{core,1}$ remains extremely large for a standard $\Lambda$CDM cosmology, which favours centrally cuspy NFW profiles \citep[][]{Navarro_1997}. Such cored DM haloes have been reported for over two decades in many strong lensing galaxy cluster systems \citep[see e.g.][]{Sand_2004, Newman_2013,Limousin_2022,Lagattuta_2023,Natarajan+2024,Cerny_2026}.

\textbf{\begin{figure*}
	\includegraphics[width=\textwidth]{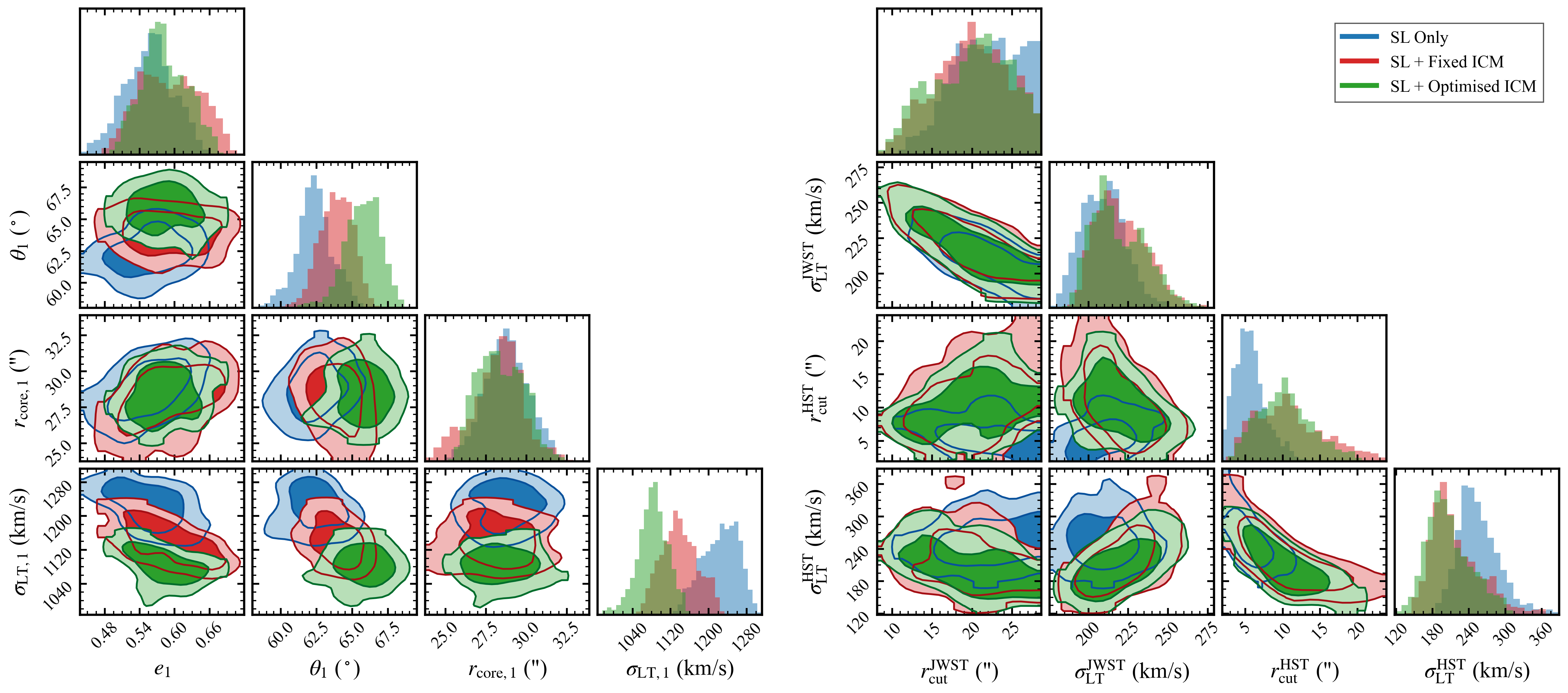}
    \caption{1D and 2D marginalised posterior probability distributions illustrating the constraints on key mass-modelling parameters for three different modelling choices: strong-lensing only (SL Only; blue), strong lensing combined with a fixed ICM model (SL + Fixed ICM; red), and strong lensing combined with a simultaneously optimised ICM model (SL + Optimised ICM; green). The left panels show the DM Halo 1 parameters ellipticity ($e_{1}$), position angle ($\theta_{1}$), core radius ($r_{core,1}$), and \textsc{LensTool} velocity dispersion ($\sigma_{LT,1}$). The right panels illustrate the scaling parameters for the galaxy-scale components, including the truncation radii ($r_{cut}$) and velocity dispersions ($\sigma_{LT}$) under both \textit{JWST} and \textit{HST} frameworks. Shaded 2D contours represent the $1\sigma$ and $2\sigma$ credible regions.}
    \label{fig:ICM_model_variations_posteriors}
\end{figure*}}

While the four secondary DM haloes do not dominate the macro-scale potential, their structural parameters are nonetheless sensitive to the inclusion of the ICM. In Table \ref{tab:halo_median_shifts}, we detail the relative shifts in total mass ($\Delta M$), core radius ($\Delta r_{core}$), and ellipticity ($\Delta e$) for each secondary halo between the lensing-only model and our fiducial fixed-ICM model. As with Fig. \ref{fig:ICM_model_variations_posteriors}, we only compare the Spec model here. In contrast to the primary halo, the secondary haloes exhibit only modest, non-systematic mass shifts. We attribute these localised variations to the freedom granted to the secondary haloes once the ICM is decoupled from the collisionless DM: separating the gas relieves parameter degeneracies and allows each subhalo to fit its local substructure more independently, rather than being dominated by the primary halo. If we consider the aggregate mass of the central DM haloes, the net DM mass in the cluster core exhibits a decrease of $\sim 9\%$. This confirms that, although separating these components increases the flexibility of the local DM potentials, the total DM budget is correctly reduced to accommodate the ICM, aligning with physical expectations.

Extending our analysis beyond the central core, we note that the response of individual haloes to the ICM is not straightforwardly predictable from their spatial proximity to the X-ray gas. Halo 3 provides an instructive example: it overlaps spatially with the South-East ICM mass (as shown in Fig. \ref{fig:mass_map_spec}), which might naively be expected to drive a localised reduction in its DM mass. In practice, however, we find its mass and core radius are essentially unchanged (Table \ref{tab:halo_median_shifts}), indicating that at the level of individual substructures the interplay between the gas and DM components is more complex than a simple one-to-one anti-correlation.

\begin{table}
\centering
\setlength{\tabcolsep}{4pt}
\caption{Marginalised median masses and structural parameter shifts comparing the strong lensing-only model to the fiducial strong lensing + fixed ICM model (Spec model). Parameter shifts are expressed as the statistical tension ($\Delta / \sigma$) between the two models. Errors denote the $1\sigma$ credible interval.}
\label{tab:halo_median_shifts}
\begin{tabular}{cccccc}
\toprule
Halo & $M_{\rm{SL}}$ & $M_{\rm{SL+ICM}}$ & $\Delta M / \sigma_M$ & $\Delta r_{\mathrm{core}} / \sigma_{r}$ & $\Delta e / \sigma_e$ \\
& \multicolumn{2}{c}{($10^{13} \, \mathrm{M_{\odot}}$)} & & & \\
\midrule
1  & $43.31^{+2.60}_{-3.07}$ & $37.20^{+2.84}_{-2.47}$ & $-1.55$ & $-0.17$ & $+0.46$ \\
2  & $10.90^{+1.03}_{-0.93}$ & $10.41^{+1.99}_{-1.44}$ & $-0.25$ & $-0.26$ & $+0.13$ \\
3  & $5.78^{+1.09}_{-0.60}$ & $5.88^{+1.35}_{-0.95}$ & $+0.07$ & $+0.13$ & $-0.15$ \\
4  & $7.87^{+1.90}_{-1.85}$ & $8.39^{+1.27}_{-1.35}$ & $+0.23$ & $+0.20$ & $-0.41$ \\
5  & $3.36^{+1.01}_{-0.96}$ & $2.95^{+0.95}_{-0.62}$ & $-0.32$ & $-0.33$ & $-0.18$ \\
\bottomrule
\end{tabular}
\end{table}

The right panel of Fig. \ref{fig:ICM_model_variations_posteriors} illustrates the impact of the three models on the galaxy-scale mass distribution. By comparing the inferred scaling relation parameters ($r_{cut}$ and $\sigma_{LT}$) for both the \textit{JWST} and \textit{HST} cluster member catalogues, we demonstrate that - unlike the macro-scale potential - these subhalo mass estimates remain highly robust and largely insensitive to the chosen ICM configuration.

\begin{table}
\centering
\caption{Summary of the statistical performance for the three Spec mass models. We report the minimum $\chi^2$, the image-plane root-mean-square (RMS) offset of the multiple images, and the BIC and AICc. For the optimised ICM model the $\chi^2_{min}$ refers to the lensing contribution only, and its information criteria are omitted (see text).}
\label{tab:model_statistics}
\begin{tabular}{lcccc}
\toprule
Model & $\chi_{min}^2$ & $\Delta_{\mathrm{RMS}}$(") & BIC & AICc \\
\midrule
Lensing Only  & $92.16$ & $0.57$ & $378.40$ & $340.58$ \\
Fixed ICM     & $88.06$ & $0.56$ & $374.30$ & $336.48$ \\
\textit{Optimised ICM} & \textit{$90.02$} & \textit{$0.57$} & - & - \\
\bottomrule
\end{tabular}
\end{table}

Table \ref{tab:model_statistics} summarises the statistical performance of the three macroscopic mass models. Following the Jeffreys scale \citep[][]{Kass_1995}, the inclusion of a fixed ICM component improves the fit relative to both the lensing-only and optimised ICM models. For the optimised ICM model we report only the lensing contribution to the $\chi^2$ and RMS, and we deliberately omit its BIC and AICc. This model is not directly comparable to the other two as its ICM parameters are optimised simultaneously with the lens model through an additional X-ray likelihood term (e.g. Section \ref{sec:intracluster_medium_modelling}). Any information criterion computed for it would therefore not lie on a common scale with the lensing-only and fixed-ICM values, and a formal ranking would be ill-defined. Therefore, the $\chi^2_{min}$ is measured only from the lensing constraints. Taken together, these results motivate our choice of the fixed ICM model as the fiducial configuration. Of the three, it marginally provides the best statistical description of the lensing data, being favoured over the lensing-only model by both the BIC and AICc. Fixing the ICM mass distribution thus allows us to incorporate the physically motivated gas component while keeping the dimensionality of the parameter space computationally feasible.

\subsection{Cluster Dynamical State Insights}
\label{sec: cluster_dynamical_state}

Fig. \ref{fig:mass_map_spec} reveals a pronounced elongation of the cluster's mass distribution along the North-West to South-East axis, strongly suggesting a post-merger dynamical state \citep[][]{Smith_2005, Richard_2010}. This aligns with the findings of \citet{DAddona_2024}, and is consistent with previous \citep{Bagchi_2011, Bonafede_2014}, and more recent \citep{Gitti_2025, Rajpurohit_2025} X-ray and radio analyses, as well as the weak-lensing analysis based on Subaru and \textit{HST} data by \citet{Finner_2017}. Crucially, these radio observations reveal the presence of peripheral radio relics — synchrotron emission features that trace the outward propagation of merger-driven shock waves, clearly indicating a recent and energetic collision. These independent analyses collectively indicate that the merger axis lies primarily within the plane of the sky, a scenario that naturally accounts for the highly elongated morphology observed in our mass map (and in the magnification maps - Fig. \ref{fig:magnification_power_comparison_plot}). Further evidence of this disturbed dynamical state is the presence of the two BCGs, forming a bimodal core at the cluster centre. This work represents the first strong-lensing analysis to explicitly incorporate this bimodal mass distribution. As demonstrated in Section \ref{sec:comparison_of_mass_models}, accounting for this physical complexity yields a significantly improved fit to the observed lensing constraints. Furthermore, the necessity of incorporating a fifth, northernmost DM halo into the mass models highlights the heavily disturbed, multi-component nature of this ongoing merger. 

\textbf{\begin{figure*}
	\includegraphics[width=\textwidth]{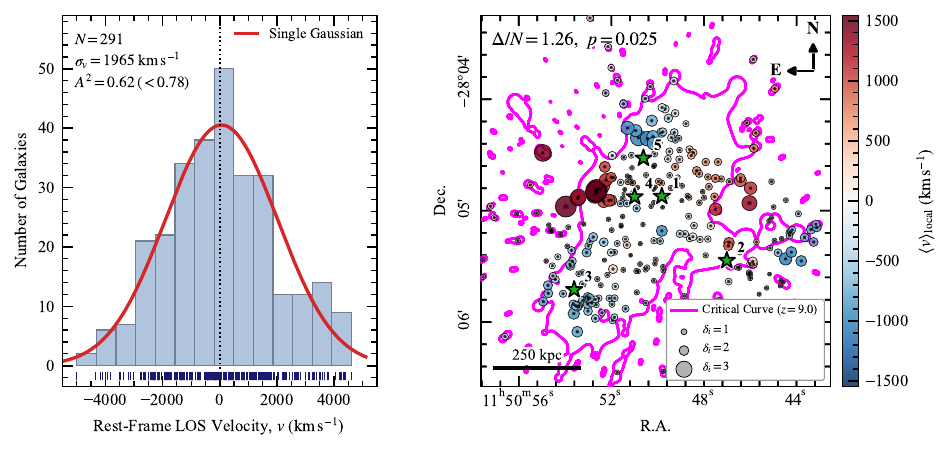}
    \caption{Kinematics of \mbox{J1150$-$2805} from the $291$ spectroscopic cluster members. Left panel: Rest-frame LOS velocity distribution (blue histogram; individual galaxies as the lower rug in dark blue) with the single Gaussian defined by the systemic velocity and the biweight dispersion (solid red curve). Right panel: Dressler--Shectman bubble plot over the critical curves (pink) for a source at $z_{s} = 9.0$. Each galaxy is a circle of area $\propto e^{\delta_{i}}$ (equation~\ref{eq:ds}) coloured by the local mean LOS velocity (blue: approaching; red: receding). Green stars mark median positions of the five strong lensing DM haloes ($1$-$5$). The DS statistic $\Delta/N=1.26$ has a Monte-Carlo significance $p=0.025$ ($5000$ velocity-reshuffling realisations).}
    \label{fig:DS_Bubble_plot}
\end{figure*}}

In order to probe the internal dynamics of J1150$-$2805, we exploit the publicly available redshift catalogue provided by \citet[][]{DAddona_2024}. As described in Section \ref{sec:cluster_members}, we find $N_{\rm CM}=291$ within the catalogue. From this, we calculate an exceptionally high rest-frame LOS velocity dispersion of $\sigma_{v}=1960^{+84}_{-92}\,\mathrm{km\,s^{-1}}$ (left panel of Fig. \ref{fig:DS_Bubble_plot}). This is among the largest dispersions measured for any radio relic cluster and is fully consistent within the differing member selections, with the $\sigma_{v} = 1756 \pm 74\,\mathrm{km\,s^{-1}}$ reported by \citet[][]{Golovich_2019}. Such an extreme dispersion presents a fascinating geometric puzzle. For a single virialised halo, assuming standard scaling relations \citep[e.g.,][]{Evrard_2008}, this dispersion would imply an unphysical dynamical mass of order $\sim10^{16}M_\odot$ - roughly a factor of 10 greater than our strong lensing and X-ray estimates. Within systems such as MACS J1206 \citep[][]{Richard_2021}, similarly inflated velocity dispersions paired with relatively undisturbed X-ray morphologies are explained by mergers occurring along the line of sight. However, as previously established, the merger axis for J1150$-$2805 appears to be in the plane of the sky. The coexistence of a `mixed' X-ray morphological classification \citep{Campitiello_2022}, an apparent plane of sky merger and a large LOS velocity dispersion points to a complex, multi-axis accretion scenario.

The one-dimensional velocity distribution alone offers little direct evidence of this apparent complexity. To quantitatively test for departures from a normal distribution, we apply the Anderson-Darling \citep[AD,][]{Anderson_Darling_1952} test, finding the kinematics to be statistically indistinguishable from a single Gaussian (left panel of Fig.~\ref{fig:DS_Bubble_plot}; $A^{2}=0.62$, below the $5 \%$ critical value of $0.75$). This is, however, the expected signature of a merger proceeding largely in the plane of the sky \citep[the generic behaviour of radio-relic mergers found by][]{Golovich_2019}. A purely one-dimensional analysis would, therefore, be blind to the substructure that the radio, X-ray and lensing data imply. The Dressler--Shectman (DS) test \citep{Dressler_Shectman_1988} is designed precisely for this regime, because it is sensitive to localised, spatially-coherent departures in the kinematics rather than to the global velocity histogram. For each galaxy, $i$, we identify its $N_{\rm{nn}}=[\sqrt{N_{CM}}]=17$ nearest neighbours on the sky and compute the local mean velocity $\bar{v}_{\rm local}$ and local dispersion $\sigma_{\rm local}$, which we compare to the global values $\bar{v}$ and $\sigma$ through the deviation

\begin{equation}
  \delta_{i}^{2} = \frac{N_{\rm nn}+1}{\sigma^{2}}
  \left[\left(\bar{v}_{\rm local}-\bar{v}\right)^{2}
  + \left(\sigma_{\rm local}-\sigma\right)^{2}\right].
  \label{eq:ds}
\end{equation}

The cumulative statistic $\Delta=\sum_{i}\delta_{i}$ measures the total kinematic substructure; in the absence of any correlation between position and velocity, one expects $\Delta\simeq N$. We find $\Delta/N=1.26$. To assess its significance, we performed $5000$ Monte-Carlo realisations in which the velocities are randomly shuffled among the fixed galaxy positions, thereby destroying any spatial-kinematic correlation while preserving both the velocity and the spatial distributions \citep{Pinkney_1996}. Only $2.5\%$ of the random realisations reach the observed $\Delta$, giving $p=0.025$ and rejecting the null hypothesis of no substructure at the
$\sim$$97.5\%$ confidence level.

The spatial-kinematic structure underlying this result is shown in the right hand panel of Fig. \ref{fig:DS_Bubble_plot}, where each galaxy is plotted as a circular bubble whose area scales as $e^{\delta_{i}}$ and whose colour encodes the local mean velocity. To illustrate the structure of the lensing signal, we plot the critical curves (pink) evaluated at a reference source redshift of $z_{s}=9.0$. Additionally, we mark the optimised centres of the five cluster-scale dark matter haloes (green stars) established by the Spec model. Rather than a single random scatter, the high-$\delta$ galaxies organise into four spatially segregated, kinematically-coherent groupings distributed along the North-West to South-East axis, alternating between redshifted and blueshifted local velocities. 

The most prominent deviation ($\delta_{i}\gtrsim3$) manifests as a compact, strongly redshifted group located East of the cluster core (right panel of Fig. \ref{fig:DS_Bubble_plot}). Furthermore, we identify three distinct blueshifted subgroups situated to the North, South-East, and West of the core. Strikingly, the spatial coordinates of these dynamical substructures are highly coincident with the optimised positions of the cluster-scale DM haloes (in particular, the four secondary DM haloes 2, 3, 4, and 5) recovered independently through our joint strong lensing and X-ray modelling (Section \ref{sec:comparison_of_mass_models}). The DS substructures provide an independent, purely dynamical confirmation of the complex multi-component mass distribution implied through the strong lensing. This striking spatial concordance between the kinematic groupings and the modelled DM haloes is highly unlikely to arise by chance and, therefore, provides further compelling justification for our decision to model the cluster with a bimodal core and an additional northernmost halo. We therefore report the detection of two new substructures within the central core of J1150$-$2805, which we associate to both the Eastern DM halo (Halo 4) and, more tentatively, to the Northern DM halo (Halo 5). Such association between observed galaxy groups and the strong lensing inferred DM haloes avoids the need for a dark clump in the north of the models. Additionally, the detection of these substructures via our two distinct methods provides more confidence in their physicality. Overall, the identification of these surviving subhaloes provides critical missing context for the assembly history of the cluster \cite{Natarajan+2004, Natarajan+2007,Natarajan_2009, Natarajan_2017}.

The large-scale $\sim$4~Mpc double radio relics \citep{Bagchi_2011, Bonafede_2014}, giant radio halo \citep{Rajpurohit_2025}, and North-West X-ray shock \citep{Gitti_2025} are signatures of a primary, cluster-wide collision driven primarily in the plane of the sky. By contrast, our lensing and kinematic findings probe the dense inner core of the cluster. Zooming in on this smaller scale, our work captures a complex 3D accretion geometry. The elongated central mass distribution (Figs. \ref{fig:mass_map_spec} and \ref{fig:magnification_power_comparison_plot}) clearly traces the primary merging axis in the plane of the sky, whereas the inflated velocity dispersion and the identified substructures provide independent evidence for secondary merging events occurring along the line of sight. Ultimately, our lensing and kinematic findings contribute to a cohesive, multi-wavelength picture of the J1150$-$2805 galaxy cluster as an ongoing, multiple merger simultaneously accreting across different scales and axes.

\subsection{Lensing Power and Magnification}
\label{sec:lensing_power_and_magnification}

We quantify the lensing power of J1150$-$2805 through its effective Einstein radius, $\theta_{\mathrm{E}} \equiv \sqrt{A/\pi}$, where
\textit{A} is the area enclosed by the tangential critical curve \citep[][]{Zitrin_2015, Meneghetti_2017}.

Fig. \ref{fig:einstein_radius_comparison_plot} shows how $\theta_{\mathrm{E}}$ varies as a function of $z_{s}$ for J1150$-$2805 (solid dark blue curve). For the sake of comparison, we also plot the same relation for all of the Hubble Frontier Fields \citep[HFF - dashed lines,][]{Lotz_2017} clusters, which provide the natural benchmark for assessing cluster lensing strength, having been selected as among the most efficient known lenses and modelled to high precision by several independent teams \citep[][]{Meneghetti_2017}. For our analysis, we use the publicly available v4 and v4.1 convergence and shear maps produced by the CATS team using \textsc{LensTool} \citep[][]{Richard_2014, Jauzac_2014, Jauzac_2015, Jauzac_2016, Lagattuta_2017, Jauzac_2018, Mahler_2018, Lagattuta_2019}. As the profiles derived from the Spec model and the Phot model are virtually superimposed, we plot only the Spec model in Fig. \ref{fig:einstein_radius_comparison_plot} for clarity.

\textbf{\begin{figure}
	\includegraphics[width=\columnwidth]{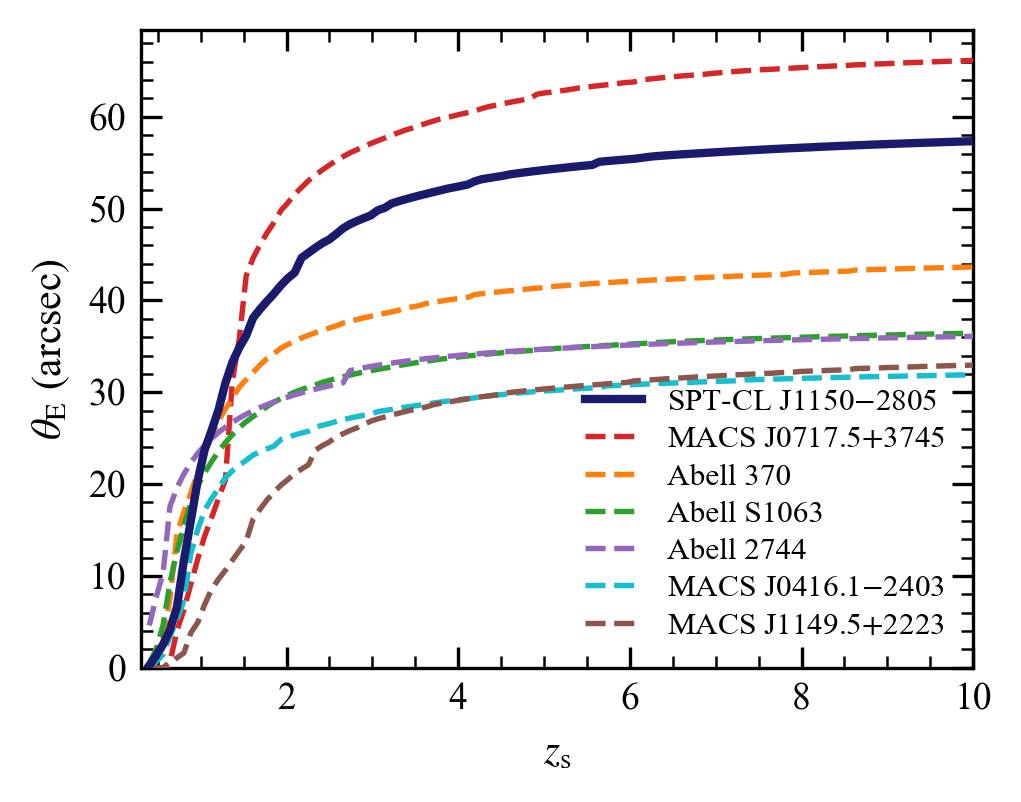}
    \caption{Effective Einstein radius ($\theta_{\mathrm{E}}$) of J1150$-$2805 (solid blue curve) as a function of source redshift ($z_{s}$). Hubble Frontier Fields (HFF) clusters are also plotted for comparative purposes (dashed lines).}
    \label{fig:einstein_radius_comparison_plot}
\end{figure}}

For an arbitrary source at $z_{s}=2$, we measure $\theta_{\mathrm{E}} = 42.41^{+0.05}_{-0.04}\arcsec$, enclosing a projected mass $M(< \theta_{\mathrm{E}}) = 3.35^{+0.04}_{-0.03} \times 10^{14} M_{\odot}$, and the curve rises to $\theta_{\mathrm{E}} = 57.34^{+0.06}_{-0.07}\,\mathrm{''}$, enclosing $M(< \theta_{\mathrm{E}}) = 5.14^{+0.04}_{-0.06} \times 10^{14} M_{\odot}$ as $z_{s} \rightarrow \infty$. As expected from standard lensing geometry, the cross-section asymptotes and remains largely flat for sources beyond $z_{s} \simeq 2$. This is what makes intermediate-redshift, high-mass clusters, such as J1150$-$2805, efficient magnifiers of the entire high-z background sky rather than of a narrow redshift slice.

It is important to stress that lensing strength is not governed by total mass alone. Using more than 100 cluster models, \citet[][]{Remolina_Gonzalez_2021} demonstrated that the inner slope of the projected mass-density profile correlates more strongly with lensing efficiency than does the large-scale halo mass. Furthermore, simulations by \citet[][]{Meneghetti_2017} highlight the critical roles of projected concentration, triaxiality, and substructure. A dynamically disturbed, elongated mass distribution - such as the post-merger configuration of J1150$-$2805 - can substantially enlarge the tangential critical curve at fixed mass, as the superposed deflection fields of the merging components act to extend the critical curve. This naturally accounts for a lensing cross-section that is large relative to relaxed clusters of comparable mass, and underscores the value of merging clusters for high-redshift magnification studies \citep[][]{Kneib_and_Natarajan_2011, Coe_2019}.

A complementary measure of a cluster's lensing strength - one particularly relevant for high-redshift galaxy surveys - is its cumulative magnification power: the area in the source plane that is magnified above a given factor, $\mu$ \citep[][]{Johnson_2014}. Practically, each image-plane pixel of solid angle, $A_{\mathrm{pix}}$, de-lenses to a source-plane area defined as $A_{\mathrm{pix}}/\lvert \mu \rvert$, so that

\begin{equation}
\label{eq: magnification_area_eqn}
A(>\mu) = \sum_{|\mu_i| > \mu} \frac{A_{\mathrm{pix}}}{|\mu_i|}
\end{equation}

which at high magnification approaches the universal fold-caustic limit $A(>\mu) \propto \mu^{-2}$ \citep[][]{Blandford_1986}. Multiplying the source-plane solid angle by the comoving volume of a thin redshift shell converts $A(>\mu)$ into the comoving volume surveyed above a given magnification \citep[][]{Johnson_2014}. 

\begin{figure*}
	\includegraphics[width=\textwidth]{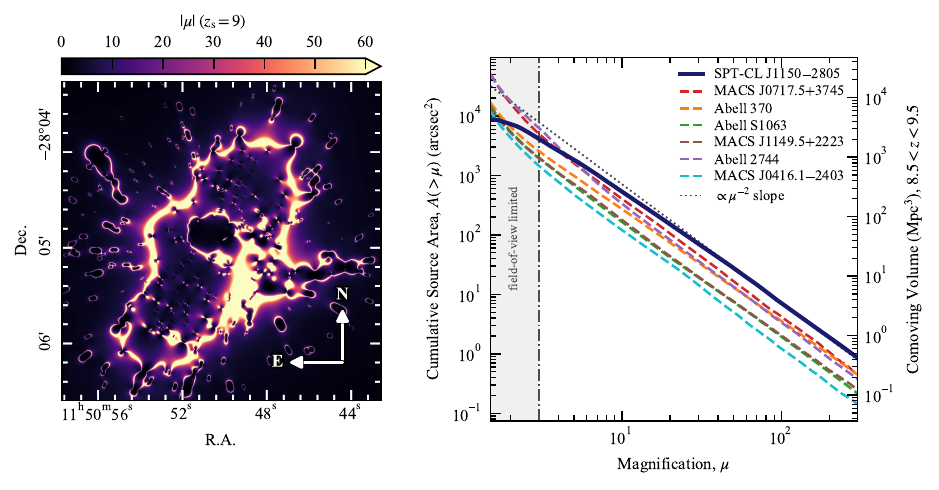}
    \caption{Lensing strength and magnification properties of J1150$-$2805 for high-redshift sources. \textit{Left panel}: Absolute magnification map ($|\mu|$) of the cluster evaluated for a source at $z_{s}=9.0$. The colour scale saturates at $\mu=60$ to highlight the high-magnification regions following the critical curves. \textit{Right panel}: Cumulative source-plane area, A($>\mu$), and the corresponding comoving volume (evaluated over the redshift interval $8.5<z<9.5$) as a function of magnification, $\mu$. The solid dark blue curve represents J1150$-$2805, while the dashed curves show the six Hubble Frontier Fields (HFF) clusters for comparison. The dotted gray line illustrates the theoretical $A(>\mu) \propto \mu^{-2}$ relation expected in the high-magnification regime. The dashed-dotted line and grey shaded region mark the field-of-view limited regime at $\mu > 3$.}
    \label{fig:magnification_power_comparison_plot}
\end{figure*}

Fig. \ref{fig:magnification_power_comparison_plot} presents the cumulative magnification power of  J1150$-$2805 for a source at $z_{s}=9$, computed directly from our \textsc{LensTool} magnification maps using Equation \ref{eq: magnification_area_eqn}. Because the profiles derived from our Spec and Phot models are virtually identical - consistent with our findings in Fig. \ref{fig:einstein_radius_comparison_plot} - we display only the Spec model for clarity (solid dark blue curve). We benchmark J1150$-$2805 against the six HFF clusters (dashed lines), and include an arbitrarily normalised reference slope to illustrate the theoretical fold-caustic $A(>\mu) \propto \mu^{-2}$ scaling. We note that the low-magnification end ($\mu \sim1-3$) is an artifact of the magnification map's total field of view and is therefore not used for comparison. Only the high-magnification regime ($\mu \geq 3$ and particularly $\mu \geq 10-30$) constitutes a robust, field-of-view-independent measure of the true lensing power. Quantitatively, we find high-magnification cross-sections of $A(>10) \simeq 0.15~\mathrm{arcmin}^{2}$ and $A(>30) \simeq 0.020~\mathrm{arcmin}^{2}$ (524 and 70~arcsec$^{2}$). For a direct comparison, we evaluate $A(>\mu)$ for both J1150$-$2805 and the six HFF clusters assuming a common source redshift of $z_{s}=9$ within a standard $100\arcsec$ radial aperture. As shown in Fig. \ref{fig:magnification_power_comparison_plot}, J1150$-$2805 has the largest cross-section of the seven at every threshold $\mu \geq 10$. It exceeds the HFF median by a factor of $3.6$ at $\mu = 30$ and the strongest HFF lens, MACS J0717, by a factor of $2.0$. In terms of survey yield, this corresponds to a comoving volume of $34~\mathrm{Mpc}^{3}$ magnified above $\mu = 30$ in a shell $8.5 < z < 9.5$, against an HFF median of $10~\mathrm{Mpc}^{3}$. 

The physical origin of this exceptional lensing power is visually apparent in the magnification map (left panel of Fig. \ref{fig:magnification_power_comparison_plot}): the tangential critical curve forms an extended, multi-lobed structure spanning a maximum extent of $\sim 190\arcsec$. Its total length is measured at $L \sim 920\arcsec$, against a median of $401 \pm 184\arcsec$ for the six HFF clusters. Crucially, J1150$-$2805 achieves this with an effective Einstein radius $\theta_{\rm E} = 57''$ that is smaller than that of MACS J0717 ($66\arcsec$). Comparing the 2D mass distributions and convergence maps shows that 
J1150$-$2805 has the least concentrated mass distribution in the sample. A flatter mass distribution spreads the deflection field over a larger region, so that at fixed $\theta_{\rm E}$ the region above a given magnification is larger in the source plane. Its advantage lies not in the area enclosed by the critical curve but in the curve's length $L/2\pi\theta_{\rm E} = 2.56$ against $1.89$ for MACS J0717. This elongated morphology is characteristic of a major merger (Section \ref{sec: cluster_dynamical_state}) - a dynamical configuration similarly responsible for the extreme lensing power of MACS J0717 \citep[][]{Limousin_2016, Jauzac_2018}. In such systems, the superposed deflection fields of the merging components stretch the critical curves, drastically enlarging the high-magnification area. At the highest magnifications, the profile of J1150$-$2805 correctly asymptotes to the theoretical fold-caustic slope of $-2$ \citep[][]{Blandford_1986}, verifying the physical behaviour of our modelling near the critical curves.

The cumulative magnification power is the quantity that most directly establishes a cluster's value as a gravitational telescope, as it determines the effective source-plane volume accessible at extreme magnifications and, consequently, the limiting intrinsic luminosity reachable in a fixed exposure \citep[][]{Johnson_2014, Richard_2014}. This metric has become paramount in the \textit{JWST} era. Ultra-deep programmes such as UNCOVER \citep[][]{Bezanson_2024}, GLIMPSE \citep[][]{Atek_2023_proposal} and CANUCS \citep[][]{Willott_2023} exploit cluster lenses to push $\geq2$ magnitudes below blank-field limits, enabling the spectroscopic confirmation of galaxies at $z>9$ \citep[][]{Atek_2023, Furtak_2023} and demonstrating that faint, lensed dwarf galaxies likely dominated the reionising photon budget \citep[][]{Atek_2024}. Currently, the principal limitation of such studies is no longer instrumental depth, but rather the systematics of the magnification models themselves. These systematics dominate the error budget at the faint-end of the ultraviolet luminosity function and differ substantially between reconstruction techniques \citep[][]{Atek_2018}. Enlarging the sample of well-constrained cluster lenses beyond the six HFF fields - particularly by identifying rare, exceptionally massive lenses from within wide-area SZ surveys like SPT \citep[][]{Bleem_2015, Bleem_2020} - addresses two critical needs. It substantially increases the high-magnification source-plane volume available for discovering rare targets, and it supplies independent lines of sight to help calibrate magnification systematics. Demonstrating a high-$\mu$ cross-section that exceeds those of the HFF benchmark clusters, J1150$-$2805 represents an exceptionally powerful cosmic telescope and stands as a premier target for future deep, multi-band \textit{JWST} observations - akin to that of UNCOVER and GLIMPSE - to push the frontiers of the high-redshift universe \citep[][]{Kneib_and_Natarajan_2011, Coe_2019}.

\subsection{Cluster Member Robustness Testing}
\label{sec:cluster_member_robustness}

As discussed in Sections \ref{sec:cluster_members} and \ref{sec:cluster_member_modelling}, the large spatial extent of J1150$-$2805 necessitated the construction of distinct cluster member catalogues using both \textit{JWST} and \textit{HST} data. Methodologically, this was implemented by defining the \textit{JWST} and \textit{HST} subhalo populations as two independent cluster member catalogues within the \textsc{LensTool} models. This configuration allowed us to optimise the empirical scaling relations (equations. \ref{eq: sigma_scaling_relation_equation} and \ref{eq: cut_radius_scaling_relation_equation}) for each catalogue separately. Isolating the two datasets in this manner allows us to directly compare the global mass scaling parameters (\textit{$\sigma_\mathrm{0}^\mathrm{CM}$} and \textit{$r_\mathrm{cut}^\mathrm{CM}$}) driven by differing photometry. 

Fig. \ref{fig:scaling_relations_plot} shows how these parameters behave as a function of F322W2 apparent magnitude. As the cluster member catalogues predominantly consist of early-type galaxies that reside on a well-defined red-sequence (e.g. Fig. \ref{fig:red_sequence}), we apply a constant colour correction to project the \textit{HST} photometry onto the \textit{JWST} axis. By subtracting the median colour offset of $\Delta m=0.787$ from the F814W observations, both sets of scaling relations can be evaluated against an F322W2 defined baseline. The upper and lower panels display the relations derived from the Spec model and the Phot model, respectively. Within each panel, the scaling relations optimised for the \textit{HST} and \textit{JWST} cluster member catalogues are denoted by red and blue lines, accompanied by their corresponding $1-3\sigma$ credible intervals shaded in the respective colours. We note that the \textit{HST} scaling relations exhibit broader credible intervals than those of the \textit{JWST} catalogue. This is a direct consequence of the spatial distribution of the subhaloes; the \textit{HST} catalogue extends to larger cluster-centric radii where strong lensing multiple images are sparse, resulting in weaker constraints on the mass parameters of these outer members. Additionally, the individually optimised cluster member galaxies are overplotted to illustrate their scatter relative to the global relations. Circular and square markers denote members originally belonging to the \textit{JWST} and \textit{HST} catalogues, respectively. The marker colours indicate the projected physical distance (in kpc) of each member from the cluster core (defined by BCG-1), while the associated error bars denote the $1\sigma$ uncertainties extracted from the MCMC posteriors.

\textbf{\begin{figure*}
	\includegraphics[width=\textwidth]{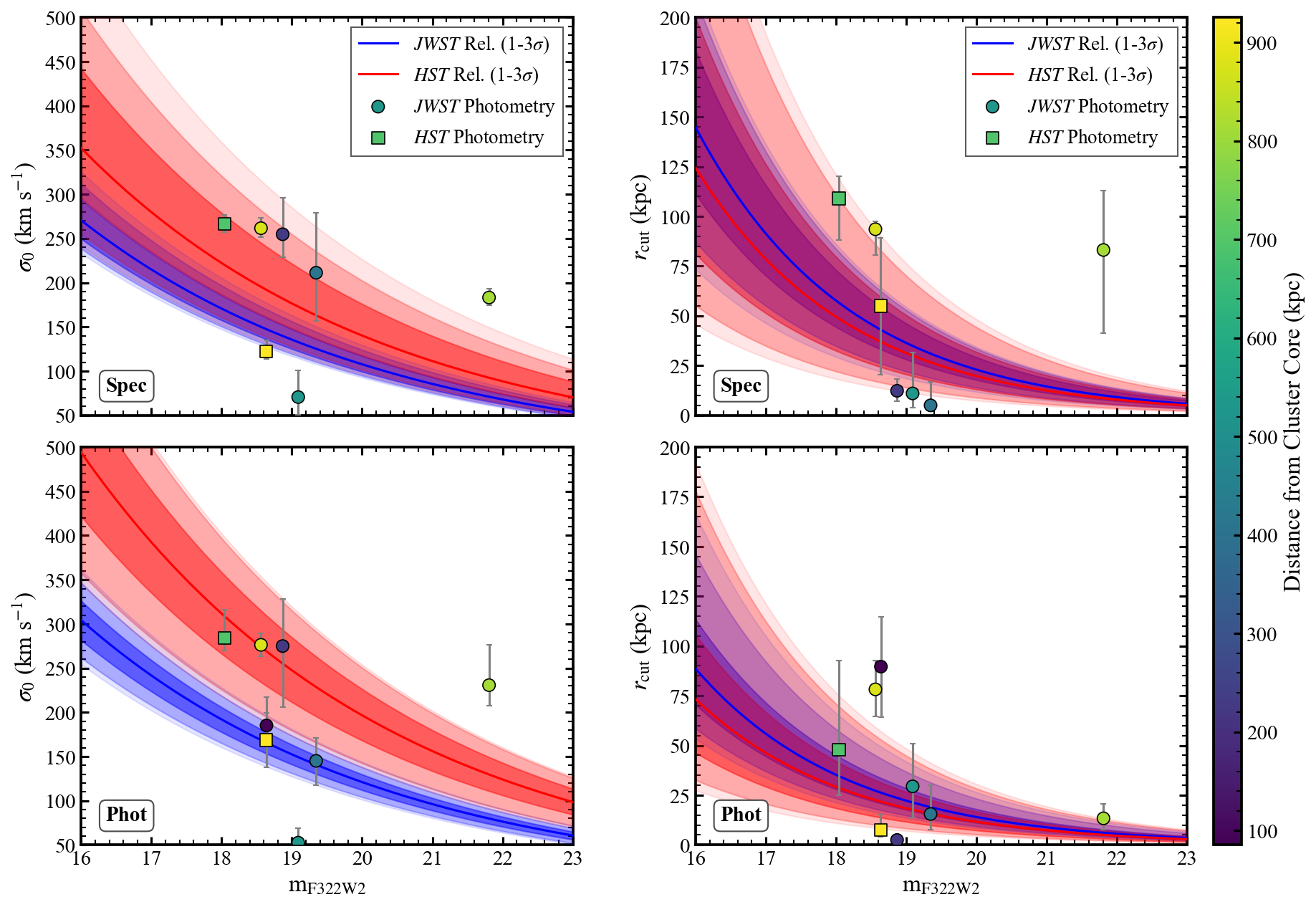}
    \caption{Galaxy-scale component scaling relations for J1150$-$2805 under two modelling configurations: the Spec model (\textit{top row}) and the Phot model (\textit{bottom row}). \textit{Left column}: The central velocity dispersion ($\sigma_0$) as a function of the reference magnitude ($m_{\mathrm{F322W2}}$). \textit{Right column}: The corresponding truncation radius ($r_{\mathrm{cut}}$) as a function of the reference magnitude ($m_{\mathrm{F322W2}}$). In all panels, the solid blue and red lines represent the median Faber-Jackson scaling relations for the \textit{JWST} and \textit{HST} cluster member populations, respectively, with shaded regions denoting the nested 1, 2, and 3$\sigma$ confidence intervals. Individual cluster members with freely optimised potentials are overlaid as data points (circles for \textit{JWST} photometry, squares for \textit{HST} photometry) representing their $1\sigma$ marginalised uncertainties. The markers are colour-coded by their projected physical distance from the cluster core}
    \label{fig:scaling_relations_plot}
\end{figure*}}

In both model variations, the \textit{$r_\mathrm{cut}^\mathrm{CM}$} relations derived from the two catalogues are fully consistent with each other, while the \textit{$\sigma_\mathrm{0}^\mathrm{CM}$} relations show a clear, systematic offset that is reproduced in both the Spec and Phot models. Both relations are calibrated against the measured luminosity in a given band, so, in principle, neither relation is independent of the photometric band adopted. The luminosity itself is an imperfect and band-dependent tracer of the underlying stellar (and total subhalo) mass, and the position angle and ellipticity entering the dPIE profiles are likewise measured on the corresponding image and can, therefore, differ between catalogues. Nevertheless, we interpret the consistency between the \textit{$r_\mathrm{cut}^\mathrm{CM}$} relations as evidence that our mass modelling is internally consistent across the whole set of cluster members regardless of the photometric catalogue used, which is itself a useful validation of our method.

The fact that a systematic offset is visible in \textit{$\sigma_\mathrm{0}^\mathrm{CM}$} but not in \textit{$r_\mathrm{cut}^\mathrm{CM}$} is better understood as a consequence of the different power-law slopes relating the two parameters to luminosity (e.g. Section \ref{sec:cluster_member_modelling}). For a given fractional bias or scatter in $L$ introduced by the choice of band, the induced variation in $\sigma_\mathrm{0}$ is roughly half, in logarithmic terms, of that induced in $r_\mathrm{cut}$. \textit{$\sigma_\mathrm{0}$} is consequently the tighter, more sensitive tracer of the two: a genuine offset in the underlying luminosities between the \textit{JWST} and \textit{HST} catalogues is more likely to remain visible in the \textit{$\sigma_\mathrm{0}^\mathrm{CM}$} relation, while the intrinsically larger scatter of \textit{$r_\mathrm{cut}^\mathrm{CM}$} is more likely to absorb the same bias, blending it into the relation rather than producing a resolvable offset. This offers a natural explanation for why both relations can be affected by the choice of photometric band, yet only one shows a resolvable offset in practice.

The offset itself is plausibly linked, at least in part, to the differing mass-to-light properties traced by the \textit{JWST} and \textit{HST} photometric bands: \textit{JWST} mid-infrared observations trace older stellar populations and are, in general, a somewhat more direct proxy for stellar mass, whereas \textit{HST} optical/near-infrared imaging is more sensitive to recent star formation and more susceptible to dust attenuation \citep[][]{Madau_2014}. As \textit{$\sigma_\mathrm{0}$} acts as the normalisation connecting observed luminosity to total subhalo mass, such a systematic difference in mass-to-light ratio between bands would be expected to shift its calibration, consistent with what we observe.

An important caveat when interpreting these results is the degeneracy between \textit{$\sigma_\mathrm{0}$}
and \textit{$r_\mathrm{cut}$} inherent to the dPIE mass profile \citep[][]{Bergamini_2019}. Since the total mass of a dPIE subhalo scales as $M\propto\sigma^2r_\mathrm{cut}$, an increase in the central velocity dispersion can be mathematically offset by a proportional decrease in the truncation radius to yield an identical total halo mass. Consequently, it can be difficult for the models to independently decouple these two parameters, because the strong lensing deflection field primarily constrains the total enclosed mass. Fig. \ref{fig:scaling_relations_plot} shows that, despite this degeneracy, the majority of individually optimised cluster members remain in reasonable agreement with the global scaling relations, even though they are freed from these constraints during the modelling; the BCGs are excluded from this comparison, as their distinct formation histories and extended stellar envelopes make them unrepresentative of the general cluster member population that the scaling relations are intended to describe \citep[][]{Bernardi_2007}. A number of individual members do scatter away from both relations, which is to be expected given the simplifying assumptions underlying the relations - such as a common mass-to-light behaviour and a single dPIE parametrisation applied across the population - that are necessarily neglected at the level of any individual member. Nonetheless, the agreement seen for the bulk of the population indicates that the global relations provide a fair representation of the cluster members: it suggests that individual subhaloes are reasonably constrained by the localised strong lensing data, and that the global scaling relations provide a fair representation of the cluster member population.

To ensure a robust comparison, the selection of individually optimised cluster members was kept consistent between the Spec and Phot models. The sole deviation occurs in the Phot model, where an additional member was freed from the scaling relations to account for localised mass perturbations near the Silver class multiple images as discussed in Section \ref{sec:cluster_member_modelling}. Given that nine of the ten individually optimised members (including the BCGs) are subject to identical local strong lensing constraints in both models, they serve as an ideal control sample to test the consistency of the mass decomposition. Theoretically, their physical masses should remain conserved between the models. We therefore calculate the total physical mass of these subhaloes via:

\begin{equation}
M_{\text{tot}} = \frac{\pi \sigma_0^2 r_{\text{cut}}}{G},
\label{eq:subhalo_mass}
\end{equation}

and present a direct, one-to-one comparison of the resulting masses derived from each model in Fig. \ref{fig:individually_optimised_cluster_member_masses}. Subhaloes whose masses remain robustly consistent across both models will fall along the diagonal grey dashed line of unity. Any significant deviation from this relation indicates that the mass convergence of that specific subhalo was forced to adjust to changes in the macroscopic cluster potential, despite its own localised constraints remaining identical. Visually, Fig. \ref{fig:individually_optimised_cluster_member_masses} shows good agreement between the two models. To quantitatively assess this, we calculate the statistical tension of each individually optimised member relative to the 1-1 relation, factoring in the combined uncertainties of both models. We find that 5 out of 9 are consistent within $1 \sigma$, 8 within $2 \sigma$ and all 9 are consistent within $3 \sigma$. Furthermore, we calculate the root-mean-square error in log space (RMSE) and the reduced chi-square $(\chi^{2}_{\nu})$ against the 1-1 line where $\chi^{2}_{\nu}$ is calculated as the distance of each point from the line divided by the combined uncertainty $(\sigma^{2}_{x}+\sigma^{2}_{y})$. We find RMSE = 0.36 dex, indicating that the total subhalo masses are conserved within a factor of $10^{0.36} \approx 2.3$ range and $\chi^{2}_{\nu} = 1.765$, indicating only slight intrinsic scatter from the 1-1 line. Therefore, the two models are statistically consistent with each other in their mass reconstruction of the individually optimised cluster member galaxy subhaloes.

\textbf{\begin{figure}
	\includegraphics[width=\columnwidth]{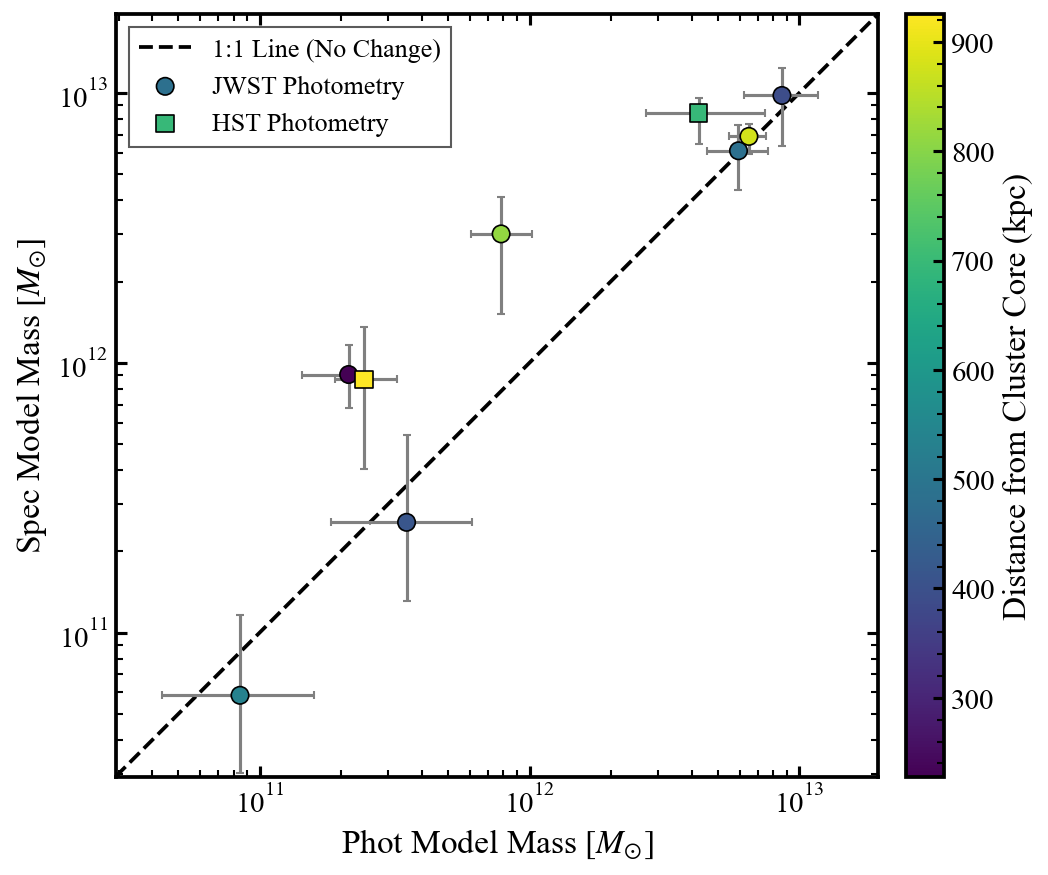}
    \caption{One-to-one comparison of the derived masses for individually optimised cluster members between the Phot model ($x$-axis) and the Spec model ($y$-axis) of J1150$-$2805. The dashed black line indicates a 1:1 relation. Data points correspond to individually optimised subhaloes and are plotted with their $1\sigma$ marginalised uncertainties. The marker shape denotes the origin of the subhalo's photometry (circles for \textit{JWST}, squares for \textit{HST}), while the colour scale represents the projected physical distance of each member from the cluster core in kpc.}
    \label{fig:individually_optimised_cluster_member_masses}
\end{figure}}

\subsection{GGSL in the Context of Scaling Relations}
\label{sec:GGSL_section}

Independent kinematic studies have robustly established that intrinsic variance exists among cluster members \citep{Bergamini_2019, Granata_2022, Bergamini_2023}. When applying global scaling relations to cluster mass models, any scatter in the mass profiles of the member populations is automatically lost. Although we establish in Section \ref{sec:cluster_member_robustness} that these scaling relations provide a robust approximation for the overall cluster member population of J1150$-$2805, \citet{DAddona_2024} demonstrated that this cluster exhibits a higher-than-expected probability of galaxy-galaxy strong lensing \citep[GGSL;][]{Meneghetti_2020,Tokayer+2024,Dutra+2025,Chiang+2026,Natarajan+2026,Roche+2026a}. The GGSL cross-section is fundamentally tied to the mass distribution of individual cluster members. Therefore, J1150$-$2805 presents an ideal opportunity to investigate whether enforcing rigid scaling relations artificially biases the predicted strong lensing cross-section of the cluster. Following \citet{Meneghetti_2020,Meneghetti_2022} we define:

\begin{equation}
\label{eq:p_ggsl}
P_{\mathrm{GGSL}}(z_s)
  = \frac{\sigma_{\mathrm{GGSL}}(z_s)}{A_s(z_s)},
\end{equation}

where $\sigma_{\mathrm{GGSL}}$ is the total source plane area enclosed by all galaxy-scale tangential caustics with Einstein radii in the $0.5\arcsec\!<\!\theta_{\mathrm{E}}\!<\!3\arcsec$ range, and $A_s$ is the image-plane field of view de-lensed through Equation \ref{eq:lens_equation} onto the source plane \citep[][Eq.~S7]{Meneghetti_2020}. For each realisation of the scattered scaling relation, we re-derive the cluster deflection maps with \textsc{LensTool} and recompute both $\sigma_{\mathrm{GGSL}}$ and $A_s$ on the same $200\arcsec \times 200\arcsec$ image-plane field, evaluated at nine source redshifts spanning $z_s \in [1,\,7]$.

We parametrise the scatter as a single number $\sigma_{\log M}$, the $1\sigma$ width of a per-galaxy log-normal perturbation applied independently to every cluster member. For each Monte~Carlo realisation of the best-fit baseline, the total mass of each member is multiplied by $10^{\delta_i}$ with $\delta_i \sim \mathcal{N}(0,\,\sigma_{\log M}^2)$. Because the dPIE truncation radius is held fixed and the dPIE total mass satisfies $M \propto \sigma_v^2\,r_{\mathrm{cut}}$, the perturbation reduces exactly to a kinematic scatter:

\begin{equation}
\sigma_{\log \sigma_0} \;=\; \frac{1}{2}\,\sigma_{\log M},
\end{equation}

i.e.\ each member's velocity dispersion is rescaled by $10^{\delta_i / 2}$. $\sigma_{\log M} = 0$ recovers the model's rigid scaling relation baseline. We sweep $\sigma_{\log M} \in \{0.10, \allowbreak 0.20, \allowbreak 0.30, \allowbreak 0.40, \allowbreak 0.50, \allowbreak 0.60\}$\,dex with 50 independent realisations per level, corresponding to $\sigma_{\log \sigma_v} \in \{0.05, \allowbreak 0.10, \allowbreak 0.15, \allowbreak 0.20, \allowbreak 0.25, \allowbreak 0.30\}$\,dex. From the MUSE spectroscopic calibration of the Faber-Jackson relation on $\sim\!60$ HFF cluster members, \citet{Bergamini_2019} report a residual scatter $\sigma_{\log \sigma_v} \approx 0.05 - 0.10$\,dex, and \citet{Bergamini_2023} obtain a comparable value for the Abell S1063 sample. Our lower two sweep points ($\sigma_{\log M} = 0.10,\,0.20$\,dex) therefore sit exactly on the lower and upper \citet{Bergamini_2019} bounds, and the upper four points ($0.30$--$0.60$\,dex) extend the test up to a factor of three above the spectroscopic upper limit, deliberately probing scatter levels for which there is no physical motivation in order to expose any latent sensitivity of $P_{\mathrm{GGSL}}$ to the assumption.

In Fig. \ref{fig:GGSL_comparison_plot}, we show the effect of introducing kinematic scatter to both the Spec (blue) and Phot (orange) models on the calculated GGSL probability of the cluster. At physically motivated levels, the response of $P_{\rm GGSL}$ is small. Within $\sigma_{\log M} \le 0.20$\,dex the redshift-averaged change from the baseline is $-4.5 \pm 1.1\,\%$ and $-10.1 \pm 1.8\,\%$ at $\sigma_{\log M} = 0.10$ and $0.20$\,dex for the Phot model, and $+1.8 \pm 1.4\,\%$ and $+3.4 \pm 1.7\,\%$ for the Spec model. Pushed into an unphysical regime, the models then diverge. The Phot model declines monotonically, reaching $-19\, \%$ $\sigma_{\log M} = 0.60$\,dex. The Spec model is consistent with no change up to $0.30$\,dex and then declines gently to $-8\,\%$ at $\sigma_{\log M} = 0.60$\,dex. These declines are driven by the largest cluster members being pushed over the allowed secondary critical curve size limit ($0.5\arcsec\!<\!\theta_{\mathrm{E}}\!<\!3\arcsec$) as all the cluster members are given more freedom to scatter. Overall, both models experience declines in $P_{\rm GGSL}$, but neither is significant. At the levels supported by spectroscopic calibrations, it moves, at most, by $\sim 10\,\%$; far less than the factor of $\sim 2$ separating the Phot and Spec models. This contrast between the two models demonstrates that the predicted GGSL cross-section is sensitive to the exact calibration of the cluster member scaling relations. Crucially, it highlights that two models can be virtually indistinguishable on the macroscopic cluster scale, yet exhibit fundamentally different mass distributions on the small scales responsible for galaxy-galaxy strong lensing. From these results, it is clear that accurately recovering the small-scale mass distribution requires a more standardised approach to treating cluster member populations. Securing  spectroscopic data from instruments like MUSE is vital to properly calibrate these scaling relations and mitigate the biases introduced by using the lensing constraints alone.

\textbf{\begin{figure}
	\includegraphics[width=\columnwidth]{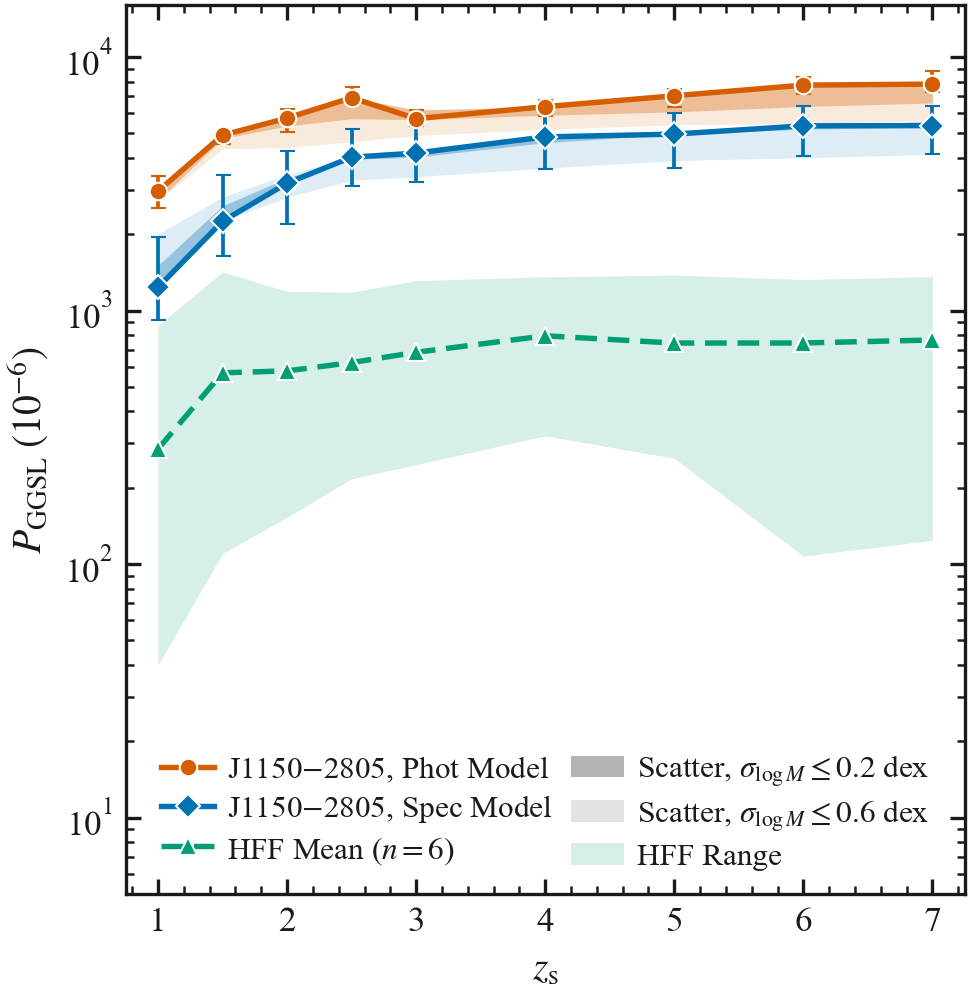}
    \caption{Galaxy--galaxy strong-lensing probability versus source redshift for J1150$-$2805 (Orange circles: Phot model; blue diamonds: Spec model), compared with the six HFF clusters (green dashed: sample mean; green band: full range across the six). Error bars are statistical $1\sigma$ intervals from posterior sampling of the lens model parameters. Shaded orange and blue bands show the systematic effect of scatter in the cluster member scaling relation, for $\sigma_{\log M} \le 0.20$\,dex (darker) and $\le 0.60$\,dex (lighter).}
    \label{fig:GGSL_comparison_plot}
\end{figure}}

Additionally, we note a substantial excess in lensing efficiency relative to the HFF sample (green in Fig. \ref{fig:GGSL_comparison_plot}). For the Phot model, J1150$-$2805 lies a factor of $\sim 8-11$ above the HFF mean at every $z_{\rm s}$ tested; for the Spec model, the factor is $\sim 4-7$. From the Phot model we obtain $P_{\mathrm{GGSL}}(z_{\rm s}=3) = 5.71^{+0.47}_{-0.39} \times 10^{-3}$ and $P_{\mathrm{GGSL}}(z_{\rm s}=6) = 7.75^{+0.63}_{-0.59} \times 10^{-3}$, and from the Spec model $P_{\mathrm{GGSL}}(z_{\rm s}=3) = 4.18^{+1.18}_{-0.96} \times 10^{-3}$ and $P_{\mathrm{GGSL}}(z_{\rm s}=6) = 5.35^{+1.05}_{-1.28} \times 10^{-3}$, compared with HFF means of $6.8 \times 10^{-4}$ and $7.5 \times 10^{-4}$. Even the strongest HFF clusters fall a factor of $\sim 3-4$ below either model at $z_{\rm s} = 3$. J1150$-$2805 is therefore an exceptionally powerful gravitational lens.

Our lensing efficiency is nevertheless significantly lower than previous estimates for this cluster. \citet{DAddona_2024} report
$P_{\mathrm{GGSL}}^{\mathrm{D24}}(z_{\rm s}=6) = 1.7^{+0.5}_{-0.2} \times 10^{-2}$, whereas we obtain $P_{\mathrm{GGSL}}(z_{\rm s}=6) = 7.75^{+0.63}_{-0.59} \times 10^{-3}$ from our Phot model and $5.35^{+1.05}_{-1.28} \times 10^{-3}$ from our Spec model. These are lower by factors of 2.2 and 3.2, and the difference is significant at $\gtrsim 4\sigma$ for both models. Even with our lower estimates, J1150$-$2805 appears to be a distinct outlier among observed clusters - an excess that we have shown cannot be resolved by introducing scatter into the scaling relations. Resolving the physical origin of this extreme lensing efficiency will be the subject of future work, further motivating the continued study of this system as a premier cosmic telescope.

\section{Conclusions} \label{sec:conclusions}

In this work, we have presented a new, high-precision strong-lensing $+$ X-ray mass model of the merging galaxy cluster SPT-CLJ1150$-$2805 ($z=0.383$), built upon the exquisite imaging delivered by the \textit{JWST} SLICE programme. By combining an unprecedented sample of multiple-image constraints with an independent, X-ray-derived intracluster medium (ICM) mass component, we have reconstructed the cluster potential to sub-arcsecond accuracy and characterised its mass distribution, dynamical state, and lensing power. To assess the robustness of our reconstruction, we constructed two models in parallel: a Spec model, constrained solely by gold (spectroscopically confirmed) multiple images, and a Phot model, which additionally incorporates Silver (photometric) systems. Our principal findings are summarised below.

\begin{enumerate}

    \item The cluster is best described by a multi-component mass distribution comprising five cluster-scale DM haloes and three cluster-scale ICM potentials. We find decisive statistical preference for a bimodal core in which each of the two BCGs hosts its own distinct DM halo (Haloes 1 and 4), together with a fifth, North-East halo (Halo 5) which we are able to tentatively associate with a baryonic substructure. Relative to the three-halo baseline, the full five-halo parametrisation is strongly favoured by the Bayesian evidence alongside minimised information criteria. This represents the first strong-lensing analysis to explicitly model and resolve the bimodal core of J1150$-$2805, and we robustly recover the secondary West (Halo 2) and South-East (Halo 3) structures, whose positions coincide with the haloes identified in the weak-lensing analysis of \citet{Finner_2017}. 

    \item An independent kinematic analysis of 291 spectroscopic cluster members corroborates the disturbed, multi-component dynamical state of the core inferred from strong lensing. We measure an exceptionally high rest-frame line-of-sight velocity dispersion of $\sigma_v=1960^{+84}_{-92}\,\mathrm{km\,s^{-1}}$, among the largest known for a radio-relic cluster and consistent with previous estimates for this system. Our Dressler-Shectman test reveals multiple substructures ($\Delta/N=1.26$, $p=0.025$). The four kinematically coherent subgroups identified spatially coincide with the secondary DM haloes recovered through lensing, providing a purely dynamical confirmation of the cluster's multi-component nature and the detection of two new core substructures (the East Halo 4 and North-East Halo 5). Our results add to a coherent story of an ongoing multiple merger accreting simultaneously across different scales and axes: a primary collision proceeding largely in the plane of the sky at large scale, with the distinct substructures in the core pointing to additional, secondary mergers along the line of sight.

    \item Our fiducial models reproduce the observed multiple image positions with an image-plane scatter of ${\mathrm{RMS}}=0.56''$ in both cases. The Spec model uses 118 multiple images as constraints while the Phot model uses 149. These results are a substantial improvement over previous reconstructions of this cluster \citep[$0.75''$ from 47 images;][]{DAddona_2024}. Crucially, the inclusion of the 31 additional Silver images does not change the RMS value, and every optimised parameter agrees between the two models within $2\sigma$. The two reconstructions are statistically indistinguishable across $\simeq95\%$ of the field. This strong agreement demonstrates the robustness of our modelling.

    \item Our mass reconstruction confirms J1150$-$2805 occupies the extreme high-mass end of the cluster mass function, with $M(<1~\mathrm{Mpc}) = 1.86 \pm 0.02 \times 10^{15} M_{\odot}$. The Spec and Phot models recover total enclosed-mass profiles that agree within $1\sigma$.

    \item Additionally, J1150$-$2805 is an exceptionally powerful gravitational lens. Its effective Einstein radius ranges from $\theta_{\mathrm{E}}=42.41^{+0.05}_{-0.04}{''}$ at $z_{\rm s}=2$ to $57.34^{+0.06}_{-0.07}{''}$ as $z_{\rm s}\rightarrow\infty$. Its high-magnification cross-sections ($A(>10)\simeq0.15\,\mathrm{arcmin}^2$, $A(>30)\simeq0.02\,\mathrm{arcmin}^2$) exceed those of all six HFF clusters, surpassing the strongest HFF lens, MACS~J0717, by a factor of $2.0$ at $\mu=30$. This exceptional efficiency is driven by the superposed deflection fields of the merging components, which stretch the tangential critical curve into an extended, multi-lobed structure spanning $\sim190\arcsec$, establishing the cluster as a premier cosmic telescope.

    \item The cluster produces an anomalous GGSL cross-section that surpasses the HFF mean by up to a factor of 11. Our tests reveal this excess is insensitive to intrinsic scatter being introduced into the cluster member population. However, the $P_{\mathrm{GGSL}}$ statistic proves highly sensitive to the scaling relation calibration, highlighting that models agreeing on the macroscopic scale can diverge significantly on the subhalo scales responsible for GGSL. Ultimately, we find that the standard scaling relations used in cluster mass modelling offer a fair approximation to the population but need reliable calibration to minimise the effects of bias.

    \item Finally, we find that incorporating an X-ray-constrained ICM component is statistically favoured (yielding the lowest $\chi^2$, BIC, and AICc) over lensing-only models. Decoupling the collisional gas from the collisionless DM  reduces the mass of the primary DM halo (as expected) - decreasing the net DM mass in the cluster core by $\sim 9\%$ - without altering its core radius or density slope. Furthermore, while the primary macro-potential is sensitive to this mass redistribution, the galaxy-scale scaling relations remain highly robust and insensitive to the chosen ICM configuration.
    
\end{enumerate}

Taken together, these results establish SPT$-$CLJ1150$-$2805 as one of the most massive and powerful gravitational lenses known, and as a dynamically young, multiple-merger system. Our analysis demonstrates the value of combining deep \textit{JWST} imaging, an enlarged multiple-image catalogue, an X-ray-informed ICM component, and independent kinematic testing to construct a self-consistent, multi-wavelength portrait of a complex cluster. Furthermore, we have shown how strong lensing can be uniquely leveraged to resolve fine-grained substructure within complex merging systems. With such a high-magnification cross-section, J1150$-$2805 is a compelling target for future multi-band \textit{JWST} follow-up of the distant Universe. A natural extension of this work would also be to enlarge the spectroscopic footprint beyond the central $<1$~Mpc probed here, enabling the large-scale merger geometry to be mapped directly and the surviving subhaloes to be traced into the cluster outskirts.

\section*{Acknowledgements}

JRB gratefully acknowledges financial support from the Department of Physics at Durham University through the Dowthwaite Scholarship and the Arthur Browne Scholarship.
JRB, BB, MJ and SW acknowledge support from the United Kingdom Research and Innovation (UKRI) Future Leaders Fellowship `Using Cosmic Beasts to uncover the Nature of Dark Matter' (grant number MR/X006069/1).
BB acknowledges the Swiss National Science Foundation (SNSF) for supporting this work. 
ML acknowledges the Centre National de la Recherche Scientifique (CNRS) and the Centre National des Etudes Spatiale (CNES) for support.
P.N. gratefully acknowledges funding from the Department of Energy grant DE-SC001766; support from the John Templeton 
Foundation via grant 126613; and support from NASA GLIMPSE JWST GO-03293.026 for this work. I.D. acknowledges support from 
NASA under award No. 80NSSC25K0311 under the NASA FINESST programme.
This work is primarily based on observations taken by the SLICE program (Program ID: GO-5594; PI: G. Mahler), made with the NASA/ESA/CSA \textit{JWST}. Additionally, our work utilises observations taken by the RELICS Treasury Programme (GO 14096) with the NASA/ESA \textit{HST}. These data were obtained from the Mikulski Archive for Space Telescopes (MAST) at the Space Telescope Science Institute, which is operated by the Association of Universities for Research in Astronomy, Inc., under NASA contract NAS 5-26555. The scientific results reported in this article are also based in part on data obtained from the \textit{Chandra} Data Archive (DOI: \href{https://doi.org/10.25574/cdc.584}{10.25574/cdc.584}), and this research has made use of software provided by the \textit{Chandra} X-ray Center (CXC) in the application package CIAO. Furthermore, this work is based on observations collected at the European Southern Observatory under ESO programme 0102.A-0640(A), and data obtained from the ESO Science Archive Facility (DOI: \url{https://doi.org/10.18727/archive/41}). We acknowledge the use of the \textsc{LensTool} software package \citep{Jullo_2007} for the optimisation of the mass models. Additionally, this research made use of \textsc{SExtractor} \citep{Bertin_1996} and \textsc{Photutils}, an \textsc{Astropy} package \citep{Bradley_2025}, for the detection and photometry of astronomical sources.

\section*{Data Availability}

Our mass models and associated products are made publicly available for download at the Strong Lensing Cluster Atlas Data Base, which is hosted at Laboratoire d'Astrophysique de Marseille (\url{https://data.lam.fr/sl-cluster-atlas/home}). These files can be interpreted with the X-ray version of \textsc{LensTool}. Supplementary models used for comparative purposes in this work can also be shared by the authors upon reasonable request. The spectroscopic catalogue used in this work is publicly available in \citet{DAddona_2024}



\bibliographystyle{mnras}
\bibliography{bib} 




\appendix

\section{Input Parameters and Priors}
\label{sec:input_parameters_and_priors}

In this section we detail the input parameters and priors for the two lens models. Table \ref{tab:lens_params_priors} lists all of the fixed and free parameters. The corresponding optimised parameters can be found in Table \ref{tab:optimised_model_parameters_table}.

\begin{table*}
\centering
\caption{Input parameters and uniform priors used in the strong-lensing optimisation of SPT-CLJ1150$-$2805. The upper half lists the Spec model, the lower half the Phot model.}
\label{tab:lens_params_priors}
\footnotesize
\setlength{\tabcolsep}{3pt}
\renewcommand{\arraystretch}{1.0}
\begin{tabular}{llccccccc}
\toprule
Category & Component & $\Delta x$ & $\Delta y$ & $e$ & $\theta$ & $\sigma_{\rm LT}$ & $r_{\rm core}$ & $r_{\rm cut}$ \\
 & & [arcsec] & [arcsec] & & [$^{\circ}$] & [km\,s$^{-1}$] & [kpc] & [kpc] \\
\midrule
\multicolumn{9}{c}{Spec model} \\
\midrule
Cluster-scale DM & Halo 1 & $\mathcal{U}(-5,\,5)$ & $\mathcal{U}(-5,\,5)$ & $\mathcal{U}(0.001,\,0.9)$ & $\mathcal{U}(-90,\,90)$ & $\mathcal{U}(380,\,1800)$ & $\mathcal{U}(0,\,180)$ & $[2200.00]$ \\
 & Halo 2 & $\mathcal{U}(33,\,53)$ & $\mathcal{U}(-33,\,-13)$ & $\mathcal{U}(0.001,\,0.9)$ & $\mathcal{U}(-90,\,90)$ & $\mathcal{U}(180,\,1500)$ & $\mathcal{U}(0,\,150)$ & $[2200.00]$ \\
 & Halo 3 & $\mathcal{U}(-51,\,-31)$ & $\mathcal{U}(-64,\,-44)$ & $\mathcal{U}(0.001,\,0.9)$ & $\mathcal{U}(-90,\,90)$ & $\mathcal{U}(180,\,1500)$ & $\mathcal{U}(0,\,150)$ & $[2200.00]$ \\
 & Halo 4 & $\mathcal{U}(-21,\,-11)$ & $\mathcal{U}(2,\,12)$ & $\mathcal{U}(0.001,\,0.9)$ & $\mathcal{U}(-90,\,90)$ & $\mathcal{U}(380,\,1500)$ & $\mathcal{U}(0,\,150)$ & $[2200.00]$ \\
 & Halo 5 & $\mathcal{U}(-10,\,10)$ & $\mathcal{U}(24,\,48)$ & $\mathcal{U}(0.001,\,0.9)$ & $\mathcal{U}(-90,\,90)$ & $\mathcal{U}(180,\,1500)$ & $\mathcal{U}(0,\,150)$ & $[2200.00]$ \\
\addlinespace
Foreground galaxies & NE Foregr. & $[-77.20]$ & $[51.76]$ & $[0.10]$ & $[-76.32]$ & $\mathcal{U}(100,\,400)$ & $[0.001]$ & $\mathcal{U}(0,\,150)$ \\
 & W Foregr. & $[46.25]$ & $[-18.95]$ & $[0.80]$ & $[-171.46]$ & $\mathcal{U}(100,\,400)$ & $[0.001]$ & $\mathcal{U}(0,\,150)$ \\
 & NW Foregr. & $[45.14]$ & $[59.21]$ & $[0.07]$ & $[-2.87]$ & $\mathcal{U}(100,\,400)$ & $[0.001]$ & $\mathcal{U}(0,\,150)$ \\
 & SE Foregr. & $[-21.80]$ & $[-39.67]$ & $[0.23]$ & $[-64.69]$ & $\mathcal{U}(100,\,400)$ & $[0.001]$ & $\mathcal{U}(0,\,150)$ \\
\addlinespace
BCGs & BCG 1 (ID 328) & $[0.00]$ & $[0.00]$ & $[0.42]$ & $[-121.31]$ & $\mathcal{U}(100,\,450)$ & $[0.001]$ & $\mathcal{U}(0,\,150)$ \\
 & BCG 2 (ID 559) & $[-16.75]$ & $[7.02]$ & $[0.47]$ & $[-136.84]$ & $\mathcal{U}(100,\,450)$ & $[0.001]$ & $\mathcal{U}(0,\,150)$ \\
\addlinespace
Cluster members & Gal 1745 & $[29.55]$ & $[-28.45]$ & $[0.16]$ & $[-157.32]$ & $\mathcal{U}(50,\,400)$ & $[0.001]$ & $\mathcal{U}(0,\,125)$ \\
 & Gal 637 & $[16.02]$ & $[9.01]$ & $[0.01]$ & $[-157.89]$ & $\mathcal{U}(20,\,250)$ & $[0.001]$ & $\mathcal{U}(0,\,100)$ \\
 & Gal 1684 & $[-38.03]$ & $[32.07]$ & $[0.35]$ & $[-71.99]$ & $\mathcal{U}(50,\,350)$ & $[0.001]$ & $\mathcal{U}(0,\,100)$ \\
 & Gal 631 & $[-0.23]$ & $[31.96]$ & $[0.44]$ & $[-47.50]$ & $\mathcal{U}(50,\,350)$ & $[0.001]$ & $\mathcal{U}(0,\,100)$ \\
 & Gal 1322 & $[67.17]$ & $[-8.26]$ & $[0.15]$ & $[-157.50]$ & $\mathcal{U}(50,\,400)$ & $[0.001]$ & $\mathcal{U}(0,\,125)$ \\
 & Gal 7 & $[77.46]$ & $[-16.16]$ & $[0.22]$ & $[-11.43]$ & $\mathcal{U}(50,\,350)$ & $[0.001]$ & $\mathcal{U}(0,\,100)$ \\
 & Gal 1272 & $[86.14]$ & $[-18.86]$ & $[0.37]$ & $[-144.76]$ & $\mathcal{U}(50,\,350)$ & $[0.001]$ & $\mathcal{U}(0,\,100)$ \\
\addlinespace
X-ray potentials & X-ray Halo 1 & $[15.90]$ & $[19.38]$ & $[0.33]$ & $[126.10]$ & $[321.72]$ & $[207.42]$ & $[2200.00]$ \\
 & X-ray Halo 2 & $[17.30]$ & $[14.74]$ & $[0.85]$ & $[39.25]$ & $[382.00]$ & $[534.29]$ & $[2200.00]$ \\
 & X-ray Halo 3 & $[-48.03]$ & $[-44.77]$ & $[0.58]$ & $[115.33]$ & $[338.28]$ & $[494.57]$ & $[2200.00]$ \\
\addlinespace
Scaling relations & \textit{JWST} & ($m^{\star}=16.83$) & & & & $\mathcal{U}(100,\,450)$ & $[0.001]$ & $\mathcal{U}(10,\,150)$ \\
 & \textit{HST} & ($m^{\star}=18.83$) & & & & $\mathcal{U}(50,\,400)$ & $[0.001]$ & $\mathcal{U}(10,\,125)$ \\
\midrule
\multicolumn{9}{c}{Phot model} \\
\midrule
Cluster-scale DM & Halo 1 & $\mathcal{U}(-5,\,5)$ & $\mathcal{U}(-5,\,5)$ & $\mathcal{U}(0.001,\,0.9)$ & $\mathcal{U}(-90,\,90)$ & $\mathcal{U}(380,\,1800)$ & $\mathcal{U}(0,\,180)$ & $[2200.00]$ \\
 & Halo 2 & $\mathcal{U}(33,\,53)$ & $\mathcal{U}(-33,\,-13)$ & $\mathcal{U}(0.001,\,0.9)$ & $\mathcal{U}(-90,\,90)$ & $\mathcal{U}(180,\,1500)$ & $\mathcal{U}(0,\,150)$ & $[2200.00]$ \\
 & Halo 3 & $\mathcal{U}(-51,\,-31)$ & $\mathcal{U}(-64,\,-44)$ & $\mathcal{U}(0.001,\,0.9)$ & $\mathcal{U}(-90,\,90)$ & $\mathcal{U}(180,\,1500)$ & $\mathcal{U}(0,\,150)$ & $[2200.00]$ \\
 & Halo 4 & $\mathcal{U}(-21,\,-11)$ & $\mathcal{U}(2,\,12)$ & $\mathcal{U}(0.001,\,0.9)$ & $\mathcal{U}(-90,\,90)$ & $\mathcal{U}(380,\,1500)$ & $\mathcal{U}(0,\,150)$ & $[2200.00]$ \\
 & Halo 5 & $\mathcal{U}(-10,\,10)$ & $\mathcal{U}(24,\,48)$ & $\mathcal{U}(0.001,\,0.9)$ & $\mathcal{U}(-90,\,90)$ & $\mathcal{U}(180,\,1500)$ & $\mathcal{U}(0,\,150)$ & $[2200.00]$ \\
\addlinespace
Foreground galaxies & NE Foregr. & $[-77.20]$ & $[51.76]$ & $[0.10]$ & $[-76.32]$ & $\mathcal{U}(100,\,400)$ & $[0.001]$ & $\mathcal{U}(0,\,150)$ \\
 & W Foregr. & $[46.25]$ & $[-18.95]$ & $[0.80]$ & $[-171.46]$ & $\mathcal{U}(100,\,400)$ & $[0.001]$ & $\mathcal{U}(0,\,150)$ \\
 & NW Foregr. (ID 13) & $[45.14]$ & $[59.21]$ & $[0.07]$ & $[-2.87]$ & $\mathcal{U}(100,\,400)$ & $[0.001]$ & $\mathcal{U}(0,\,150)$ \\
 & SE Foregr. (ID 1397) & $[-21.80]$ & $[-39.67]$ & $[0.23]$ & $[-64.69]$ & $\mathcal{U}(100,\,400)$ & $[0.001]$ & $\mathcal{U}(0,\,150)$ \\
\addlinespace
BCGs & BCG 1 (ID 328) & $[0.00]$ & $[0.00]$ & $[0.42]$ & $[-121.31]$ & $\mathcal{U}(100,\,450)$ & $[0.001]$ & $\mathcal{U}(0,\,150)$ \\
 & BCG 2 (ID 559) & $[-16.75]$ & $[7.02]$ & $[0.47]$ & $[-136.84]$ & $\mathcal{U}(100,\,450)$ & $[0.001]$ & $\mathcal{U}(0,\,150)$ \\
\addlinespace
Cluster members & Gal 1745 & $[29.55]$ & $[-28.45]$ & $[0.16]$ & $[-157.32]$ & $\mathcal{U}(50,\,400)$ & $[0.001]$ & $\mathcal{U}(0,\,125)$ \\
 & Gal 637 & $[16.02]$ & $[9.01]$ & $[0.01]$ & $[-157.89]$ & $\mathcal{U}(20,\,250)$ & $[0.001]$ & $\mathcal{U}(0,\,100)$ \\
 & Gal 1684 & $[-38.03]$ & $[32.07]$ & $[0.35]$ & $[-71.99]$ & $\mathcal{U}(50,\,350)$ & $[0.001]$ & $\mathcal{U}(0,\,100)$ \\
 & Gal 631 & $[-0.23]$ & $[31.96]$ & $[0.44]$ & $[-47.50]$ & $\mathcal{U}(50,\,350)$ & $[0.001]$ & $\mathcal{U}(0,\,100)$ \\
 & Gal 1322 & $[67.17]$ & $[-8.26]$ & $[0.15]$ & $[-157.50]$ & $\mathcal{U}(50,\,400)$ & $[0.001]$ & $\mathcal{U}(0,\,125)$ \\
 & Gal 7 & $[77.46]$ & $[-16.16]$ & $[0.22]$ & $[-11.43]$ & $\mathcal{U}(50,\,350)$ & $[0.001]$ & $\mathcal{U}(0,\,100)$ \\
 & Gal 1272 & $[86.14]$ & $[-18.86]$ & $[0.37]$ & $[-144.76]$ & $\mathcal{U}(50,\,350)$ & $[0.001]$ & $\mathcal{U}(0,\,100)$ \\
 & Gal 1337 & $[-67.86]$ & $[38.27]$ & $[0.12]$ & $[-74.13]$ & $\mathcal{U}(50,\,400)$ & $[0.001]$ & $\mathcal{U}(0,\,125)$ \\
\addlinespace
X-ray potentials & X-ray Halo 1 & $[15.90]$ & $[19.38]$ & $[0.33]$ & $[126.10]$ & $[321.72]$ & $[207.42]$ & $[2200.00]$ \\
 & X-ray Halo 2 & $[17.30]$ & $[14.74]$ & $[0.85]$ & $[39.25]$ & $[382.00]$ & $[534.29]$ & $[2200.00]$ \\
 & X-ray Halo 3 & $[-48.03]$ & $[-44.77]$ & $[0.58]$ & $[115.33]$ & $[338.28]$ & $[494.57]$ & $[2200.00]$ \\
\addlinespace
Scaling relations & \textit{JWST} & ($m^{\star}=16.83$) & & & & $\mathcal{U}(100,\,450)$ & $[0.001]$ & $\mathcal{U}(10,\,150)$ \\
 & \textit{HST} & ($m^{\star}=18.83$) & & & & $\mathcal{U}(50,\,400)$ & $[0.001]$ & $\mathcal{U}(10,\,125)$ \\
\bottomrule
\end{tabular}
\begin{flushleft}\textit{Notes.} Coordinates $\Delta x,\,\Delta y$ are angular offsets from the reference position (RA, Dec) = (177.7089950, $-28.0821673$); $\Delta x$ increases West of the reference. All radii are reported in proper kpc at the lens redshift $z_{\rm lens}=0.383$ (1\,arcsec $\approx$ 5.23\,kpc). Ellipticity is $e = (a^2-b^2)/(a^2+b^2)$; the position angle $\theta$ is measured East of North. Free parameters use uniform priors $\mathcal{U}(\mathrm{min},\,\mathrm{max})$. Values in square brackets were held fixed at the listed initial value. The galaxy-scale scaling relation rows give the reference parameters, $\sigma^{\star}\equiv\sigma_{\rm LT}$ and $r_{\rm cut}^{\star}\equiv r_{\rm cut}$, at the reference magnitude $m^{\star}$ for the scaling relations $\sigma_i = \sigma^{\star}\,10^{0.4(m^{\star}-m_i)/\alpha_\sigma}$, $r_{{\rm cut},i} = r_{{\rm cut}}^{\star}\,10^{0.4(m^{\star}-m_i)/\alpha_{r_{\rm cut}}}$. The slopes $\alpha_\sigma = 0.25$ (Faber-Jackson) and $\alpha_{r_{\rm cut}} = 0.5$ are held fixed for both models.\end{flushleft}
\end{table*}

\section{Model Predicted Redshifts}
\label{sec:model_predicted_redshifts}

\begin{figure*}
	\includegraphics[width=\textwidth]{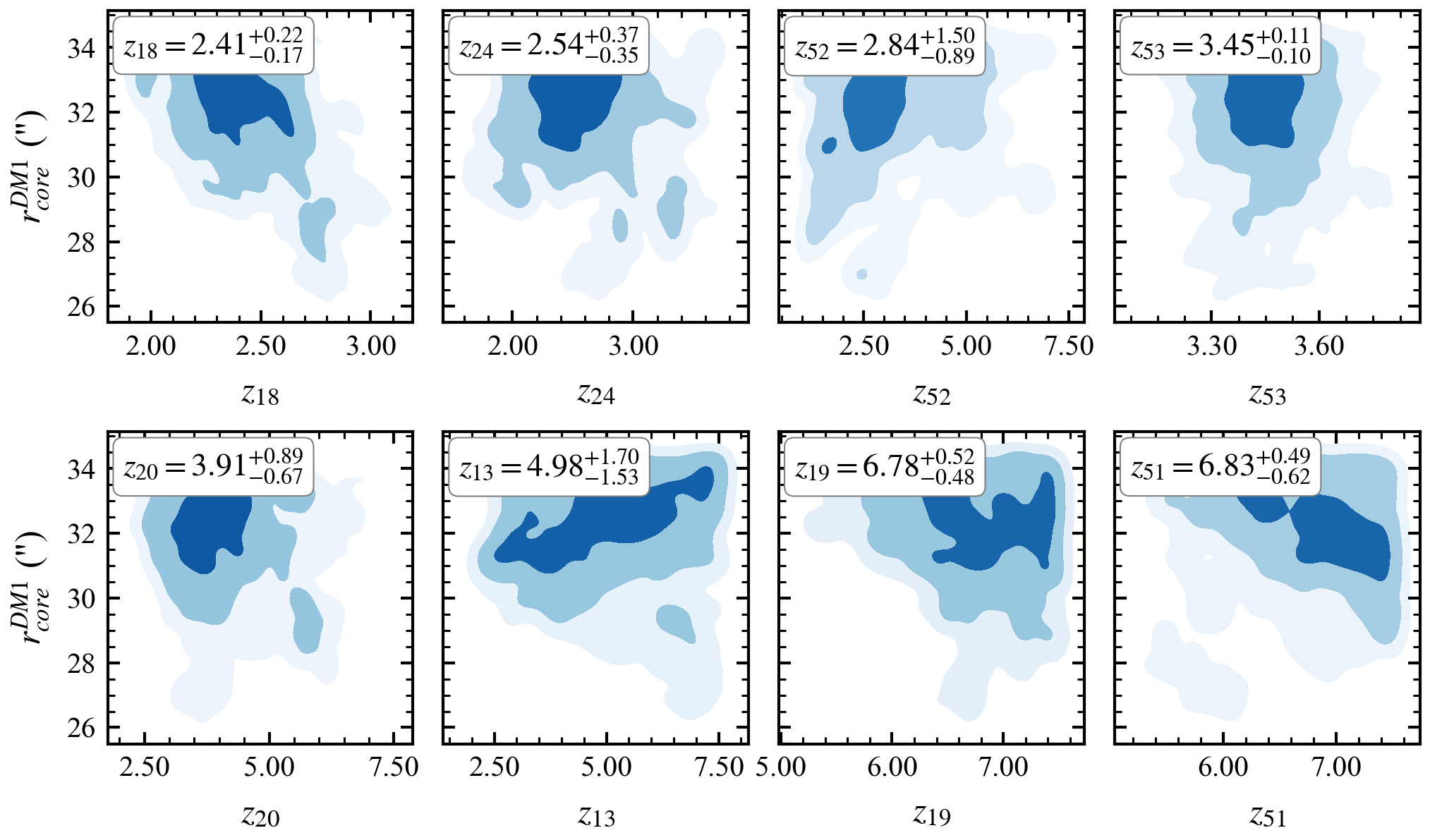}
    \caption{2D marginalised posterior probability distributions between the core radius of DM Halo 1 ($r_{core}^{DM1}$) and the inferred redshifts of all photometric sources optimised in the Phot model. The dark, medium, and light blue shaded regions correspond to the $1\sigma$ ($68\%$), 2$\sigma$ ($95\%$), and 3$\sigma$ ($99.7\%$) confidence intervals, respectively. The median redshift value along with its $1\sigma$ uncertainties are reported in the top left of each panel.}
    \label{fig:redshift_corner_plot}
\end{figure*}

In the Phot model we optimise the redshifts of a total of $8$ systems ($31$ multiple images) that do not have reliable spectroscopic measurements associated with them. In Fig. \ref{fig:redshift_corner_plot} we show the model predicted median redshift values for these systems paired with the 2D marginalised posterior probability distributions between $r_{core}^{DM1}$ and the optimised redshifts. Notably, we predict redshift values for two high-redshift lensed candidates. These are systems 19 and 51. System 19 was previously identified in \citet[][]{Salmon_2020} during their search for high-redshift objects in the RELICS survey. They measured a photometric redshift of $6.8_{-0.2}^{+0.4}$, which our model exactly replicates with a median predicted redshift value from the Phot model of $z_{19} = 6.78_{-0.48}^{+0.52}$. The system consists of 5 multiple images spread across the image-plane. Our model does not predict further multiple image locations. The other high-redshift system (system 51) has been newly identified within this work thanks to the depth of the \textit{JWST} NIRCam SLICE imaging and the predictive power of the model. As shown in Table \ref{tab:multiple_images}, this system receives very high magnification from the lensing field of the cluster. The presence of such interesting systems, once again, establishes J1150$-$2805 as an excellent target for high-redshift studies.

\section{Multiple Images}
\label{sec:multiple_images_appendix}

We provide the full multiple image catalogue in this section where the details of the images are in Table \ref{tab:multiple_images}. Images in bold denote new additions to the catalogue as a result of this work. We provide magnification and parity information, as well as image-plane RMS information for each image separately if they have been included in the modelling.

\begin{table*}
\centering
\caption{Catalogue of lensed multiple images for SPT-CLJ1150$-$2805. Identifiers follow the SLICE naming convention (system.clump.image). Columns 2 and 3 provide the RA and Dec. of each multiple image, while column 4 provides the corresponding spectroscopic redshifts from the \citet{DAddona_2024} catalogue. Multiple images are ranked based on the three-tier Gold, Silver, and Bronze system defined in Section \ref{sec:multiple_images}. System IDs in bold denote multiple images newly discovered in this work. Magnification, $\mu$, and positional image-plane RMS are model outputs for the images used as constraints. Reported magnifications include image parity, where the absolute value denotes the physical magnification factor and a negative value indicates an image with inverted parity. Quoted uncertainties represent the $1\sigma$ credible intervals derived from the posterior distribution.}
\label{tab:multiple_images}
\begin{tabular*}{\textwidth}{@{\extracolsep{\fill}} lcccccccc @{}}
\toprule
ID & R.A. (J2000) & Dec. (J2000) & $z$ & Ranking & $\mu_{\rm Spec}$ & $\mu_{\rm Phot}$ &  $\mathrm{RMS_{Spec}}$ ($''$) & $\mathrm{RMS_{Phot}} ($''$)$\\
\midrule
1.1.1 & 177.706016 & -28.083892 & 3.768 & Gold & $42.7^{+11.9}_{-9.2}$ & $62.9^{+28.4}_{-7.0}$ & 0.48 & 0.72 \\
1.1.2 & 177.704382 & -28.084657 & 3.768 & Gold & $-56.5^{+1.8}_{-13.4}$ & $-53.7^{+3.0}_{-3.9}$ & 0.47 & 0.89\\
c1.1.3 & 177.703703 & -28.085055 & 3.768 & Bronze & -- & -- & -- & --\\
\textbf{c1.1.4} & 177.724367 & -28.070453 & 3.768 & Bronze & -- & -- & -- & --\\
2.1.1 & 177.725834 & -28.097792 & 1.720 & Gold & $35.7^{+28.5}_{-4.6}$ & $30.4^{+9.5}_{-14.5}$ & 0.09 & 0.05\\
2.1.2 & 177.725230 & -28.098367 & 1.720 & Gold & $-50.8^{+13.2}_{-14.9}$ & $-41.6^{+1.0}_{-11.9}$ & 0.36 & 0.32\\
2.1.3 & 177.720165 & -28.100895 & 1.720 & Gold & $14.4^{+2.8}_{-3.6}$ & $13.5^{+3.8}_{-1.8}$ & 0.08 & 0.11\\
2.1.4 & 177.720356 & -28.101080 & 1.720 & Bronze & -- & -- & -- & --\\
2.2.1 & 177.725755 & -28.097865 & 1.720 & Gold & $56.0^{+120.9}_{-7.4}$ & $48.0^{+307.3}_{-47.2}$ & 0.13 & 0.07\\
2.2.2 & 177.725314 & -28.098289 & 1.720 & Gold & $-59.3^{+17.7}_{-21.2}$ & $-50.1^{+1.4}_{-14.6}$ & 0.21 & 0.22\\
\textbf{2.2.3} & 177.720090 & -28.100924 & 1.720 & Gold & $13.6^{+2.4}_{-3.2}$ & $12.9^{+3.3}_{-1.6}$ & 0.16 & 0.04\\
2.3.1 & 177.725643 & -28.097973 & 1.720 & Gold & $77.1^{+14.0}_{-328.3}$ & $70.0^{+399.7}_{-313.8}$ & 0.17 & 0.04\\
2.3.2 & 177.725377 & -28.098233 & 1.720 & Gold & $-81.6^{+30.3}_{-42.3}$ & $-68.5^{+2.3}_{-19.1}$ & 0.12 & 0.04\\
2.3.3 & 177.720017 & -28.100960 & 1.720 & Gold & $13.4^{+2.2}_{-3.1}$ & $12.7^{+3.2}_{-1.6}$ & 0.32 & 0.14\\
2.4.1 & 177.726020 & -28.097592 & 1.720 & Gold & $23.5^{+10.6}_{-2.5}$ & $19.8^{+4.0}_{-5.7}$ & 0.14 & 0.08\\
2.4.2 & 177.724994 & -28.098572 & 1.720 & Gold & $-30.4^{+4.9}_{-6.1}$ & $-25.5^{+0.6}_{-6.8}$ & 0.50 & 0.38\\
2.4.3 & 177.720308 & -28.100841 & 1.720 & Gold & $15.3^{+3.4}_{-4.1}$ & $14.2^{+4.4}_{-2.0}$ & 0.07 & 0.09\\
3.1.1 & 177.714773 & -28.094305 & 2.005 & Gold & $-45.3^{+3.5}_{-5.5}$ & $-40.0^{+3.4}_{-4.9}$ & 0.34 & 0.25\\
3.1.2 & 177.713296 & -28.095074 & 2.005 & Gold & $47.0^{+34.8}_{-24.2}$ & $46.0^{+10.4}_{-22.7}$ & 0.45 & 0.31\\
3.2.1 & 177.714285 & -28.094495 & 2.005 & Gold & $-175^{+8}_{-34}$ & $-129^{+10}_{-12}$ & 0.36 & 0.28\\
3.2.2 & 177.713594 & -28.094851 & 2.005 & Gold & $82.7^{+424.7}_{-728.9}$ & $72.7^{+253.9}_{-156.9}$ & 0.64 & 0.46\\
c4.1.1 & 177.711816 & -28.095223 & -- & Bronze & -- & -- & -- & --\\
c4.1.2 & 177.713592 & -28.094857 & -- & Bronze & -- & -- & -- & --\\
c4.1.3 & 177.726928 & -28.084564 & -- & Bronze & -- & -- & -- & --\\
\textbf{c4.1.4} & 177.711001 & -28.095789 & -- & Bronze & -- & -- & -- & --\\
5.1.1 & 177.712031 & -28.075342 & 1.409 & Gold & $-14.3^{+7.3}_{-5.4}$ & $-25.1^{+1.7}_{-6.4}$ & 0.51 & 0.14\\
5.1.2 & 177.713760 & -28.075280 & 1.409 & Gold & $17.6^{+1.6}_{-1.7}$ & $17.6^{+1.7}_{-1.2}$ & 0.49 & 1.47\\
5.1.3 & 177.696844 & -28.087799 & 1.409 & Gold & $120^{+28}_{-150}$ & $50.8^{+51.1}_{-20.3}$ & 0.70 & 0.17\\
5.1.4 & 177.716168 & -28.081573 & 1.409 & Gold & $-6.20^{+0.44}_{-0.22}$ & $-5.62^{+0.37}_{-0.61}$ & 0.64 & 0.76\\
5.2.1 & 177.711050 & -28.075333 & 1.409 & Gold & $-7.89^{+3.37}_{-2.59}$ & $-9.63^{+1.01}_{-0.63}$ & 1.12 & 0.41\\
5.2.2 & 177.714419 & -28.075068 & 1.409 & Gold & $21.6^{+6.4}_{-2.4}$ & $26.3^{+5.0}_{-2.9}$ & 0.74 & 0.22\\
5.2.3 & 177.696864 & -28.086842 & 1.409 & Gold & $16.0^{+3.1}_{-1.4}$ & $16.1^{+2.1}_{-1.0}$ & 0.31 & 0.39\\
5.2.4 & 177.716170 & -28.081908 & 1.409 & Gold & $-7.41^{+0.30}_{-0.13}$ & $-6.57^{+0.36}_{-0.55}$ & 0.69 & 0.44\\
6.1.1 & 177.697056 & -28.078136 & 4.347 & Gold & $-13.4^{+1.1}_{-1.3}$ & $-12.0^{+0.4}_{-0.6}$ & 0.31 & 0.25\\
6.1.2 & 177.695811 & -28.081732 & 4.347 & Gold & $17.3^{+0.8}_{-3.0}$ & $19.9^{+1.5}_{-1.8}$ & 0.43 & 0.39\\
6.1.3 & 177.697514 & -28.089132 & 4.347 & Gold & $-17.4^{+0.5}_{-1.1}$ & $-22.8^{+0.6}_{-1.3}$ & 0.92 & 0.27\\
6.1.4 & 177.700447 & -28.089029 & 4.347 & Gold & $-9.77^{+0.40}_{-0.89}$ & $-6.81^{+1.00}_{-1.34}$ & 0.51 & 0.59\\
\textbf{6.1.5} & 177.699780 & -28.093653 & 4.347 & Gold & $14.9^{+1.8}_{-0.6}$ & $16.8^{+0.8}_{-0.4}$ & 0.56 & 0.33\\
6.1.6 & 177.710729 & -28.090203 & 4.347 & Gold & $-9.76^{+0.16}_{-0.56}$ & $-9.99^{+0.15}_{-0.22}$ & 0.26 & 0.58\\
\textbf{6.1.7} & 177.723308 & -28.071080 & 4.347 & Gold & $4.19^{+0.20}_{-0.10}$ & $4.00^{+0.11}_{-0.09}$ & 0.28 & 0.56\\
\textbf{6.2.1} & 177.696802 & -28.078544 & 4.347 & Gold & $-4.99^{+0.17}_{-0.15}$ & $-3.83^{+0.04}_{-0.07}$ & 0.38 & 0.46\\
\textbf{6.2.2} & 177.695903 & -28.080976 & 4.347 & Gold & $20.2^{+1.5}_{-4.5}$ & $22.3^{+2.2}_{-1.8}$ & 0.57 & 0.49\\
\textbf{6.2.3} & 177.697612 & -28.089452 & 4.347 & Gold & $-13.6^{+0.5}_{-0.9}$ & $-15.8^{+0.9}_{-0.4}$ & 0.91 & 0.57\\
\textbf{6.2.4} & 177.700649 & -28.088866 & 4.347 & Gold & $-12.7^{+0.5}_{-1.0}$ & $-10.2^{+1.1}_{-1.5}$ & 0.61 & 0.89\\
\textbf{6.2.5} & 177.699920 & -28.093531 & 4.347 & Gold & $16.1^{+2.1}_{-0.5}$ & $18.0^{+0.8}_{-0.5}$ & 0.44 & 0.40\\
\textbf{6.2.6} & 177.710521 & -28.090189 & 4.347 & Gold & $-10.3^{+0.2}_{-0.6}$ & $-10.5^{+0.2}_{-0.2}$ & 0.43 & 0.75\\
\textbf{6.2.7} & 177.723345 & -28.070971 & 4.347 & Gold & $4.46^{+0.22}_{-0.10}$ & $4.29^{+0.13}_{-0.09}$ & 0.40 & 0.50\\
\textbf{6.3.1} & 177.696955 & -28.078405 & 4.347 & Gold & $-20.3^{+1.7}_{-2.5}$ & $-21.3^{+0.7}_{-1.3}$ & 0.21 & 0.22\\
\textbf{6.3.2} & 177.695923 & -28.081548 & 4.347 & Gold & $18.6^{+0.8}_{-3.8}$ & $21.3^{+1.8}_{-1.9}$ & 0.51 & 0.43\\
\textbf{6.3.3} & 177.697671 & -28.089266 & 4.347 & Gold & $-16.2^{+0.4}_{-1.1}$ & $-20.0^{+0.8}_{-1.0}$ & 0.87 & 0.31\\
\textbf{6.3.4} & 177.700449 & -28.088813 & 4.347 & Gold & $-11.8^{+0.5}_{-0.9}$ & $-8.95^{+1.03}_{-1.37}$ & 0.59 & 0.83\\
\textbf{6.3.5} & 177.700021 & -28.093617 & 4.347 & Gold & $15.8^{+2.0}_{-0.5}$ & $17.6^{+0.8}_{-0.5}$ & 0.61 & 0.45\\
\textbf{6.3.6} & 177.710543 & -28.090275 & 4.347 & Gold & $-10.3^{+0.2}_{-0.6}$ & $-10.6^{+0.2}_{-0.2}$ & 0.25 & 0.54\\
\textbf{6.3.7} & 177.723407 & -28.071075 & 4.347 & Gold & $4.16^{+0.20}_{-0.09}$ & $3.98^{+0.11}_{-0.09}$ & 0.33 & 0.52\\
\bottomrule
\end{tabular*}
\end{table*}

\begin{table*}
\centering
\contcaption{Catalogue of lensed multiple images for SPT-CLJ1150$-$2805.}
\begin{tabular*}{\textwidth}{@{\extracolsep{\fill}} lcccccccc @{}}
\toprule
ID & R.A. (J2000) & Dec. (J2000) & $z$ & Ranking & $\mu_{\rm Spec}$ & $\mu_{\rm Phot}$ &  $\mathrm{RMS_{Spec}}$ ($''$) & $\mathrm{RMS_{Phot}} ($''$)$\\
\midrule
7.1.1 & 177.705204 & -28.073224 & 3.342 & Gold & $-9.76^{+0.47}_{-0.73}$ & $-10.2^{+0.5}_{-0.8}$ & 0.19 & 0.18\\
7.1.2 & 177.715415 & -28.088051 & 3.342 & Gold & $-5.91^{+0.24}_{-0.14}$ & $-6.06^{+0.09}_{-0.21}$ & 0.12 & 0.03\\
7.1.3 & 177.695742 & -28.094908 & 3.342 & Gold & $5.30^{+0.29}_{-0.22}$ & $5.60^{+0.19}_{-0.13}$ & 0.15 & 0.83\\
7.1.4 & 177.720306 & -28.073212 & 3.342 & Gold & $10.6^{+5.8}_{-5.1}$ & $8.52^{+0.61}_{-1.44}$ & 0.17 & 0.39\\
7.2.1 & 177.705270 & -28.073155 & 3.342 & Gold & $-10.1^{+0.5}_{-0.8}$ & $-10.7^{+0.6}_{-0.9}$ & 0.27 & 0.19\\
7.2.2 & 177.715436 & -28.087973 & 3.342 & Gold & $-5.96^{+0.24}_{-0.15}$ & $-6.12^{+0.09}_{-0.21}$ & 0.15 & 0.03\\
7.2.3 & 177.695641 & -28.094883 & 3.342 & Gold & $5.34^{+0.29}_{-0.23}$ & $5.65^{+0.20}_{-0.13}$ & 0.13 & 0.75\\
7.2.4 & 177.720219 & -28.073158 & 3.342 & Gold & $7.44^{+136.41}_{-22.78}$ & $6.78^{+0.25}_{-0.51}$ & 0.22 & 0.43\\
7.3.1 & 177.705153 & -28.073302 & 3.342 & Gold & $-9.65^{+0.48}_{-0.71}$ & $-10.0^{+0.5}_{-0.8}$ & 0.45 & 0.26\\
7.3.2 & 177.715408 & -28.088127 & 3.342 & Gold & $-5.77^{+0.24}_{-0.14}$ & $-5.89^{+0.09}_{-0.21}$ & 0.17 & 0.14\\
7.3.3 & 177.695838 & -28.094946 & 3.342 & Gold & $5.38^{+0.30}_{-0.22}$ & $5.70^{+0.18}_{-0.13}$ & 0.33 & 0.89\\
7.3.4 & 177.720906 & -28.073868 & 3.342 & Gold & $13.2^{+2.0}_{-1.2}$ & $9.44^{+1.65}_{-1.64}$ & 1.03 & 1.10\\
\textbf{7.4.1} & 177.705206 & -28.073279 & 3.342 & Gold & $-9.65^{+0.48}_{-0.71}$ & $-10.0^{+0.5}_{-0.8}$ & 0.58 & 0.34\\
\textbf{7.4.2} & 177.715437 & -28.088104 & 3.342 & Gold & $-5.83^{+0.24}_{-0.14}$ & $-5.97^{+0.09}_{-0.21}$ & 0.28 & 0.20\\
\textbf{7.4.3} & 177.695800 & -28.094952 & 3.342 & Gold & $5.34^{+0.30}_{-0.22}$ & $5.65^{+0.19}_{-0.13}$ & 0.44 & 0.95\\
\textbf{7.4.4} & 177.720839 & -28.073850 & 3.342 & Gold & $20.3^{+1.5}_{-0.8}$ & $11.9^{+3.6}_{-3.2}$ & 1.02 & 1.16\\
\textbf{7.5.1} & 177.705135 & -28.073223 & 3.342 & Gold & $-9.73^{+0.44}_{-0.73}$ & $-10.1^{+0.5}_{-0.8}$ & 0.12 & 0.42\\
\textbf{7.5.2} & 177.715367 & -28.088067 & 3.342 & Gold & $-5.84^{+0.24}_{-0.14}$ & $-5.98^{+0.09}_{-0.21}$ & 0.13 & 0.29\\
\textbf{7.5.3} & 177.695772 & -28.094884 & 3.342 & Gold & $5.41^{+0.30}_{-0.23}$ & $5.74^{+0.18}_{-0.13}$ & 0.16 & 1.04\\
\textbf{7.5.4} & 177.720408 & -28.073384 & 3.342 & Gold & $-7.07^{+0.72}_{-0.41}$ & $-13.8^{+21.3}_{-4.7}$ & 0.25 & 0.17\\
8.1.1 & 177.719749 & -28.080098 & 1.169 & Gold & $53.8^{+6.9}_{-5.3}$ & $64.2^{+78.2}_{-20.0}$ & 0.42 & 0.30\\
8.1.2 & 177.713682 & -28.086919 & 1.169 & Gold & $-16.5^{+1.6}_{-0.5}$ & $-18.3^{+1.5}_{-1.6}$ & 0.48 & 0.26\\
8.1.3 & 177.706619 & -28.089447 & 1.169 & Gold & $11.6^{+0.7}_{-0.2}$ & $11.5^{+0.4}_{-0.3}$ & 1.07 & 0.13\\
\textbf{8.2.1} & 177.719790 & -28.080699 & 1.169 & Gold & $19.3^{+1.4}_{-2.0}$ & $18.4^{+3.2}_{-1.8}$ & 0.12 & 0.30\\
\textbf{8.2.2} & 177.713996 & -28.086845 & 1.169 & Gold & $-18.2^{+1.7}_{-0.6}$ & $-20.5^{+1.7}_{-1.8}$ & 0.47 & 0.15\\
\textbf{8.2.3} & 177.706555 & -28.089597 & 1.169 & Gold & $11.1^{+0.7}_{-0.2}$ & $11.0^{+0.3}_{-0.3}$ & 0.53 & 0.47\\
10.1.1 & 177.687251 & -28.083995 & 1.409 & Gold & $8.50^{+5.20}_{-0.80}$ & $7.79^{+2.01}_{-0.93}$ & 0.21 & 0.14\\
10.1.2 & 177.687200 & -28.084484 & 1.409 & Gold & $-14.3^{+3.2}_{-2.4}$ & $-9.99^{+2.29}_{-3.55}$ & 0.05 & 0.05\\
10.1.3 & 177.687456 & -28.085151 & 1.409 & Gold & $7.59^{+1.25}_{-0.64}$ & $6.59^{+0.45}_{-0.14}$ & 0.60 & 0.43\\
11.1.1 & 177.683808 & -28.085618 & 1.329 & Gold & $5.92^{+0.21}_{-0.49}$ & $6.83^{+0.57}_{-0.23}$ & 0.47 & 0.21\\
11.1.2 & 177.683686 & -28.085959 & 1.329 & Gold & $13.1^{+11.3}_{-5.4}$ & $32.5^{+7.8}_{-2.9}$ & 0.11 & 0.47\\
11.1.3 & 177.683583 & -28.086275 & 1.329 & Gold & $-21.9^{+3.0}_{-2.3}$ & $-8.20^{+1.98}_{-2.66}$ & 0.33 & 0.11\\
11.1.4 & 177.684146 & -28.088168 & 1.329 & Gold & $3.77^{+0.11}_{-0.11}$ & $3.78^{+0.18}_{-0.05}$ & 0.29 & 0.45\\
12.1.1 & 177.682085 & -28.088558 & 3.079 & Gold & $4.64^{+0.25}_{-0.35}$ & $4.23^{+0.28}_{-0.19}$ & 0.58 & 0.25\\
12.1.2 & 177.681641 & -28.087183 & 3.079 & Gold & $-8.37^{+0.70}_{-1.94}$ & $-3.17^{+1.54}_{-1.95}$ & 0.18 & 0.07\\
12.1.3 & 177.681651 & -28.087088 & 3.079 & Gold & $56.3^{+56.7}_{-27.3}$ & $16.3^{+5.2}_{-11.9}$ & 0.09 & 0.21\\
13.1.1 & 177.728764 & -28.070278 & -- & Silver & -- & $4.25^{+0.27}_{-0.20}$ & -- & 0.59\\
13.1.2 & 177.730353 & -28.071949 & -- & Silver & -- & $19.0^{+7.8}_{-2.8}$ & -- & 0.45\\
13.1.3 & 177.730665 & -28.072040 & -- & Silver & -- & $-3.38^{+0.64}_{-0.38}$ & -- & 0.24\\
13.1.4 & 177.733053 & -28.073812 & -- & Bronze & -- & -- & -- & --\\
\textbf{13.2.1} & 177.729112 & -28.070577 & -- & Silver & -- & $4.70^{+0.39}_{-0.20}$ & -- & 0.39\\
\textbf{13.2.2} & 177.730174 & -28.071887 & -- & Silver & -- & $6.81^{+1.27}_{-0.11}$ & -- & 0.23\\
\textbf{13.2.3} & 177.730932 & -28.072196 & -- & Silver & -- & $-1.95^{+0.31}_{-0.17}$ & -- & 0.63\\
\textbf{13.2.4} & 177.732683 & -28.073635 & -- & Bronze & -- & -- & -- & --\\
c14.1.1 & 177.732283 & -28.067779 & -- & Bronze & -- & -- & -- & --\\
c14.1.2 & 177.732769 & -28.068307 & -- & Bronze & -- & -- & -- & --\\
c14.1.3 & 177.733773 & -28.068697 & -- & Bronze & -- & -- & -- & --\\
c15.1.1 & 177.710272 & -28.065197 & -- & Bronze & -- & -- & -- & -- \\
\textbf{c15.1.2} & 177.709311 & -28.065263 & -- & Bronze & -- & -- & -- & --\\
c16.1.1 & 177.724757 & -28.104880 & -- & Bronze & -- & -- & -- & --\\
c16.1.2 & 177.724625 & -28.104961 & -- & Bronze & -- & -- & -- & --\\
c16.1.3 & 177.723253 & -28.105424 & -- & Bronze & -- & -- & -- & --\\
c17.1.1 & 177.688896 & -28.085733 & 1.060 & Bronze & -- & -- & -- & -- \\
c17.1.2 & 177.688877 & -28.082965 & 1.060 & Bronze & -- & -- & -- & -- \\
18.1.1 & 177.688634 & -28.087687 & -- & Silver & -- & $-12.3^{+3.7}_{-2.6}$ & -- & 1.00\\
18.1.2 & 177.688910 & -28.086413 & -- & Silver & -- & $16.8^{+4.1}_{-1.3}$ & -- & 0.18\\
18.1.3 & 177.688565 & -28.089303 & -- & Silver & -- & $7.30^{+1.39}_{-0.50}$ & -- & 0.42\\
\textbf{18.2.1} & 177.688618 & -28.087890 & -- & Silver & -- & $-12.9^{+3.9}_{-2.7}$ & -- & 0.52\\
\textbf{18.2.2} & 177.688952 & -28.086269 & -- & Silver & -- & $13.7^{+2.4}_{-0.8}$ & -- & 0.11\\
\textbf{18.2.3} & 177.688573 & -28.089162 & -- & Silver & -- & $8.51^{+2.10}_{-0.66}$ & -- & 1.06\\
\bottomrule
\end{tabular*}
\end{table*}

\begin{table*}
\centering
\contcaption{Catalogue of lensed multiple images for SPT-CLJ1150$-$2805.}
\begin{tabular*}{\textwidth}{@{\extracolsep{\fill}} lcccccccc @{}}
\toprule
ID & R.A. (J2000) & Dec. (J2000) & $z$ & Ranking & $\mu_{\rm Spec}$ & $\mu_{\rm Phot}$ &  $\mathrm{RMS_{Spec}}$ ($''$) & $\mathrm{RMS_{Phot}} ($''$)$\\
\midrule
19.1.1 & 177.704985 & -28.070739 & -- & Silver & -- & $-5.56^{+0.15}_{-0.33}$ & -- & 1.16\\
19.1.2 & 177.719912 & -28.071548 & -- & Silver & -- & $5.41^{+0.16}_{-0.08}$ & -- & 0.62\\
19.1.3 & 177.716374 & -28.087856 & -- & Silver & -- & $-5.81^{+0.07}_{-0.18}$ & -- & 0.29\\
19.1.4 & 177.692520 & -28.095865 & -- & Silver & -- & $4.33^{+0.32}_{-0.08}$ & -- & 0.31\\
19.1.5 & 177.710156 & -28.081649 & -- & Silver & -- & $2.89^{+0.22}_{-0.17}$ & -- & 0.71\\
20.1.1 & 177.710478 & -28.083575 & -- & Silver & -- & $36.9^{+3.8}_{-5.8}$ & -- & 0.12\\
20.1.2 & 177.710956 & -28.084205 & -- & Silver & -- & $-36.8^{+21.9}_{-56.1}$ & -- & 0.19\\
22.1.1 & 177.706590 & -28.080849 & 2.932 & Gold & $23.5^{+9.7}_{-9.9}$ & $-51.6^{+16.5}_{-40.5}$ & 0.23 & 0.77\\
22.1.2 & 177.707795 & -28.093930 & 2.932 & Gold & $29.8^{+3.1}_{-4.1}$ & $29.8^{+2.1}_{-3.2}$ & 1.74 & 1.53\\
22.1.3 & 177.711264 & -28.092271 & 2.932 & Gold & $-14.1^{+0.5}_{-0.9}$ & $-13.1^{+0.3}_{-0.2}$ & 0.94 & 1.09\\
c22.1.4 & 177.7069520 & -28.0810990 & 2.932 & Silver & -- & -- & -- & -- \\
c22.1.5 & 177.7071200 & -28.0812050 & 2.932 & Silver & -- & -- & -- & --\\
24.1.1 & 177.708240 & -28.068160 & -- & Silver & -- & $51.3^{+184.4}_{-440.7}$ & -- & 0.07\\
24.1.2 & 177.707212 & -28.068375 & -- & Silver & -- & $-6.86^{+0.17}_{-0.59}$ & -- & 0.15\\
35.1.1 & 177.6836779 & -28.0863022 & 5.467 & Silver & -- & -- & -- & -- \\
35.1.2 & 177.6854785 & -28.0863885 & 5.467 & Silver & -- & -- & -- & -- \\
35.1.3 & 177.6845906 & -28.0838651 & 5.467 & Silver & -- & -- & -- & -- \\
35.1.4 & 177.6850326 & -28.0902189 & 5.467 & Silver & -- & -- & -- & -- \\
36.1.1 & 177.722052 & -28.089918 & 5.391 & Gold & $-8.97^{+0.26}_{-0.74}$ & $-7.39^{+0.86}_{-0.63}$ & 1.05 & 0.83\\
36.1.2 & 177.724230 & -28.087357 & 5.391 & Gold & $14.8^{+1.0}_{-1.0}$ & $14.6^{+1.4}_{-1.2}$ & 0.95 & 0.71\\
37.1.1 & 177.700820 & -28.077693 & 5.184 & Bronze & -- & -- & -- & --\\
37.1.2 & 177.703351 & -28.079865 & 5.184 & Gold & $-50.3^{+5.9}_{-17.6}$ & $-102^{+16}_{-76}$ & 0.25 & 0.54\\
37.1.3 & 177.704574 & -28.080340 & 5.184 & Gold & $930^{+336}_{-190}$ & $-58.3^{+12.9}_{-10.2}$ & 0.12 & 0.58\\
37.1.4 & 177.711266 & -28.093481 & 5.184 & Bronze & -- & -- & -- & --\\
38.1.1 & 177.723432 & -28.083088 & 3.567 & Gold & $15.3^{+2.2}_{-1.4}$ & $17.6^{+1.2}_{-2.5}$ & 1.24 & 0.63\\
38.1.2 & 177.719720 & -28.087612 & 3.567 & Gold & $-10.9^{+0.4}_{-0.7}$ & $-11.0^{+0.5}_{-0.5}$ & 0.37 & 0.23\\
38.1.3 & 177.708331 & -28.076208 & 3.567 & Gold & $-8.32^{+0.69}_{-1.56}$ & $-9.28^{+0.82}_{-1.31}$ & 0.51 & 0.27\\
38.1.4 & 177.709749 & -28.079203 & 3.567 & Gold & $6.63^{+0.68}_{-0.74}$ & $7.31^{+0.59}_{-0.61}$ & 0.72 & 0.67\\
39.1.1 & 177.708953 & -28.072944 & 2.931 & Gold & $-1.87^{+0.87}_{-1.69}$ & $-7.15^{+3.56}_{-2.59}$ & 0.17 & 0.07\\
39.1.2 & 177.709196 & -28.074017 & 2.931 & Gold & $-4.15^{+0.64}_{-0.65}$ & $-4.33^{+1.06}_{-0.72}$ & 0.69 & 0.47\\
39.1.3 & 177.712560 & -28.080287 & 2.931 & Gold & $3.58^{+1.23}_{-0.83}$ & $4.67^{+1.92}_{-0.68}$ & 0.16 & 0.51\\
39.1.4 & 177.719036 & -28.085032 & 2.931 & Gold & $-12.8^{+0.4}_{-0.6}$ & $-12.6^{+0.2}_{-0.3}$ & 0.20 & 0.45\\
39.1.5 & 177.693896 & -28.095507 & 2.931 & Gold & $3.84^{+0.13}_{-0.16}$ & $4.17^{+0.27}_{-0.06}$ & 1.20 & 1.22\\
\textbf{39.1.6} & 177.718108 & -28.074601 & 2.931 & Bronze & -- & -- & -- & --\\
40.1.1 & 177.699102 & -28.074610 & 4.334 & Gold & $-9.44^{+0.30}_{-0.59}$ & $-10.3^{+0.2}_{-0.9}$ & 0.23 & 0.29\\
40.1.2 & 177.694214 & -28.082444 & 4.334 & Gold & $10.9^{+0.9}_{-0.8}$ & $12.5^{+0.5}_{-1.2}$ & 1.09 & 1.18\\
40.1.3 & 177.694944 & -28.088317 & 4.334 & Gold & $-7.91^{+0.20}_{-1.08}$ & $-5.92^{+0.17}_{-0.34}$ & 1.14 & 1.18\\
40.1.4 & 177.696382 & -28.093203 & 4.334 & Gold & $11.1^{+1.1}_{-0.8}$ & $12.3^{+0.3}_{-0.3}$ & 0.50 & 0.58\\
40.1.5 & 177.712167 & -28.088760 & 4.334 & Gold & $-6.01^{+0.17}_{-0.19}$ & $-6.04^{+0.15}_{-0.16}$ & 0.34 & 0.52\\
41.1.1 & 177.707796 & -28.073011 & 4.517 & Gold & $-8.01^{+0.23}_{-0.68}$ & $-6.66^{+0.18}_{-0.32}$ & 0.16 & 0.57\\
41.1.2 & 177.710593 & -28.080308 & 4.517 & Gold & $3.27^{+0.32}_{-0.27}$ & $3.38^{+0.20}_{-0.34}$ & 0.29 & 0.32\\
41.1.3 & 177.718442 & -28.087553 & 4.517 & Gold & $-8.26^{+0.26}_{-0.43}$ & $-8.93^{+0.29}_{-0.30}$ & 0.23 & 0.55\\
42.1.1 & 177.694041 & -28.088634 & 4.877 & Gold & $-16.6^{+1.9}_{-2.2}$ & $-14.2^{+1.6}_{-1.7}$ & 0.64 & 0.48\\
42.1.2 & 177.693439 & -28.082370 & 4.877 & Gold & $10.2^{+0.9}_{-0.5}$ & $11.5^{+0.5}_{-1.1}$ & 1.11 & 1.00\\
42.1.3 & 177.695204 & -28.093087 & 4.877 & Gold & $9.92^{+0.89}_{-0.77}$ & $11.0^{+0.3}_{-0.2}$ & 1.03 & 1.07\\
\textbf{42.1.4} & 177.712588 & -28.088321 & 4.877 & Bronze & -- & -- & -- & --\\
43.1.1 & 177.696311 & -28.076989 & 5.975 & Gold & $-29.7^{+6.6}_{-6.2}$ & $-32.5^{+2.6}_{-3.8}$ & 0.57 & 0.23\\
43.1.2 & 177.694610 & -28.080284 & 5.975 & Gold & $14.4^{+1.0}_{-1.9}$ & $15.6^{+1.0}_{-1.0}$ & 0.80 & 0.82\\
43.1.3 & 177.695484 & -28.090450 & 5.975 & Gold & $-70.8^{+26.3}_{-12.2}$ & $-117^{+45}_{-201}$ & 0.44 & 0.47\\
43.1.4 & 177.696713 & -28.093145 & 5.975 & Gold & $13.8^{+1.6}_{-1.1}$ & $15.5^{+0.4}_{-0.4}$ & 0.80 & 0.79\\
44.1.1 & 177.700784 & -28.081167 & 3.460 & Bronze & -- & -- & -- & --\\
44.1.2 & 177.697979 & -28.080816 & 3.460 & Bronze & -- & -- & -- & --\\
c45.1.1 & 177.707602 & -28.093686 & 0.604 & Bronze & -- & -- & -- & --\\
c45.1.2 & 177.707333 & -28.093069 & 0.604 & Bronze & -- & -- & -- & --\\
47.1.1 & 177.724811 & -28.084751 & 4.132 & Bronze & -- & -- & -- & --\\
47.1.2 & 177.724551 & -28.085286 & 4.132 & Bronze & -- & -- & -- & --\\
47.1.3 & 177.720899 & -28.089290 & 4.132 & Bronze & -- & -- & -- & --\\
48.1.1 & 177.712278 & -28.096467 & 3.393 & Bronze & -- & -- & -- & --\\
48.1.2 & 177.711485 & -28.096485 & 3.393 & Bronze & -- & -- & -- & --\\
\bottomrule
\end{tabular*}
\end{table*}

\begin{table*}
\centering
\contcaption{Catalogue of lensed multiple images for SPT-CLJ1150$-$2805.}
\begin{tabular*}{\textwidth}{@{\extracolsep{\fill}} lcccccccc @{}}
\toprule
ID & R.A. (J2000) & Dec. (J2000) & $z$ & Ranking & $\mu_{\rm Spec}$ & $\mu_{\rm Phot}$ &  $\mathrm{RMS_{Spec}}$ ($''$) & $\mathrm{RMS_{Phot}} ($''$)$\\
\midrule
49.1.1 & 177.726182 & -28.094406 & 4.521 & Gold & $303^{+392}_{-163}$ & $27064^{+64}_{-222}$ & 0.09 & 0.17\\
49.1.2 & 177.726002 & -28.094582 & 4.521 & Gold & $-128^{+29}_{-73}$ & $-83.3^{+251.2}_{-179.4}$ & 0.15 & 0.46\\
49.1.3 & 177.723506 & -28.096399 & 4.521 & Bronze & -- & -- & -- & --\\
49.2.1 & 177.726611 & -28.094121 & 4.521 & Gold & $53.8^{+151.3}_{-72.2}$ & $110^{+38}_{-154}$ & 0.49 & 0.15\\
49.2.2 & 177.725650 & -28.094947 & 4.521 & Gold & $-40.9^{+7.3}_{-7.0}$ & $-38.0^{+7.8}_{-51.0}$ & 0.62 & 0.88\\
\textbf{50.1.1} & 177.709354 & -28.065151 & -- & Bronze & -- & -- & -- & --\\
\textbf{50.1.2} & 177.709867 & -28.065161 & -- & Bronze & -- & -- & -- & --\\
\textbf{51.1.1} & 177.697037 & -28.080494 & -- & Silver & -- & $-147^{+17}_{-54}$ & -- & 0.16\\
\textbf{51.1.2} & 177.697012 & -28.081627 & -- & Silver & -- & $89.6^{+53.8}_{-18.4}$ & -- & 0.91\\
\textbf{52.1.1} & 177.695029 & -28.066973 & -- & Silver & -- & $-30.4^{+29.1}_{-3.4}$ & -- & 0.07\\
\textbf{52.1.2} & 177.694675 & -28.067155 & -- & Silver & -- & $23.3^{+23.2}_{-26.3}$ & -- & 0.13\\
\textbf{52.1.3} & 177.698186 & -28.064705 & -- & Bronze & -- & -- & -- & --\\
\textbf{52.2.1} & 177.694904 & -28.067047 & -- & Silver & -- & $-1706^{+43}_{-12}$ & -- & 0.11\\
\textbf{52.2.2} & 177.694814 & -28.067093 & -- & Silver & -- & $66.7^{+49.0}_{-54.2}$ & -- & 0.11\\
\textbf{52.2.3} & 177.698223 & -28.064687 & -- & Bronze & -- & -- & -- & --\\
\textbf{53.1.1} & 177.704988 & -28.073698 & -- & Silver & -- & $-9.99^{+0.50}_{-0.75}$ & -- & 0.33\\
\textbf{53.1.2} & 177.715576 & -28.088671 & -- & Silver & -- & $-4.72^{+0.06}_{-0.15}$ & -- & 0.35\\
\textbf{53.1.3} & 177.696137 & -28.095413 & -- & Silver & -- & $5.49^{+0.18}_{-0.14}$ & -- & 0.42\\
\textbf{53.1.4} & 177.721736 & -28.074376 & -- & Silver & -- & $5.54^{+0.14}_{-0.12}$ & -- & 0.49\\
\textbf{53.2.1} & 177.704947 & -28.073785 & -- & Bronze & -- & -- & -- & --\\
\textbf{53.2.2} & 177.715557 & -28.088733 & -- & Bronze & -- & -- & -- & --\\
\textbf{53.2.3} & 177.696242 & -28.095430 & -- & Bronze & -- & -- & -- & --\\
\textbf{53.2.4} & 177.721837 & -28.074433 & -- & Bronze & -- & -- & -- & --\\
\textbf{54.1.1} & 177.704299 & -28.073908 & -- & Bronze & -- & -- & -- & --\\
\textbf{54.1.2} & 177.715158 & -28.088896 & -- & Bronze & -- & -- & -- & --\\
\textbf{54.1.3} & 177.695589 & -28.095239 & -- & Bronze & -- & -- & -- & --\\
\textbf{54.1.4} & 177.721974 & -28.074152 & -- & Bronze & -- & -- & -- & --\\
\textbf{55.1.1} & 177.710369 & -28.096646 & -- & Bronze & -- & -- & -- & --\\
\textbf{55.1.2} & 177.708916 & -28.097296 & -- & Bronze & -- & -- & -- & --\\
\textbf{55.2.1} & 177.710342 & -28.096682 & -- & Bronze & -- & -- & -- & --\\
\textbf{55.2.2} & 177.708973 & -28.097305 & -- & Bronze & -- & -- & -- & --\\
\bottomrule
\end{tabular*}
\end{table*}


\bsp	
\label{lastpage}
\end{document}